\documentclass[twocolumn,10pt,a4paper,float,amsmath,amssymb,aps,pra]{revtex4-2}

\usepackage[utf8]{inputenc}

\usepackage{float}
\usepackage{graphicx}
\usepackage{dcolumn}
\usepackage{bm}
\usepackage{bbm}
\usepackage{mathrsfs}

\usepackage[dvipsnames]{xcolor}
\usepackage{hyperref}
\hypersetup{colorlinks=true,
    citecolor=BrickRed,
    linkcolor=RoyalBlue}

\usepackage{ulem}
\newcommand\redout{\bgroup\markoverwith
{\textcolor{red}{\rule[0.5ex]{2pt}{0.8pt}}}\ULon}

\begin{document}

\title{\large Origin of degenerate bound states in the continuum in a grating waveguide:\\ Parity symmetry breaking due to mode crossing}

\author{C.B. Reynolds, Vl.V. Kocharovsky, V.V. Kocharovsky\\
\textit{Department of Physics and Astronomy, Texas A\&M University, College Station, TX 77843, USA}
}
\date{\today}

\begin{abstract}
We explain the origin of bound states in the continuum (BICs) in a planar grating waveguide, in particular, a mechanism for formation of degenerate BICs, via the analytical theory of the infinite-grating eigenmodes.
Conventional symmetry-protected BICs are formed at normal incidence mainly by a single infinite-grating eigenmode that has an odd spatial parity on both sides of the BIC resonance. The odd parity is the reason for a cutoff from the radiation-loss channel and appearance of such BICs. The mechanism of emergence of a degenerate BIC in a vicinity of a degenerate frequency of two infinite-grating eigenmodes is different. The degenerate BIC is formed by an anti-phased coherent superposition of two crossing infinite-grating eigenmodes both of which possess a mixed parity and experience parity symmetry breaking as the frequency scans through the degeneracy point. In this case a cutoff from the radiation-loss channel and extremely high-Q narrow resonance is achieved due to the destructive interference of the two crossing eigenmodes.
Implementation of such a mechanism can be instructive for designing BICs in other photonic crystals and structures.
\end{abstract}

\maketitle

\section{Introduction: A bound state in the continuum as a resource for a solitary high-Q resonance}

\begin{figure}[h]
    \centering
    \includegraphics[width=80mm]{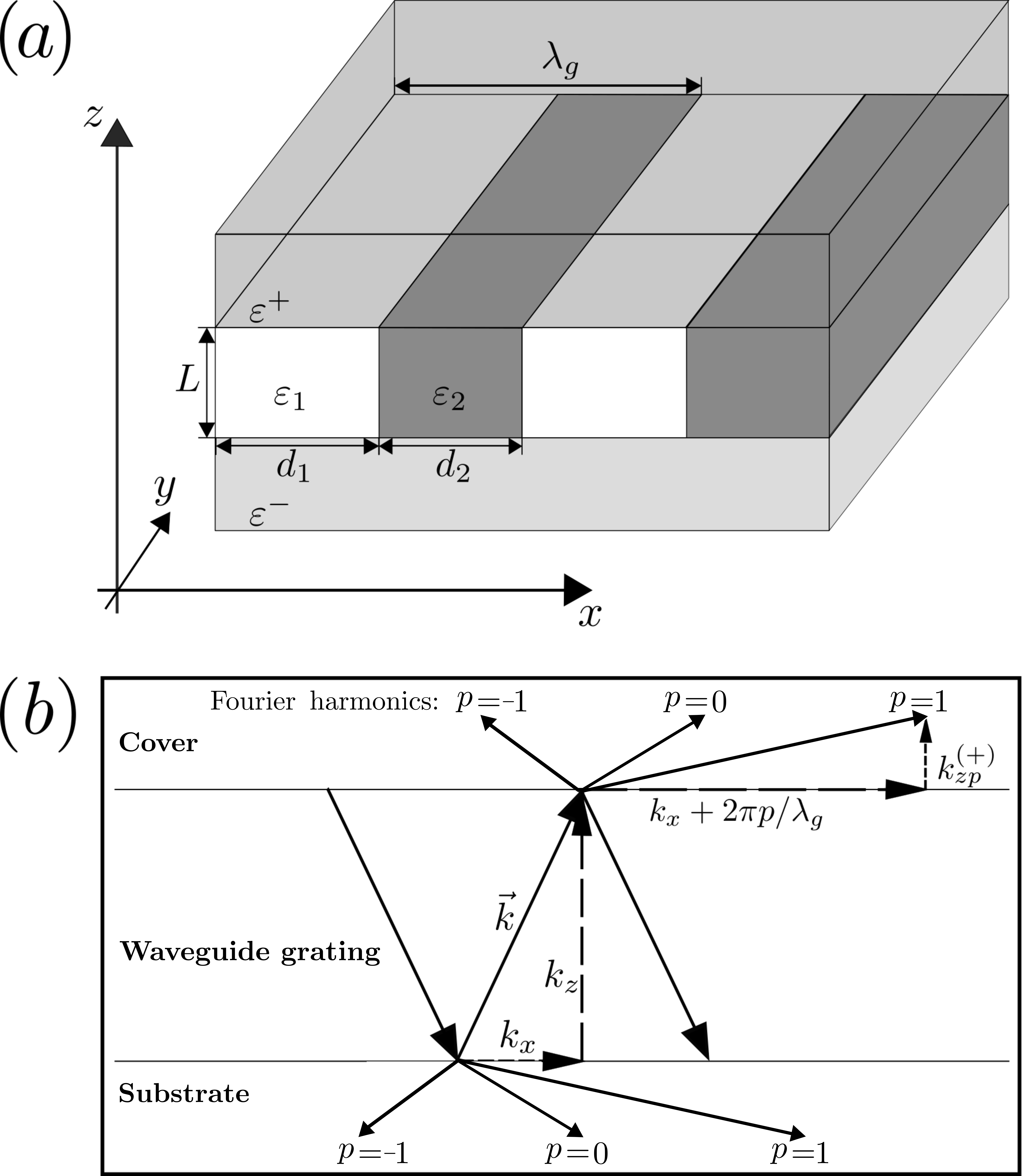}
    \caption{The planar grating waveguide. (a) Geometry: the grating of thickness $L$ and period $\lambda_g$ sandwiched between the substrate and cover with permittivities \(\varepsilon^-\) and \(\varepsilon^+\), respectively. (b) The wave vectors associated with the propagating or evanescent eigenmodes inside the grating and the plane waves (Fourier harmonics) of different Bragg diffraction orders in the substrate and cover, $p = 0, \pm 1, \ldots$.}
     \label{Fig:Geometry}
\end{figure}

By means of the analytical theory of the infinite-grating eigenmodes \cite{PRA2019}, we disclose a distinct, explicit mechanism leading to the formation of the degenerate bound states in the continuum (BICs) and an entire hierarchy of BICs in a planar grating waveguide (1D grating slab with in-plane $C_2$ symmetry, Fig. \ref{Fig:Geometry}).

\begin{figure}[ht]
\centering
\includegraphics[width=85mm]{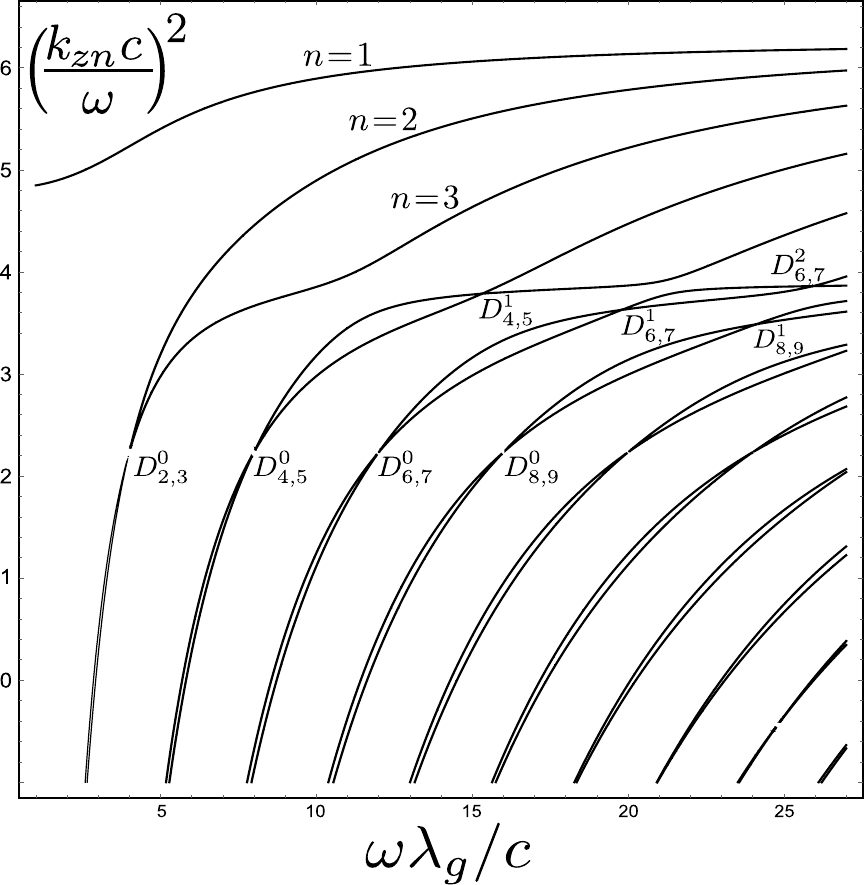}
\caption{Hierarchy of dispersion curves $k_{zn}^2(\omega)$, Eq.~(\ref{CharactEq}), and their intersections (degeneracy points) for the infinite-grating eigenmodes in the titanium-oxide grating; $\varepsilon_1 = 6.25, \varepsilon_2 = 3.9, \rho = 0.39183; k_x=0$. The symbol $D_{n,n+1}^l$ labels the $l$-th intersection of the $n$-th and $(n+1)$-th eigenmodes and signifies the associated BIC.}
\label{dispCurveMulti}
\end{figure}

Namely, far from the crossing points of the infinite-grating-eigenmode dispersion curves shown in Fig.~\ref{dispCurveMulti}, one has only symmetry-protected BICs formed at normal incidence mainly by just one relevant single infinite-grating eigenmode whose in-plane spatial profile is purely odd with respect to the in-plane mirror symmetry. Such a BIC waveguide eigenmode involves an admixture of other odd-parity infinite-grating eigenmodes due to reflection at the boundaries with the substrate and cover, but is completely disconnected from the zeroth spatial Fourier harmonic designed to be the only radiation-loss channel available in the grating waveguide.

On the contrary, a degenerate BIC arises in a vicinity of a degeneracy point where dispersion curves of two infinite-grating eigenmodes of opposite parity intersect, as at $D^0_{2,3}$ in Fig.~\ref{dispCurveMulti}. 
This couple forms an anti-phased superposition and constitutes a degenerate-BIC eigenmode of the grating waveguide due to modes' mutual reflections at the borders of the grating layer. 
Although both infinite-grating eigenmodes possess a mixed parity and undergo a parity symmetry breaking in a vicinity of the degeneracy point, their anti-phased superposition remains decoupled from
the only radiation channel -- emission of the zeroth spatial Fourier harmonic outside the grating waveguide into the cover and substrate. All other Fourier harmonics of higher Bragg diffraction orders cannot be emitted into the cover and substrate because they are evanescent there. Such a phenomenon of destructive interference producing the degenerate-BIC waveguide eigenmode decoupled from the radiation-loss channel is of general nature and could exist in a number of other photonic structures and crystals. 

The BIC waveguide eigenmode demonstrates an extremely narrow, solitary high-Q resonance despite the fact that a continuum of plane waves in a wide range of frequencies and wave vectors around the resonant ones can freely escape from the waveguide into the cover and substrate. The latter distinguishes the phenomenon of BICs from a conventional trapping of fields inside waveguides or cavities which is based on a complete prevention (say, by means of reflection) of the entire continuum of waves from propagating outside the waveguide or cavity.
The search, analysis, design, and applications of the BICs constitute a very active field of modern research in a variety of wave systems, especially in photonics, despite the BIC having first been suggested for matter waves in quantum mechanics \cite{vonNeumann1929,Feshbach1958,Friedrich1985}. 
The mechanisms of formation and properties of various BICs in photonic structures had been widely discussed in the literature (see, for example, reviews \cite{Hsu-Nature2016,Koshelev2020,Sadreev2020,Peng2020,Kildishev2021} and papers \cite{PRB2019,Yang2014,Quaranta2018,Wang1990,Rybin2017,Sadrieva2017,Doeleman2018,Gomis-Nature2017,BulgakovPRA2017,Boyd2021,Cerjan2021,Doiron2022}). In particular, the BIC in the grating layer sandwiched by the cover and substrate had been demonstrated via numerical simulations for a particular set of parameters in \cite{PRB2019}. Yet, despite the fact that the planar lamellar (that is, uniform along $y$ axis) grating waveguide is one of the most basic, simple examples of photonic structures supporting BICs, a full analytical theory of BICs, in particular, degenerate BICs, in the grating waveguide and eigenmode-based understanding of degenerate-BIC appearance have been missing until now. This approach, in particular, reveals exact analytical formulas clarifying universal features of the BIC formation which, otherwise, would stay hidden.

Apparently, the main reason for that is the fact that most works on BICs are based on {\it ad hoc} numerical simulations dominating modern analysis of photonic-crystal structures. They employ various software based on finite-element modeling like COMSOL \cite{COMSOL2016}, rigorous coupled wave analysis (RCWA) \cite{RCWA}, finite-difference time-domain (FDTD) method \cite{FDTD2010}, codes for layered periodic structures \cite{Solver2012}, finite element method (FEM) \cite{Sadrieva2017}, etc.

Another reason is that the BIC-formation mechanism in such a simple, basic photonic structure as the planar lamellar-grating waveguide turns out to be so special that it simultaneously merges almost all other known mechanisms leading to the BIC formation (see reviews \cite{Hsu-Nature2016,Koshelev2020,Sadreev2020,Peng2020,Kildishev2021}), including mechanisms of the symmetry-protected BIC, accidental BIC, single-resonance parametric BIC, coupled multiple resonances (Friedrich–Wintgen, but not Fabry–Pérot) BIC, interference-based BIC through parameter tuning, and topologically-protected BIC. A more detailed discussion of this fact requires results presented in sections II--VI and is postponed to sect. VII. 

The present paper is devoted to introducing such an analytical eigenmode approach per se, not to its thorough application for classification of all possible BICs or discovery of new BICs. We give a new interpretation of the known symmetry-protected BIC, in particular, degenerate BICs, but do not attempt, within the scope of the present paper, to design a BIC of a new type. It is remarkable how merely splitting the central dielectric layer of the planar waveguide into two alternating sections of lengths $d_1, d_2$ (with a fill factor $\rho = d_1/(d_1+d_2)$) and different permittivities $\varepsilon_1, \varepsilon_2$ converts an elementary, trivially soluble problem of the planar dielectric waveguide \cite{Kogelnik1975,Marcuse1991,Young2021} into a rich, complex problem demonstrating many generic features of optical crystals.    

Optical gratings and related waveguides are also interesting by themselves, irrespective to BICs, since they have found numerous applications and been studied for decades (see, for instance, \cite{Kogelnik1975,Marcuse1991,Popov-Gratings2012,Sirenko2010,Hussein2009,Novotny2012,Buscha2007,Sakoda2005,Whittaker1999,Tikhodeev2002,Hadij2021} and references therein). One of the most elegant methods of their studies is based on the analysis of eigenmodes that is very fruitful both for the analytical theory and numerical simulations (see \cite{PRA2019,Young2021,Popov-Gratings2012,Sirenko2010,Hussein2009} and references therein). Another, more straightforward method is the Fourier modal method which prevails in the literature and is known as the rigorous coupled wave theory in the diffractive optics community \cite{Popov-Gratings2012,Sirenko2010,Yang2014} or the scattering-matrix approach in the photonic-crystal community \cite{Novotny2012,Buscha2007,Sakoda2005,Whittaker1999,Tikhodeev2002,Hadij2021}.

In the present paper, we consider a particular example of a waveguide eigenmode -- a degenerate BIC originating due to the mechanism outlined in the second paragraph of the paper. The in-plane $x$-wavenumber of that BIC waveguide eigenmode is chosen to be near the $\Gamma$-point -- the center of the first Brillouin zone of the grating, that is $k_x \approx 0$. (We assume that $k_y = 0$ since we consider only the standard, invariant in the $y$ direction, grating problem when fields are uniform along the $y$ axis and propagate in the $xz$-plane. So, we leave the conical case of diffraction \cite{Li1993} aside.) This choice is predetermined by strongly enhanced backscattering coupling in this exceptional high-symmetry point in the momentum space so that the corresponding eigenmode looks more like a cavity eigenmode rather than a waveguide eigenmode. In fact, it has been shown by means of the representation theory that in the standard photonic crystal slabs the symmetry-protected BICs exist only at the center of the Brillouin zone \cite{Cerjan2021,Doiron2022}. At wavenumbers away from the Gamma point $k_x = 0$, the waveguide eigenmode radiates and forms a leaky resonance. The latter had been illustrated in \cite{PRB2019} by a numerical example. 

We assume that just the central, zeroth diffraction order $p=0$, spatial Fourier harmonic is emitted out of the planar grating layer into the cover and substrate, while all higher-order, $p \neq 0$, Fourier harmonics are evanescent and, hence, do not contribute to the radiation losses. According to the plane-wave dispersion relation, the $p$-th Fourier harmonic becomes evanescent in the cover (superscript ``+") or substrate (superscript ``-") when its $z$-wavenumber squared turns negative: 
\begin{equation} \label{p-nonemitted}
(k_{zp}^{\pm})^2 = \varepsilon^{\pm} \omega^2/c^2 - (k_x + pk_g)^2 < 0 . 
\end{equation}
Here $\varepsilon^+$ or $\varepsilon^-$ is the permittivity of the cover or substrate, respectively.
For modes at the center of the first Brillouin zone, $k_x = 0$, the $p$-th Fourier harmonic ceases to provide a radiation-loss channel when
\begin{equation} \label{p=0emitted}
\varepsilon^{\pm} < \Big( \frac{2\pi cp}{\omega \lambda_g} \Big)^2 .
\end{equation} 

Below we mainly focus on the degenerate BIC originating from the degeneracy point $D_{2,3}^0$ (see Fig.~\ref{dispCurveMulti}) when the condition (\ref{p=0emitted}) is satisfied for all nonzero diffraction orders $p = \pm 1, \pm 2, \ldots$. The analysis of the degenerate BICs at other mode-crossing points $D^l_{n,n+1}$ is alike.   

The content of the paper is as follows. We overview the genesis and field constituents of BICs in a planar grating waveguide in sect. II. In sect. III we present the necessary analytical formulas describing the infinite-grating eigenmodes as well as the waveguide eigenmodes. In sect. IV we disclose the universality of the behavior of the infinite-grating eigenmodes within the vicinity of a mode crossing point and a remarkable connection between the dispersion degeneracy and parity symmetry breaking of the eigenmode spatial profiles changing from odd to even, or vise versa. We elaborate both on the special case of normal incidence, $k_x=0$, and the case of a nonzero in-plane wavenumber, $k_x \neq 0$. In sect. V we explain how to design a waveguide manifesting a degenerate BIC at a given frequency. It is done on the basis of a numerical example related to a waveguide based on titanium oxide (TiO$_2$) -- an optical material known for many applications in photonics. The mechanism behind the emergence of the degenerate BIC is explained and detailed in sect. VI. It involves a parity symmetry breaking, that occurs in accord with the above universality, and a decoupling from a radiation-loss channel due to destructive interference at mode crossing. Conclusions, discussion of the simultaneous manifestation of other known BIC mechanisms in the formation of the above BIC, and other comments make up sect. VII.

\section{Genesis and field constituents of BICs in a planar grating waveguide}

We employ the eigenmode approach which, in the case of BICs in the planar grating waveguide shown in Fig.~\ref{Fig:Geometry}, is based on the hierarchy of the following three field constituents: (i) the plane waves, (ii) the infinite-grating eigenmodes, and (iii) the waveguide eigenmodes. 

The plane waves are the simplest field configurations possessing a quasi-harmonic temporal-spatial profile $\propto e^{ik_{zp} z + i(k_x + pk_g)x - i\omega t}$ satisfying a standard dispersion law, 
\begin{equation} \label{uniform_medium_disp}
k_{zp}^2 + (k_x + pk_g)^2 = \varepsilon \omega^2/c^2, 
\end{equation}
in a uniform medium. The latter means that their phase speed depends on a permittivity $\varepsilon$ of the dielectric medium in which they propagate. The prefix 'quasi' indicates that the frequency $\omega$ and the $z$-wavenumber $k_{zp}$, in general, are complex-valued. 

A single plane wave with a wave vector ${\bf k} = (k_x, k_z)$ together with its mirror counterpart with a wave vector ${\bf k'} = (k_x, -k_z)$ constitutes a waveguide eigenmode only in the trivial case of a slab waveguide with a homogeneous central dielectric layer of permittivity $\varepsilon$. A homogeneous slab waveguide can be viewed as the zeroth-order approximation for the grating waveguide if we choose its core permittivity equal to the average permittivity of two grating sections, $\varepsilon = (\varepsilon_1 d_1 + \varepsilon_2 d_2)/\lambda_g; \ \lambda_g  =d_1+d_2$. 
\begin{figure}[ht]
\centering
    \includegraphics[width=83mm]{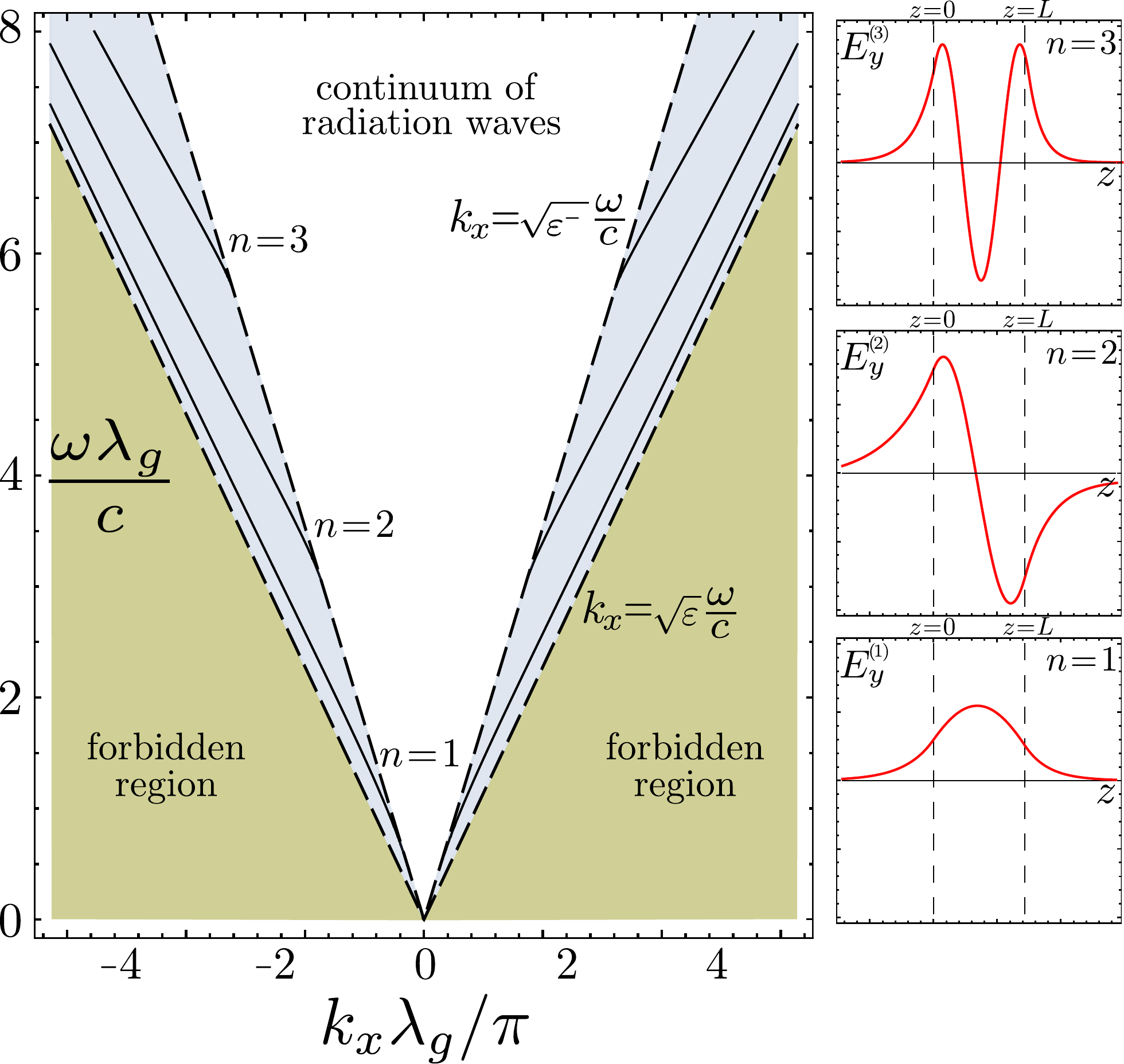}
        \caption{The first three transverse guided TE-eigenmodes of a homogeneous slab waveguide of thickness $L=0.71587\lambda_g$: Dispersion curves $k_{xn}(\omega)$ (left) and spatial $z$-profiles $E_y^{(n)}(z)$ (right). The permittivities of the central layer, substrate and cover are $\varepsilon = 4.8208, \varepsilon^- = 2$ and $\varepsilon^+ = 1$, respectively.}
    \label{slab-modes}
\end{figure}

It is immediate to find all guided TE-modes of such a slab waveguide (see, e.g., \cite{Kogelnik1975}). The dispersion curves and spatial $z$-profiles of the first three transverse eigenmodes guided by a slab waveguide are illustrated in Fig.~\ref{slab-modes}. We enumerate them by the integer $n$ equal to the number of extrema in the $z$-profile of the electric field $E_y^{(n)}(z)$. All extrema are localized within the central dielectric layer, $z \in (0,L)$. The eigenmode amplitude is constant along the $x$ axis. For a given frequency $\omega$, the field profiles $\{E_y^{(n)}(z)|n = 1, 2, \ldots\}$ constitute a series of monochromatic transverse eigenmodes associated with a discrete set of solutions for the eigen $x$-wavenumber, $\{k_{xn}(\omega)\}$, to the following dispersion equation
\begin{equation} \label{TEslab}
\begin{split}
&\frac{\omega L (\varepsilon - \varepsilon^-)}{c} = (n-1)\pi + \tan^{-1} \sqrt{\frac{b}{1-b}} + \tan^{-1} \sqrt{\frac{b+a}{1-b}}; \\ 
& a = \frac{\varepsilon^- - \varepsilon^+}{\varepsilon - \varepsilon^-}, \quad 
b = \frac{(ck_x/\omega)^2 - \varepsilon^-}{\varepsilon - \varepsilon^-}, \ n = 1, 2, 3, \ldots \ .
\end{split}
\end{equation}
The series starts from the fundamental eigenmode $n=1$.

All those guided modes in the slab have an angle of incident lying inside a sector of the total internal reflection. Hence, they are evanescent, that is, do not radiate outside the slab and ideally have an infinite Q factor. However, they are not BICs since, for a given incident angle, they have discrete frequencies lying completely outside the continuous spectrum of the leaky, radiation waves. 

\begin{figure}[ht]
\centering
    \includegraphics[width=95mm]{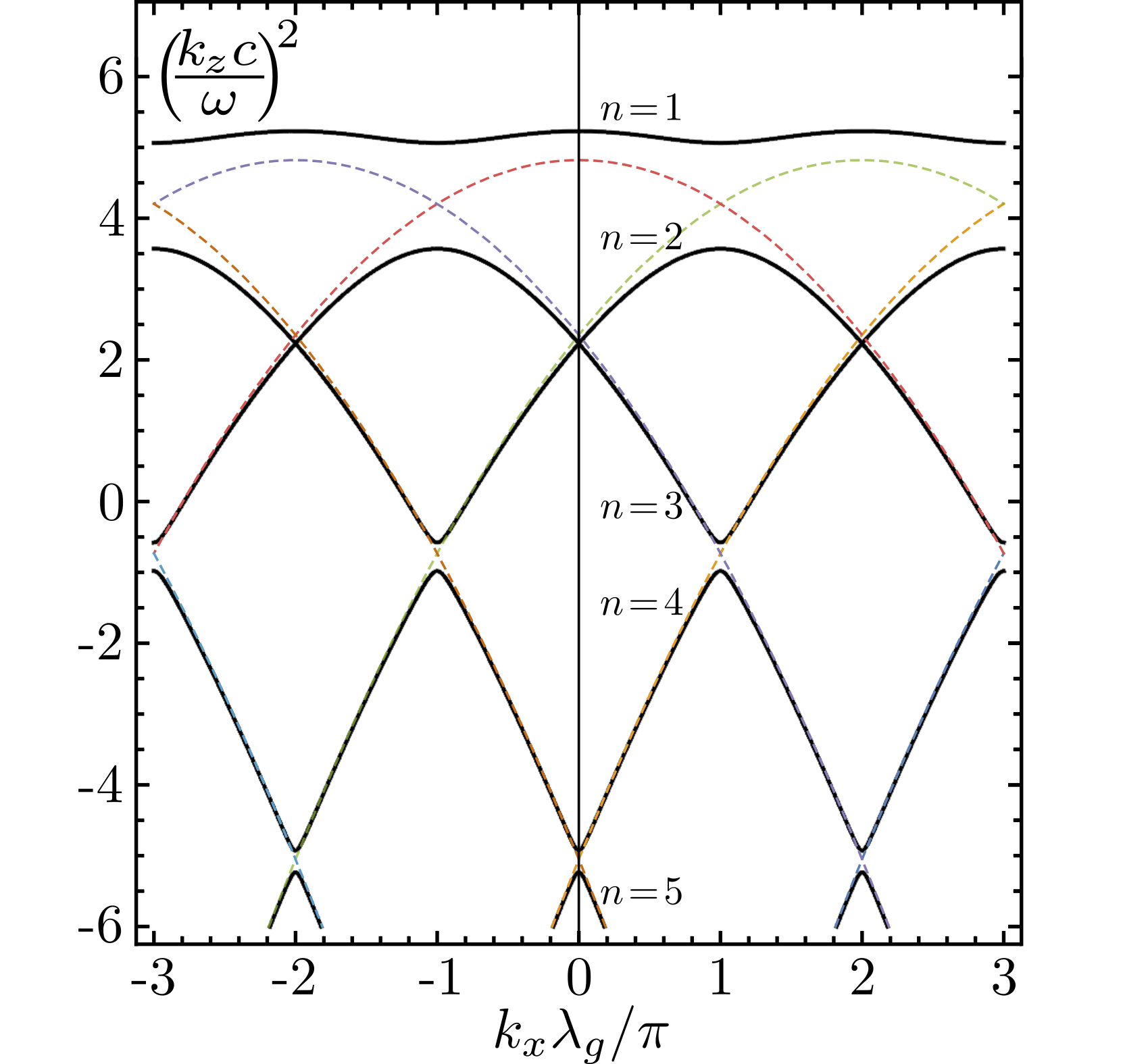}
    \caption{"Energy bands and gaps" of the 1D grating: An extended Brillouin zone representation for the dispersion curves of the infinite-grating eigenmodes, $(c/\omega)^2 k^2_{zn}(k_x)$, Eq.~(\ref{CharactEq}), (thick black curves) approximated by a superposition of the parabolic uniform-medium dispersion curves, $(c/\omega)^2 k^2_{zp}(k_x)$, Eq.~(\ref{uniform_medium_disp}), (dashed color curves) shifted due to the Bragg reflection resonances of various orders by multiples of the grating wavenumber $pk_g = 2\pi p/\lambda_g, \ p = 0, \pm 1, \pm 2, \ldots$. The permittivity of the effective uniform medium, $\varepsilon = (\varepsilon_1 d_1 + \varepsilon_2 d_2)/(d_1+d_2) = 4.8208$, is equal to the weighted sum of the permittivities $\varepsilon_1 = 6.25, \ \varepsilon_2 = 3.9$ of the two grating sections of lengths $d_1 = 0.39183 \lambda_g, \ d_2 = 0.60817 \lambda_g$, respectively; $\omega = 4c/\lambda_g$.}
    \label{Parabolas-extendedBrillouin}
\end{figure}

\begin{figure}[ht]
\centering
    \includegraphics[width=70mm]{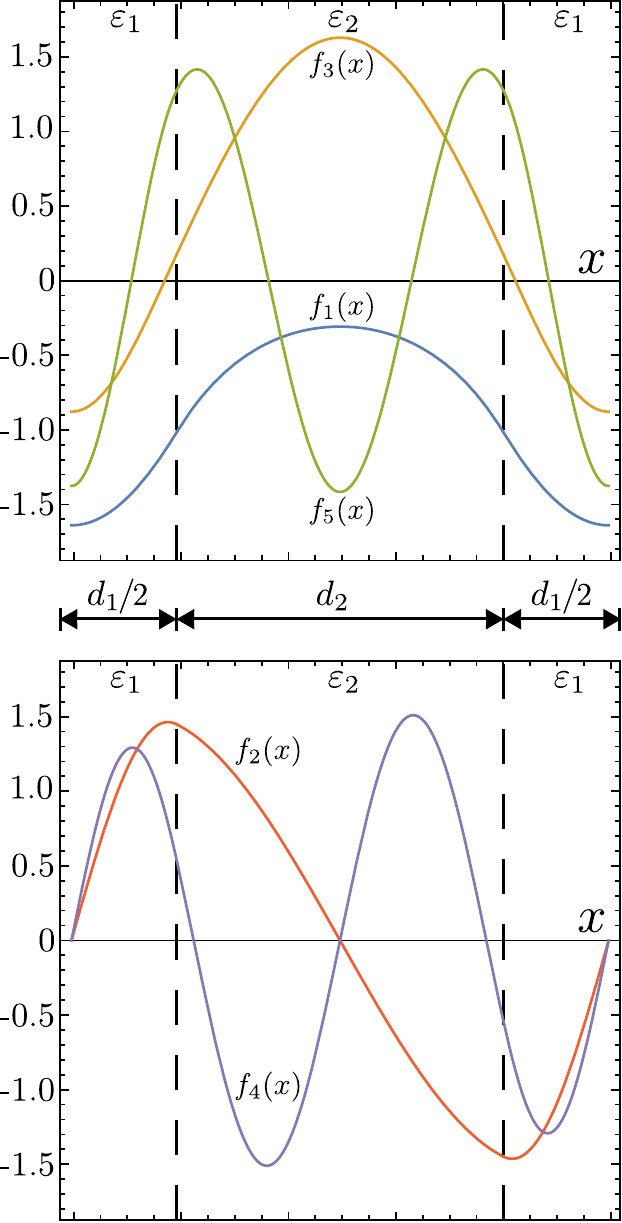}
    \caption{Longitudinal (in-plane) $x$-profiles for the first five eigenmodes of the infinite grating with two sections of lengths $d_1 = 0.39183 \lambda_g, \ d_2 = 0.60817 \lambda_g$ and permittivities $\varepsilon_1 = 6.25, \ \varepsilon_2 = 3.9$, respectively; $k_x = 0, \ \omega = \frac{5c}{\lambda_g}$. For simplicity's sake, the plotted functions $f_n(x)$ differ from those in Eq.~(\ref{f_n}) by a phase factor $e^{i\varphi_n}$ making them real-valued. Note a slight discontinuity of the derivatives $df_n/dx$ at the boundaries between the grating sections.}
    \label{fn(x)n=1-5}
\end{figure}

The infinite-grating eigenmodes \cite{PRA2019} are also closely related to the plane waves in a uniform dielectric medium. Yet, they are strongly restructured by multiple Bragg reflections on the 1D lattice of the alternating permittivities of the grating sections. Such an umklapp scattering can be visualized via a superposition of the parabolic uniform-medium dispersion curves $k^2_{zp}(k_x)$ of various diffraction orders $p = 0, \pm 1, \pm 2, \ldots$ given by Eq.~(\ref{uniform_medium_disp}) as is shown in Fig.~\ref{Parabolas-extendedBrillouin}. An effective uniform medium with the average grating permittivity $\varepsilon = (\varepsilon_1 d_1 + \varepsilon_2 d_2)/(d_1+d_2)$ yields a good zeroth-order approximation for the dispersion curves of the infinite-grating eigenmodes (cf. Fig.~\ref{dispCurveMulti}). The main nontrivial feature here is opening the gaps in the $k^2_{zn}$ spectrum near avoiding-crossing points. These gaps are induced by the Bragg-reflection resonances. The frequency for the plot in Fig.~\ref{Parabolas-extendedBrillouin} is chosen to be the degenerate frequency for the second and third infinite-grating eigenmodes, $\omega = \omega_c \equiv \frac{4c}{\lambda_g}$. In this case the gap at the intersection of the dispersion curves $k^2_{z2}$ and $k^2_{z3}$ is absent. 

Contrary to the guided modes of the homogeneous slab, the infinite-grating eigenmodes are the field configurations possessing a trivial quasi-harmonic temporal and spatial $z$-profile $\propto e^{ik_{zn} z - i\omega t}$, but a nontrivial spatial $x$-profile $f_n(x)$ (see Fig.~\ref{fn(x)n=1-5} and Eq.~(\ref{f_n})) such that it remains invariant in the course of propagation inside the infinite grating despite a persistent in-plane Bragg scattering. Such an invariance occurs only for a discrete set of eigen $z$-wavenumbers $k_{zn}$ enumerated by the integer $n = 1, 2, \ldots$ (see Fig.~\ref{dispCurveMulti}). Thus, the infinite-grating eigenmodes take care of the field boundary conditions and field transformation at all $yz$-plane boundaries between alternating, along the $x$ axis, sections of the grating, but do not take into account the grating-waveguide boundaries with the substrate and cover. For zero in-plane wavenumber $k_x = 0$, as is illustrated in Fig.~\ref{fn(x)n=1-5}, the spatial parity of the $x$-profiles $f_1, f_3, f_5, \ldots$ is purely even while the spatial parity of the $x$-profiles $f_2, f_4, \ldots$ is purely odd. Each infinite-grating eigenmode lives inside the infinite grating independently of others, is a superposition of many spatial Fourier harmonics (partial plane waves), and represents a photon dressed via Bragg, umklapp scattering. 

\begin{figure}[ht]
\centering
    \includegraphics[width=80mm]{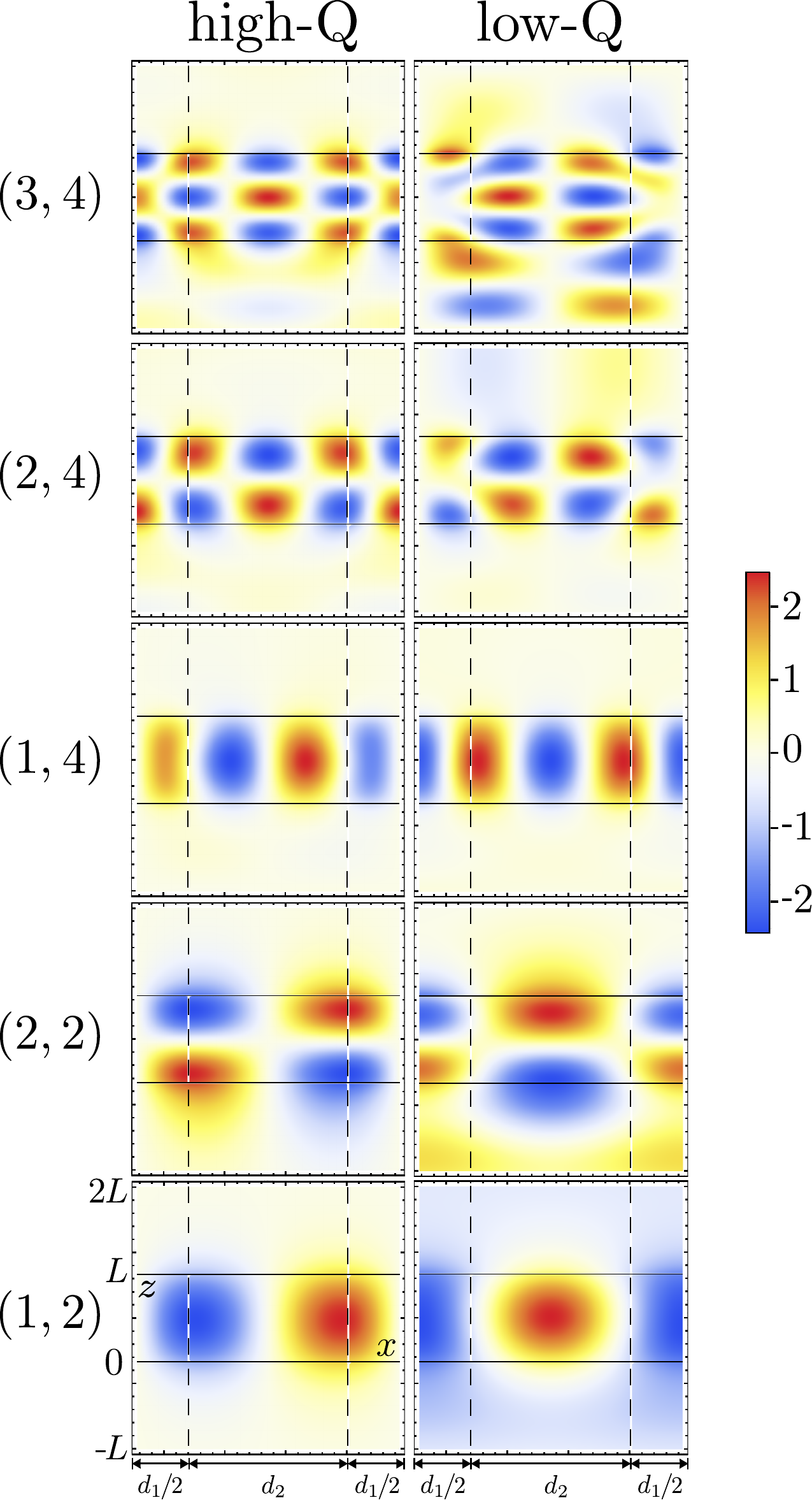}
    \caption{Two-dimensional (in the $xz$-plane) field patterns, $\text{Re}(E_y^{(m)})$, of first lower frequency grating-waveguide eigenmodes at a small $x$-wavenumber, $k_x \approx 0$. The composite index $(m_z, m_x)$ in each row labels the number of transverse ($m_z$) and longitudinal ($m_x$) field extrema per the grating period $\lambda_g$. The eigenmodes in each row pair differ by the value of the binary index, $s = 0$ or $s = 1$, labeling modes with the odd or even parity of the field's $x$-profile, that is, with the high or low Q factor. The lengths and permittivities of the two grating sections are $d_1 = 0.39183 \lambda_g, \ d_2 = 0.60817 \lambda_g$ and $\varepsilon_1 = 6.25, \ \varepsilon_2 = 3.9$, respectively. The thickness of the grating layer is $L = 0.71587 \lambda_g$. The permittivities of the substrate and cover are $\varepsilon^- = 2$ and $\varepsilon^+ = 1$, respectively. Positive and negative field values are marked in red and blue, respectively.}
    \label{2Dpatterns}
\end{figure}

At last, the waveguide eigenmodes are the field configurations which follow a quasi-harmonic evolution $\propto e^{-i\omega_m t}$ with a complex-valued eigenfrequency $\omega_m = \omega'_m - i\omega''_m$, enumerated by a composite integer $m$, while possessing a stationary (invariant in time) spatial structure. Any, say the $m$-th, waveguide eigenmode inside the grating layer is a superposition of many infinite-grating eigenmodes, coupled to each other via mutual reflections at the grating-layer borders with the cover and substrate. 

Thus, the discrete set of waveguide eigenmodes is determined by and take care of both sets of boundary planes and corresponding boundary conditions of continuity of the tangential components of the electric and magnetic fields: Two $xy$-planes, at $z=0$ and $z=L$ along the $z$ axis, which determine the series of transverse eigenmodes in the homogeneous slab waveguide as well as the infinite sequence of $yz$-planes, at $x= p\lambda_g$ and $x= d_1 + p\lambda_g, p = 0, \pm 1, \pm 2, \ldots$, along the $x$ axis, which determine the series of the infinite-grating eigenmodes. Contrary to the slab's transverse modes and infinite-grating eigenmodes, the grating waveguide eigenmodes possess two-dimensional spatial structure which is nontrivial in both transverse, $z$, and longitudinal, $x$, directions as is shown in Fig.~\ref{2Dpatterns}. 

\begin{figure}[ht]
\centering
\includegraphics[width=90mm]{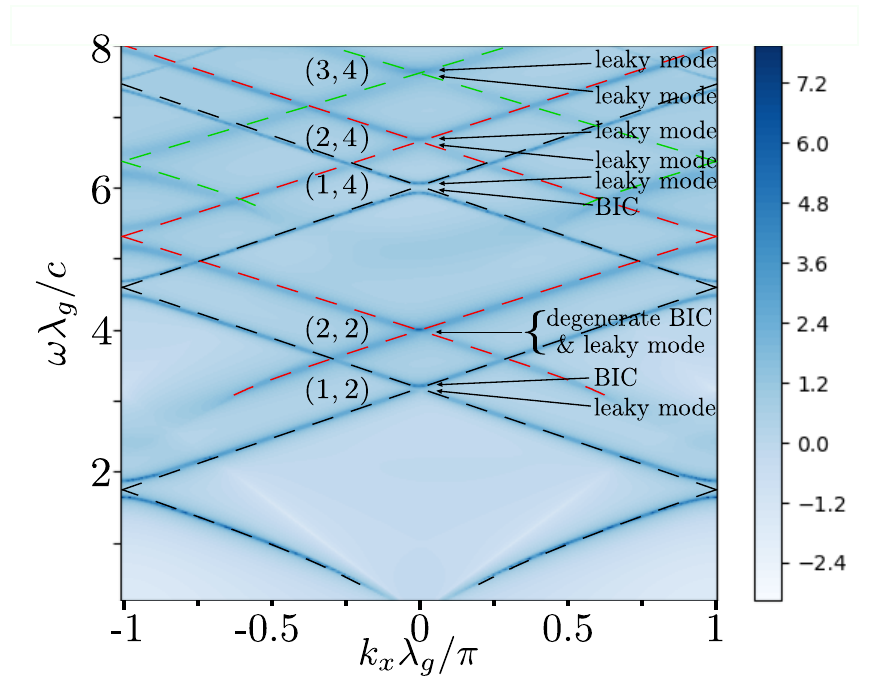}
\caption{Numerical rigorous coupled wave analysis (RCWA): Resonance response of the grating waveguide to the first five Fourier harmonics with equal (unity) amplitudes incident from the cover at $z=L$. The sum of the intensities from these harmonics outgoing into the substrate at $z=0$ (with the logarithmic $\log_{10}$ color scaling) is shown as the function of the frequency $\omega$ and in-plane wavenumber $k_x$ within the first Brillouin zone representation. It reveals the dispersion curves of the grating waveguide eigenmodes (dark blue tracks). They nicely match the dashed black, red and green curves representing the dispersion curves of the first three transverse eigenmodes of the homogeneous slab waveguide with the central layer permittivity \(\varepsilon =(\varepsilon_1d_1+\varepsilon_2d_2)/(d_1+d_2)\). The grating layer is of the thickness $L = 0.71587 \lambda_g$ and consists of two alternating sections of the lengths $d_1 = 0.39183 \lambda_g, \ d_2 = 0.60817 \lambda_g$ and permittivities $\varepsilon_1 = 6.25, \ \varepsilon_2 = 3.9$, respectively.}
    \label{fullRCWA}
\end{figure}

The waveguide eigenmodes, accordingly, can be enumerated by a composite integer $m = (m_z, m_x)$. Its first component, $m_z$, indicates the number of field extrema in the transverse direction $z$ and can be traced to the order $n$ of the corresponding transverse mode of the homogeneous slab in Eq.~(\ref{TEslab}), $m_z = n$. The second component, $m_x$, indicates the number of field extrema in the longitudinal direction $x$ per the grating period $\lambda_g$ and depends on the Brillouin zone index, $p$, of the relevant infinite-grating eigenmodes. Along with the composite index $m = (m_z, m_x)$, an additional binary index $s = 0, 1$ is needed to discern two waveguide eigenmodes with the same number of field extrema but different, somewhat shifted along the $x$ axis, field distributions associated with two different $x$-profiles in each pair of the relevant infinite-grating eigenmodes shown in Fig.~\ref{dispCurveMulti}. For a narrow vicinity of the zero longitudinal wavenumber $k_x = 0$ pertinent to BICs, the binary index value $s=0$ corresponds to the odd-parity $x$-profile, while $s=1$ - to the even-parity $x$-profile. So, there are two columns in Fig.~\ref{2Dpatterns}. 

The eigen frequencies of the waveguide eigenmodes are given by the characteristic equation (\ref{det}) and correspond to the resonant frequencies in the response of the grating waveguide driven by an appropriate monochromatic external source or incident field. The numerical approach of the rigorous coupled wave analysis (RCWA) \cite{RCWA} consists of computing such a response, for example, in the form of the radiation field emitted into the substrate or cover from the grating waveguide or some field distribution inside the grating layer (say, one or a few of the infinite-grating eigenmodes). It allows one to visualize the appearance of strong narrow resonances due to high-Q waveguide eigenmodes as well as weaker and wider resonances due to low-Q waveguide eigenmodes for a particular set of waveguide parameters and a particular choice of the source and response functions. An example of the RCWA response in shown in Fig.~\ref{fullRCWA}. It reveals the dispersion curves (dark blue tracks) of the grating waveguide eigenmodes. They closely follow the dispersion curves of the first three transverse eigenmodes of the homogeneous slab waveguide with the central layer permittivity equal to the weighted sum of the grating sections' permittivities, \(\varepsilon =(\varepsilon_1d_1+\varepsilon_2d_2)/(d_1+d_2)\). In this way, we can easily associate the RCWA resonances with a particular grating waveguide eigenmodes, their frequencies and corresponding 2D spatial patterns shown in Fig.~\ref{2Dpatterns}.

Note that the frequencies of the three lower pairs of eigenmodes in Fig.~\ref{2Dpatterns}, which are equal to the frequencies of the appropriate waveguide-eigenmode resonance branches in Fig.~\ref{fullRCWA} at $k_x = 0$, satisfy the inequality in Eq.~(\ref{p=0emitted}) for any diffraction order $p \neq 0$. These eigenmodes could leak via the only available radiation-loss channel -- the zeroth Fourier harmonic. Hence, those of them which possess the odd parity with respect to the in-plane mirror symmetry are high-Q modes, while the ones of the even parity are low-Q modes. The upper two pairs of eigenmodes in Fig.~\ref{2Dpatterns} have the frequencies of the appropriate two upper pairs of avoiding-crossing resonance branches in Fig.~\ref{fullRCWA} at $k_x = 0$. These frequencies are high enough to violate the inequality in Eq.~(\ref{p=0emitted}) for $p = \pm 1$ Fourier harmonics which, therefore, become the new radiation-loss channels, additional to the $p = 0$ Fourier-harmonic radiation-loss channel. So, among the two upper pairs of waveguide eigenmodes the odd-parity eigenmodes appear to be more leaky (low-Q) than the even-parity (relatively high-Q) eigenmodes. 

Bear in mind that we do not include in Fig.~\ref{2Dpatterns} the lowest frequency waveguide eigenmode. It corresponds to the lowest resonance branch in Fig.~\ref{fullRCWA} and is associated with the first infinite-grating eigenmode in the first Brillouin zone. Hence, it freely radiates outside the grating layer into the substrate and cover with just little reflections, like a traveling plane wave (zeroth Fourier harmonic), and is of no interest due to a very low Q-factor. 

Each waveguide eigenmode represents a real, physical, fully dressed photon living in the waveguide with the grating layer of a finite thickness $L$.

In the grating waveguide, for a given $x$-wavenumber within the first Brillouin zone $k_x \in (-k_g/2,k_g/2)$ of the lamellar grating of the period $\lambda_g \equiv 2\pi/k_g = d_1 + d_2$, the plane waves are present as an infinite series of spatial Fourier harmonics of different diffraction orders $p = 0, \pm 1, \pm 2,\ldots$ coupled to each other via Bragg scattering on the periodic, along the $x$ axis, sequence of boundaries between two alternating sections of the grating.

The genesis of the grating waveguide eigenmodes sketched above, in particular, comparative analysis of the dispersion curves, field constituents and radiation of different modes of the grating waveguide, infinite grating and homogeneous slab, provides the telltale classification and overview of all possible high-Q and low-Q eigenmodes of the grating waveguide.

In the above hierarchy of waves and eigenmodes, the plane waves (spatial Fourier harmonics), as well as the infinite-grating eigenmodes, are partial, non-physical waves. The problem of revealing BICs consists of finding the waveguide eigenmodes of vanishing decay, leakage rate $\omega''_m$. The eigenmode approach has a great advantage over other approaches since it operates with the most natural field constituents actually living in the photonic structure and, hence, minimizes the number of couplings to account for. Another great advantage of the eigenmode approach is that it elucidates the physical mechanism and interpretation of the BIC phenomenon. The difference between the eigenmode approach and other approaches (see, for instance, \cite{vonNeumann1929,Feshbach1958,Friedrich1985,Hsu-Nature2016,Koshelev2020,Sadreev2020,Peng2020,Kildishev2021,PRB2019}) is similar to the difference between the eigenmode method and the rigorous coupled-wave (Fourier-modal) method in the theory of optical gratings mentioned above. 

\begin{figure}
  \includegraphics[width=83mm]{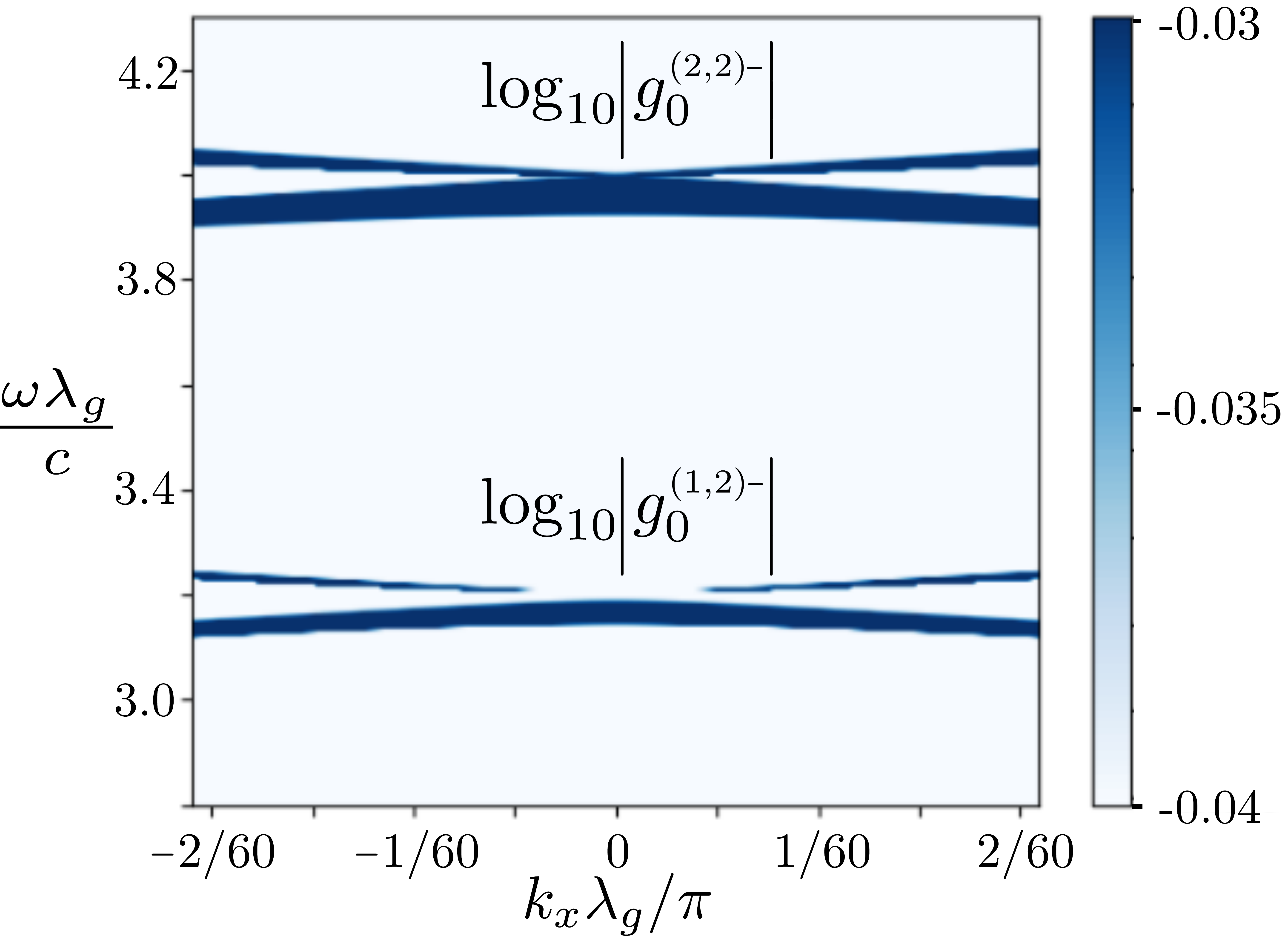}
\caption{Paradox of a seeming disappearance of the BIC at the degenerate frequency $\omega_c = 4c/\lambda_g$: Instead of the BIC's telltale feature of no response at $k_x = 0$, the RCWA shows a maximal response at $k_x = 0$. For comparison, near the frequency $\omega \approx 3.2c/\lambda_g$ (i.e., far from the degenerate frequency) the usual avoiding-crossing pair of resonant curves of the high-Q (upper branch) and low-Q (lower branch) waveguide eigenmodes clearly shows the BIC at $k_x = 0$ on the high-Q branch. Here numerical RCWA resonance response of the grating waveguide to the plane wave ($p=0$ Fourier harmonic), incident from the substrate, is shown as a density plot of the relative intensity of that same $p=0$ Fourier harmonic reflected from the substrate. The grating waveguide is the same as in Fig.~\ref{fullRCWA}.}
\label{fig:paradox}
\end{figure}

Below we focus on the analysis of the degenerate BIC and its comparison with the usual, solitary symmetry-protected BIC. Such BICs are shown in Fig.~\ref{fullRCWA} at $k_x \approx 0$ and frequencies $\omega \approx 4c/\lambda_g$ and $\omega \approx 3.2c/\lambda_g$, respectively. The degenerate BIC could be easily overlooked or not resolved by straightforward numerical approaches, like the RCWA one, even if one magnifies the resonance response in a vicinity of the mode-crossing point like in Fig.~\ref{fig:paradox}.

Disappearance of the radiation losses in a narrow vicinity of zero $x$-wavenumber, clearly seen on the upper branch of the lower pair of waveguide resonance curves in Fig.~\ref{fig:paradox}, unambiguously points to the existence of the symmetry-protected BIC at the frequency $\omega \approx 3.2c/\lambda_g$ and $x$-wavenumber $k_x \approx 0$. This BIC is an extremely high-Q waveguide eigenmode that mostly consists of just the third infinite-grating eigenmode which constitutes the maximum resonance response at this spectral point $\omega \approx 3.2c/\lambda_g, k_x \approx 0$. (This is explained in detail and justified in sect. VI and Fig.~\ref{fig:SymProtBIC}.) Contrary to the upper branch, the lower branch of the avoiding-crossing resonance at $\omega \approx 3.2c/\lambda_g$ in Fig.~\ref{fig:paradox} has relatively large radiation losses at all $x$-wavenumbers and significantly larger resonance width. Clearly, it is associated with the leaky, low-Q waveguide eigenmode which mostly consists of just the second infinite-grating eigenmode. (This follows from an absence of the lower branch in the resonance response of the third infinite-grating eigenmode in the bottom panel of Fig.~\ref{fig:SymProtBIC}.) The above attribution of the frequency-split branches in the lower pair of resonances -- the BIC (upper branch) and the low-Q eigenmode (lower branch) -- to the third and second infinite-grating eigenmodes, respectively, is fully consistent with the purely odd and even parity of the $x$-profiles (right panel in Fig.~\ref{fig:SymProtBIC}) of the third and second infinite-grating eigenmodes at $k_x=0$.

However, the radiation losses in the resonance response near the degenerate frequency $\omega_c = 4c/\lambda_g$ in a vicinity of zero $x$-wavenumber shown in Fig.~\ref{fig:paradox} do not demonstrate the telltale feature of the BIC -- the disappearance of the resonant response at normal incidence, $k_x \approx 0$. So, based on this RCWA numerical plot, one may conclude that there is no BIC at the degenerate resonant frequency $\omega_c = 4c/\lambda_g$. This would be a mistake.

Thus, one needs a better way to understand the BICs and to avoid such mistakes. Here the analytical theory of the grating-waveguide eigenmodes comes to the rescue.

A detailed explanation of this apparent paradox will be given below (see sect. VI, Fig.~\ref{fig:DegenBIC}).
The main point is that here we have a degenerate BIC whose resonant frequency coincides with the resonant frequency of the low-Q waveguide eigenmode. 
A spectrally wide response of the low-Q eigenmode to the incident plane wave in the numerical RCWA plot in Fig.~\ref{fig:paradox} simply masks the narrow telltale feature of the BIC in the response of the high-Q eigenmode at the degenerate frequency $\omega = 4c/\lambda_g$. 
Note that these two waveguide eigenmodes are different (2,2)-eigenmodes since their 2D spatial patterns are very dissimilar as is shown in Fig.~\ref{2Dpatterns}. In fact, the $x$-profiles of the main constituent of these two waveguide eigenmodes, which (say, at $k_x=0$) are \(f_2(x)-f_3(x)\) and \(f_2(x)+f_3(x)\), possess the odd and even parity, respectively, that is, are orthogonal to each other as is clear from the right panel in Fig.~\ref{fig:DegenBIC}. After this prelude, we proceed to the consistent analysis.

\section{Analytical theory of grating-waveguide eigenmodes}

In the present paper we study the high-Q transverse-electric (TE) eigenmodes of a planar grating waveguide. They can be represented by two column $N$-vectors ${\bf F^{\pm}} = (\{F_n^{\pm} |n=1,...,N \})^T$, each containing some $N$ complex amplitudes of co- or counter-propagating infinite-grating eigenmodes, respectively. (A superscript $T$ denotes the transpose operation.) The electric field of the TE waveguide eigenmode is linearly polarized along the $y$ axis and is a superposition of the infinite-grating eigenmodes:
\begin{equation} \label{E}
E_y(x,z) e^{-i\omega t} = \sum_{n=1}^N \big[ F_n^+ e^{ik_{zn}z} + F_n^- e^{ik_{zn}(L-z)} \big] f_n(x) e^{-i\omega t}, 
\end{equation}
where the non-negative factor, $\omega'' \geq 0$, in the imaginary part of the frequency $\omega = \omega' - i\omega''$ describes a mode decay and $k_{zn}$ is the $z$-wavenumber of the $n$-th infinite-grating eigenmode. The $x$-profile of the $n$-th infinite-grating eigenmode is a superposition of two counter-propagating waves,
\begin{equation} \label{f_n}
f_n(x) = \frac{B_q(x)}{\sqrt{\langle B_q^{\dagger}, B_q \rangle}}, \ B_q(x) = b_q^+ e^{ik_q(x-d_1)} + b_q^- e^{ik_q(d_1-x)}, 
\end{equation} 
where $k_q = \big[\varepsilon_q\omega^2/c^2 - k_{zn}^2 \big]^{1/2}$. The norm $\langle B_q^{\dagger}, B_q \rangle^{1/2}$ is defined in Eq. (\ref{norm}) via adjoint eigenfunctions, Eq. (\ref{adjoint f_n}). The complex amplitudes of the waves in Eq. (\ref{f_n}), 
\begin{equation} \label{b+-}
\begin{split}
& b_1^{\pm} = \pm 2e^{-ik_x\lambda_g/2} \big[ \pm ik_1\sin(k_2d_2) -k_2\cos(k_2d_2) \\
&\qquad +k_2e^{\pm ik_1d_1+ik_x\lambda_g}  \big], \\
& b_2^{\pm} = \pm 2e^{ik_x\lambda_g/2} \big[ \pm ik_2\sin(k_1d_1) +k_1\cos(k_1d_1) \\
&\qquad -k_1e^{\mp ik_2d_2-ik_x\lambda_g}  \big],
\end{split}
\end{equation}
are specified by the index $q$ equalling 1 or 2 depending on whether $x$ is inside the grating section of length $d_1$ or $d_2$ and permittivity $\varepsilon_1$ or $\varepsilon_2$, respectively (see Fig.~\ref{Fig:Geometry}). 

Below, we choose the number $N$ of the contributing infinite-grating eigenmodes based on a required accuracy of the waveguide-eigenmode representation. Usually it is enough to include just all propagating and a few lower-order evanescent eigenmodes. The number of propagating infinite-grating eigenmodes is finite. All higher-order evanescent eigenmodes exponentially decay on a path from one border to the other border of the planar waveguide. Hence, they couple very weakly to the Fourier harmonics escaping from the waveguide, while having minimal contribution to the electric field of the waveguide eigenmode inside the grating as well. At the same time, the number of the involved spatial Fourier harmonics, coupled to the essential infinite-grating eigenmodes due to Bragg scattering on the grating, also turns out to be limited to a few central harmonics. Such a low, finite dimensionality of the eigenmode approach is the main reason for its tremendous efficiency and advantage over other approaches for analysis of relatively thick gratings or other photonic structures. 

The waveguide eigenmodes can be found as the eigenvectors of an equation expressing a self-reproduction of the eigenmode amplitudes ${\bf F^+}$ after a round trip over the grating layer along the $z$ axis: (i) propagation from the grating-substrate border at $z=0$ to the grating-cover border at $z=L$, (ii) reflection into the counter-propagating eigenmodes ${\bf F^-}$ from the border at $z=L$, (iii) propagation of the eigenmode ${\bf F^-}$ back to the border at $z=0$, and (iv) reflection back into the co-propagating eigenmodes ${\bf F^+}$ from the border at $z=0$. Propagation through the grating layer along the $z$ axis is described by the diagonal transfer matrix $\Lambda = {\rm diag} \{ e^{ik_{zn}L}| n=1,...,N \}$ since each eigenmode propagates independently on the other eigenmodes in accord with its own eigen wavenumber $k_{zn}$. Reflections at the grating-layer borders are described by the matrix of reflectance coefficients $R^{\pm} = \{ (R^{\pm})_n^{n'} \}$ which depends on the permittivity $\varepsilon^{\pm}$ of the adjacent dielectric medium - cover or substrate. Thus, the $m$-th TE eigenmode of the planar grating waveguide, 
\begin{equation} \label{E^(m)}
\begin{split}
&E_y^{(m)}(x,z) e^{-i\omega_m t} = \sum_{n=1}^N e^{-i(\omega_m' - i\omega_m'') t} \Big[ F_n^{(m)} e^{ik_{zn}z} \\
&+ \Big( \sum_{n'=1}^N (R^+)_n^{n'} F^{(m)}_{n'}e^{ik_{zn'}L} \Big) e^{ik_{zn}(L-z)} \Big] f_n(x),
\end{split}
\end{equation}
is determined by the $m$-th eigenvector ${\bf F^{(m)}}$ associated with the unity eigenvalue of a matrix $M = (M_n^{n'})$ equal to the product of the aforementioned four matrices:
\begin{equation} \label{MF=F}
\begin{split}
&M {\bf F^{(m)}} = {\bf F^{(m)}}; \quad M = R^- \Lambda R^+ \Lambda , \\ 
&M_n^{n'}= \sum_{n''=1}^N (R^-)_n^{n''} (R^+)_{n''}^{n'} e^{i(k_{zn'}+k_{zn''})L} .
\end{split}
\end{equation}
This waveguide master matrix $M(\omega)$ fully describes properties of the planar grating waveguide and nontrivially depends on the complex frequency $\omega$. So, the complex eigenfrequencies $\omega_m$ of the waveguide eigenmodes are determined by the roots of the characteristic equation
\begin{equation} \label{det}
D(\omega_m) = 0; \qquad D(\omega) \equiv \det \Big[ M(\omega) - \mathbbm{1} \Big] .
\end{equation}
Hereinafter, the symbol $\mathbbm{1}$ stands for the identity matrix. 

The reflectance matrix can be found from the boundary conditions of continuity of the tangential components of the electric and magnetic fields at the borders $z=0, L$ of the grating layer. These two conditions can be written in terms of the reflectance, $R^{\pm}$, and transmittance, $T^{\pm}$, matrices as well as the basis-transformation matrix $C = (c_n^p)$ (defined below in Eq. (\ref{c_n^p})):
\begin{equation} \label{bc}
R^{\pm} = CT^{\pm} - \mathbbm{1}, \qquad C \hat{k}_z^{\pm} T^{\pm} = \hat{k}_z (\mathbbm{1} - R^{\pm}) .
\end{equation}
Here $\hat{k}_z = {\rm diag} \{ k_{zn}| n=1,2,... \}$ and $\hat{k}_z^{\pm} = {\rm diag} \{ k_{zp}^{\pm}| p=0,\pm1,\pm2,... \}$ are the diagonal matrices of the $z$-wavenumbers $k_{zn}$ of the infinite-grating eigenmodes and $z$-wavenumbers $k_{zp}^{\pm} = [\varepsilon^{\pm} \omega^2/c^2 - (k_x +pk_g)^2]^{1/2}$ of the plane waves which propagate in the cover (superscript "+") or substrate (superscript "-"); $c$ is the speed of light in vacuum. The longitudinal wavenumber of the latter $p$-th plane wave, $k_x^{(p)} = k_x +pk_g$, is the sum of the wavenumber $k_x$ within the first Brillouin zone and the $p$-th Bragg harmonic of the grating wavenumber $k_g$. The transmittance matrix $T^{\pm} = \{ (T^{\pm})_p^n \}$ is required for the evaluation of the radiation losses. Its entry $(T^{\pm})_p^n$ gives the amplitude, $G_p^{\pm} = (T^{\pm})_p^n F_n^{\pm}$, of the $p$-th spatial Fourier harmonic $G_p^{\pm} e^{i(k_x+pk_g)x + k_{zp}^{\pm} [\pm i(z-L/2)-iL/2]}$ emitted from the grating layer into the cover or substrate by the infinite-grating eigenmode $F_n^{\pm} e^{k_{zn}[\pm i(z-L/2)-iL/2]}$ incident onto the cover ($z=L$) or substrate ($z=0$) border from inside the grating, respectively.

In the general case of the complex permittivities $\varepsilon_q$ and complex frequncy $\omega = \omega' -i\omega''$, the coefficients in the Fourier expansions of (i) a spatial Fourier harmonic via the infinite-grating eigenfunctions, 
\begin{equation}   \label{c_n^p}
e^{i(k_x+p'k_g)x} = \sum_{n=1}^{\infty} c^{p'}_n f_n(x) , \ c^{p'}_n = \langle f_n^{\dagger}, e^{i(k_x+p'k_g)x} \rangle ,
\end{equation}
and (ii) an infinite-grating eigenfunction via the spatial Fourier harmonics,
\begin{equation}  \label{ticlde c_p^n}
f_n(x) = \sum_{p=-\infty}^{\infty} {\tilde c}^n_p e^{i(k_x+pk_g)x} , \ {\tilde c}^n_p = \langle e^{i(k_x+pk_g)x}, f_n \rangle,  
\end{equation}
are related to each other via the straightforward substitution $\{k_q \to k_q^{\dagger}|\,q=1,2\}$:
\begin{equation} \label{c-ticlde c}
(c_n^p)^* = \tilde{c}_p^n (\{k_q \to k_q^{\dagger}\}), \quad k_q^{\dagger} = \Big[\Big(\frac{\varepsilon_q\omega^2}{c^2}\Big)^* - (k_{zn}^2)^*\Big]^{1/2}. 
\end{equation}
It includes the following substitution of the amplitudes of the eigenmode's exponential components
\begin{equation}  \label{b^dagger}
b_q^{\pm} \to b_q^{\pm \dagger} \equiv b_q^{\pm} (\{k_q \to k_q^{\dagger} \ |q=1,2 \}) . 
\end{equation}

The set of the infinite-grating eigenfunctions $\{ f_n(x)|n=1,2,...\}$ and their adjoint eigenfunctions
\begin{equation}  \label{adjoint f_n}
f_n^{\dagger}(x) = \frac{B_q^{\dagger}(x)}{\sqrt{\langle B_q, B_q^{\dagger} \rangle}}, \ B_q^{\dagger} = b_q^{+\dagger} e^{ik_q^{\dagger}(x-d_1)} + b_q^{-\dagger} e^{ik_q^{\dagger}(d_1-x)},
\end{equation}
constitutes a biorthogonal basis, so that the product of the matrices ${\tilde C}, C$ in any order equals the identity matrix: 
\begin{equation}  \label{tildeCC}
{\tilde C}C = C{\tilde C}=\mathbbm{1} \leftrightarrow \sum_{n=-\infty}^{\infty} {\tilde c}^{n}_p c_n^{p'} = \delta_{p,p'}, \sum_{p=-\infty}^{\infty} c_n^p {\tilde c}^{n'}_p = \delta_{n,n'},
\end{equation}
where $\delta_{n,n'}$ is the Kronecker delta, $\tilde{C} = (\tilde{c}_p^n)$. (Note that here we define normalization of the entries $c_n^p$ and $\tilde{c}_p^n$ slightly different from that in \cite{PRA2019}. Namely, we move one of the norms $\langle B_q^{\dagger}, B_q \rangle^{1/2}$ from the $c_n^p$ to the $\tilde{c}_p^n$.) The norm squared is given by the standard inner product,
\begin{equation}  \label{<>}
\langle B_q^{\dagger}, B_q \rangle = \frac{1}{\lambda_g} \int_{0}^{\lambda_g} B_q^{\dagger *}(x) B_q(x) dx .
\end{equation}
The latter can be calculated explicitly by employing the biorthonormal basis 
of the eigenfunctions $f_n(x)$ and the adjoint eigenfunctions $f_n^{\dagger}(x)$ as follows
\begin{equation} \label{norm}
\begin{split}
&\langle B_q^{\dagger}, B_q \rangle \\
&=\sum_{q=1,2} \frac{d_q}{\lambda_g} \{ b_q^+[b_q^+(k_q\to k_q^{\dagger})]^* \nu_1[i(-1)^qd_q(k_q - k_q^{\dagger*})]\\ 
&\qquad \quad  + b_q^-[b_q^-(k_q \to k_q^{\dagger})]^* \nu_1[-i(-1)^qd_q(k_q - k_q^{\dagger*})] \\
&\qquad \quad + b_q^+[b_q^-(k_q \to k_q^{\dagger})]^* \nu_1[i(-1)^qd_q(k_q + k_q^{\dagger*})] \\ 
&\qquad \quad + b_q^-[b_q^+(k_q \to k_q^{\dagger})]^* \nu_1[-i(-1)^qd_q(k_q + k_q^{\dagger*})] \}.
\end{split}
\end{equation}
Here the elementary function $\nu_1(u)= (e^u-1)/u$ behaves analytically and tends to unity, $\nu_1(0) = 1$, when $u \to 0$. 
The matrix $\tilde{C} = (\tilde{c}_p^n)$ can be written explicitly as follows 
\begin{equation} \label{tildec_p^n}
\begin{split}
&{\tilde c}^n_p = \frac{ie^{-\frac{i}{2}k_x\lambda_g}}{\lambda_g} \\
&\times \sum_{q=1}^2 \frac{[(b_q^--b_q^+)k_q - (b_q^-+b_q^+)k_x^{(p)}](e^{-ik_qd_q}-e^{-ik_x^{(p)}d_q})}{\langle B_q^{\dagger}, B_q \rangle^{1/2}[(k_x^{(p)})^2-k_q^2]e^{i(1-q)(k_qd_q-k_x^{(p)}d_1)}}.    
\end{split}
\end{equation}

We find the following explicit solutions to Eq. (\ref{bc}) for the transmittance and reflectance matrices, \begin{equation}  \label{T} 
T^{\pm} = 2(\mathbbm{1}+\tilde{C}\hat{k}_z^{-1} C \hat{k}_z^j)^{-1} \tilde{C} , \ 
R^{\pm} = 2(\mathbbm{1} + \hat{k}_z^{-1} C \hat{k}_z^j \tilde{C})^{-1} - \mathbbm{1} ,
\end{equation}
suitable for their computing in the case of arbitrary finite numbers $N$ and $S=2p^*+1$ of the infinite-grating eigenmodes and spatial Fourier harmonics $p=0,\pm1,...,\pm   p^*$, respectively, chosen to approximate the field of a given waveguide eigenmode. In such an approximation the matrices $C= (c_n^p)$ and $\tilde{C} = (\tilde{c}_p^n)$ are the rectangular $N\times S$ and $S\times N$ matrices, respectively, and their inverse matrices do not exist, except the square-matrix case $N=S$. The square $N\times N$ and $S\times S$ diagonal matrices 
\begin{equation}  \label{hatk_z}
\begin{split}
&\hat{k}_z = {\rm diag} \{ k_{zn}|n=1,...,N \}, \\
&\hat{k}_z^j = {\rm diag} \{ k_{zp}^j|p=-p^*,...,-1,0,1,...,p^* \}
\end{split}
\end{equation}
of the $z$-wavenumbers $k_{zn}$ of the infinite-grating eigenmodes inside the grating layer and $z$-wavenumbers $k_{zp}^j = [\varepsilon^j \omega^2/c^2 - (k_x +pk_g)^2]^{1/2}$ of the spatial Fourier harmonics emitted into the cover ($j=+$) or substrate ($j=-$), respectively, can be easily inverted: 
\begin{equation} \label{1/k}
\hat{k}_z^{-1} = \rm{diag} \{ 1/k_{zn} \}, \quad (\hat{k}_z^j)^{-1} = \rm{diag} \{ 1/k_{zp}^j \}.
\end{equation}

The analytical formulas for the eigenmodes of an infinite lamellar grating had been found in \cite{PRA2019} for a real-valued frequency $\omega = \omega'$. However, radiation losses of the high-Q waveguide eigenmode due to emission into the cover and/or substrate of the planar grating waveguide lead to an appearance of a decay rate, $\omega''$, contributing with a small negative imaginary part to the waveguide-eigenmode frequency $\omega = \omega' -i \omega'', \ \omega''/\omega' \ll 1$. The corresponding first-order correction $\Delta k_{zn}$ to the infinite-grating eigenmode $z$-wavenumber $k_{zn} (\omega'') = k_{zn}(\omega''=0) + \Delta k_{zn}$ is easy to find from the characteristic equation for the infinite-grating eigenmodes,
\begin{equation} \label{CharactEq}
\begin{split}
&\cos(k_1 d_1)\cos(k_2 d_2) - \sin(k_1 d_1)\sin(k_2 d_2)\frac{k_1^2 + k_2^2}{2 k_1 k_2} \\ &=\cos(k_x\lambda_g); \qquad k_q = \Big(\frac{\varepsilon_q \omega^2}{c^2} - k_{zn}^2\Big)^{1/2}.
\end{split}
\end{equation}
Suppose the zeroth-order values of the $x$-wavenumber of a monochromatic eigenmode plane wave in a grating medium of the permittivity $\varepsilon_q, \ q = 1, 2$, are $k_q = k_q(\omega''=0)$. Keeping only the linear corrections to them in the characteristic equation (\ref{CharactEq}), we find
\begin{equation} \label{dk_zn}
k_{zn} (\omega'') = k_{zn}(\omega''=0) + \Delta k_{zn}, \ \Delta k_{zn} \approx \frac{i\omega'\omega''(A_1/A_0)}{c^2 k_{zn}(\omega''=0)} ;
\end{equation}
\begin{equation} \label{As}
    \begin{split}
& A_s = \frac{\varepsilon_2^sk_1^2+\varepsilon_1^sk_2^2}{k_1k_2} \big[ \cos(k_x\lambda_g) - \cos(k_1 d_1)\cos(k_2 d_2) \big] \\
&\qquad + (\varepsilon_1^s + \varepsilon_2^s) \sin(k_1 d_1)\sin(k_2 d_2) \\
&\qquad + \Big( k_2\varepsilon_1^sd_1 + \varepsilon_2^sd_2 \frac{k_1^2+k_2^2}{2k_2} \Big) \sin(k_1 d_1)\cos(k_2 d_2) \\
&\qquad + \Big( k_1\varepsilon_2^sd_2 + \varepsilon_1^sd_1 \frac{k_1^2+k_2^2}{2k_1} \Big) \cos(k_1 d_1)\sin(k_2 d_2).
\end{split}
\end{equation}
This result is valid for all infinite-grating eigenmodes everywhere except within a vicinity of points in the parameter space where $A_0=0, A_1=0$. At those points, the overall contribution of linear terms in Eq. (\ref{CharactEq}) is zero and $\Delta k_{zn}$ should be found from the characteristic equation with the quadratic corrections. Such a case is considered below and is described by a nontrivial result in Eq. (\ref{dkzn}).

\section{Spatial parity at mode crossing and universality of its establishment}

The vicinity of each degenerate point $D_{n,n+1}^l$, where the dispersion curves of the $n$-th and $(n+1)$-th infinite-grating eigenmodes intersect, requires an exceptionally detailed analysis. The reason for the latter is not just the dispersion degeneracy, but the fact that those two intersecting eigenmodes acquire, via parity symmetry breaking at this point, a certain odd or even spatial parity of the $(m_{x1},m_{x2})$ order (Eq.~(\ref{f_n})) far away from this point.

Namely, in the case of the propagating eigenmodes and $k_x=0$, the phase accumulations over each of the grating sections along the $x$ axis become multiples of $\pi$,
\begin{equation}  \label{kd=mpi}
k_qd_q = m_{xq} \pi ; \ q = 1,2; \ m_{x1} - m_{x2} = 2l, \ l=0,\pm 1,\ldots , 
\end{equation}
where $k_q = [\omega^2\varepsilon_q/c^2 - k_{zn}^2]^{1/2}$; $m_{x1},\,m_{x2} =1, 2, 3,\dots$ \ . According to Eq.~(\ref{kd=mpi}), the $z$-wavenumber $k_{zn}$ and the critical frequency $\omega_c$ at the crossing point of the spatial-parity order $(m_{x1},m_{x2})$ are given by the explicit formulas
\begin{equation} \label{cpkz}
\begin{split}
&\Big(\frac{k_{zn}c}{\omega_c}\Big)^2 = \varepsilon_2 + \frac{\varepsilon_1 - \varepsilon_2}{1 - \big[\frac{(1-\rho)m_{x1}}{\rho m_{x2}}\big]^2} \quad ,\\ 
&\Big( \frac{\omega_c \lambda_g}{c} \Big)^2 = \frac{\pi^2}{\varepsilon_1 - \varepsilon_2} \Big[\Big(\frac{m_{x1}}{\rho}\Big)^2 - \Big(\frac{m_{x2}}{1-\rho}\Big)^2 \ \Big].
\end{split}
\end{equation}

The remarkable connection (\ref{kd=mpi}) between the dispersion degeneracy and geometry (spatial parity) of the $x$-distributions of the intersecting infinite-grating eigenmodes can be proven as follows. The intersection of the two dispersion curves, $k_{zn} = k_{z(n+1)}$, at some critical frequency $\omega_c$ means that the characteristic equation (\ref{CharactEq}) admits an existence of two different derivatives $dk_{zn}/d\omega$ at this point. The latter is possible only if the Taylor expansion of the left hand side of the characteristic equation (\ref{CharactEq}) over the variables $\Delta k_{zn} = k_{zn} - k_{zn}(\omega_c)$ and $\Delta \omega = \omega - \omega_c$ does not contain linear terms in the vicinity of this point. Hence, the corresponding partial derivatives should be zero, that is, $A_0 = 0$ and $A_1 = 0$ (see Eq.~(\ref{As})). It is easy to see that, in the general case, the multiple of $\pi$ phase accumulations in Eq.~(\ref{kd=mpi}) provide the required solution to the equations $A_0 = 0$ $\&$ $A_1 = 0$.  

Therefore, the leading asymptotic corrections are quadratic and constitute a quadratic equation for $\Delta k_{zn}$. As a result, it yields exactly two solutions which describe two intersecting dispersion curves with different slopes, as it should be at the point of mode crossing. 

\begin{figure}[ht]
\centering
\includegraphics[width=85mm]{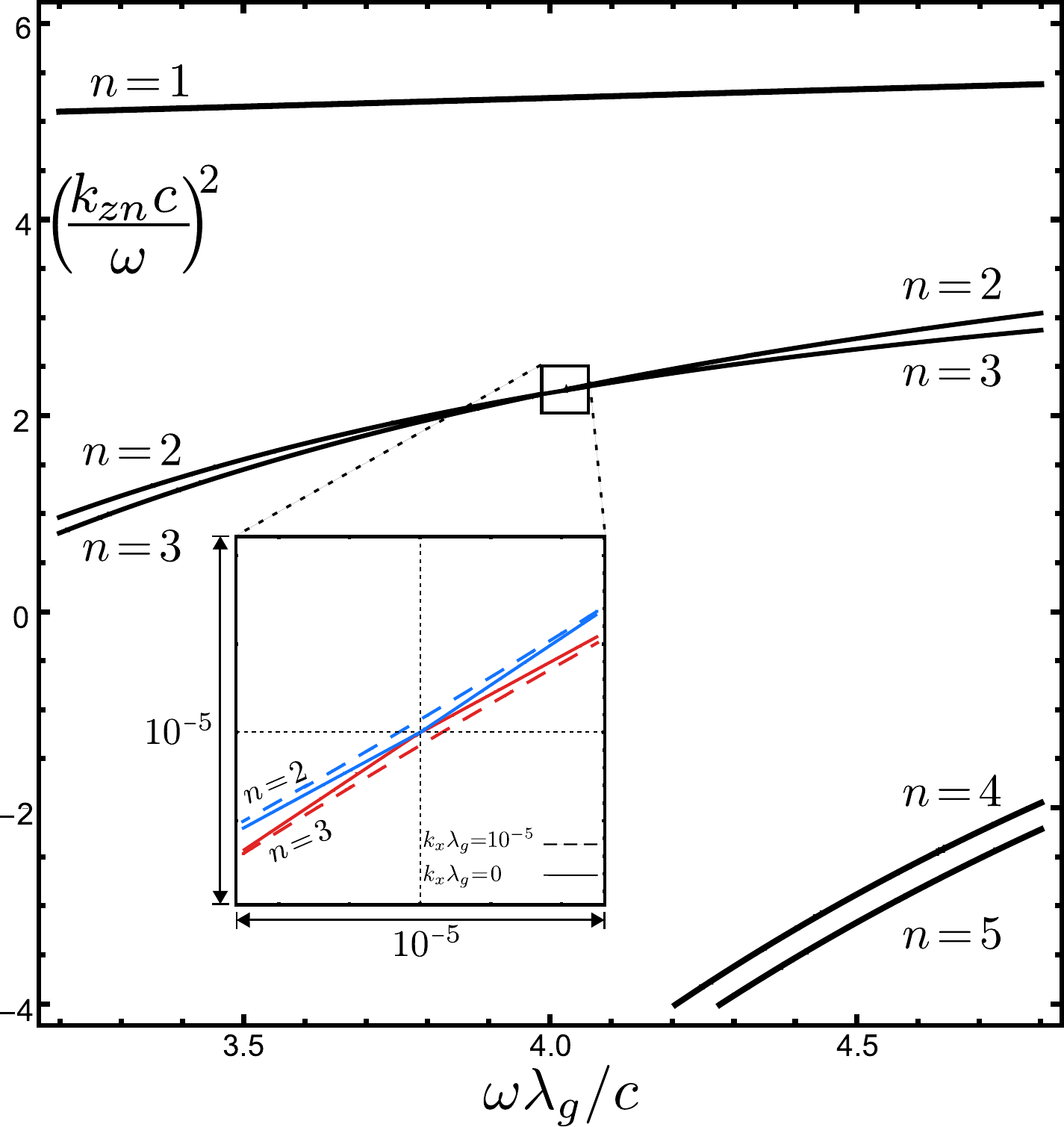}
\caption{Dispersion curves, $k_{zn}^2(\omega)$, of the infinite-grating eigenmodes in the titanium-oxide grating ($\varepsilon_1 = 6.25, \varepsilon_2 = 3.9, \rho = 0.39183$), $k_x=0$. The first three eigenmodes are propagating ($k_{zn}^2 >0$) while the higher order ones are evanescent ($k_{zn}^2 <0$) within the selected range of frequencies. The insert shows the avoiding crossing of the second and third eigenmodes in a vicinity of the degenerate frequency $\omega_c = 4c/\lambda_g$ in a general case of nonzero in-plane wavenumber, $k_x \lambda_g = 10^{-5}$ (dashed curves), as compared to the limiting case $k_x = 0$ (solid curves).}
\label{Fig:Dispersion}
\end{figure}

In the vicinity of a degenerate point, the two intersecting infinite-grating eigenmodes show the following remarkable universality in their behavior when approaching such a point. Namely, if one approaches this point along a path parameterized by a function $\nu$ of the complex frequency $\omega = \omega' -i\omega''$ and grating parameters $\{ \varepsilon_q, \ d_q|\, q=1,2\}$ reaching a value $\nu = \nu_0$ at this point, then the ratio of the related derivatives of the above phase accumulations acquires only two possible values,
\begin{equation}  \label{d(kd)/dv}
\frac{d(k_1d_1)/d\nu}{d(k_2d_2)/d\nu}\Big|_{\nu = \nu_0} = - \Big( \frac{d_1}{d_2} \Big)^{\eta_n}, \qquad \eta_n = \pm 1 , 
\end{equation}
determined by a purely geometrical parameter -- the ratio between the lengths of the grating sections, $d_1/d_2$.
The proof of the fact stated in Eq. (\ref{d(kd)/dv}) follows from the second derivative of the characteristic equation (\ref{CharactEq}) with respect to the parameter $\nu$ at $\nu = \nu_0$.

It is then easy to see that the leading contribution to the amplitudes $b^{\pm}_q$ of each of the two infinite-grating eigenmodes in Eqs. (\ref{f_n}), (\ref{b+-}) tends to zero in terms of the small deviation $\nu - \nu_0 \to 0$ either linearly or quadratically: 
\begin{equation} \label{b0}
\begin{split}
&b^{\pm}_q \sim \frac{db^{\pm}_q}{d\nu}\Big|_{\nu=\nu_0}(\nu - \nu_0) \qquad \textrm{if} \qquad \eta_n = +1, \\ &b^{\pm}_q \sim \frac{1}{2} \frac{d^2b^{\pm}_q}{d\nu^2}\Big|_{\nu=\nu_0}(\nu - \nu_0)^2 \qquad \textrm{if} \qquad \eta_n = -1.
\end{split}
\end{equation}
The two intersecting infinite-grating eigenmodes have different critical indices $\eta_{n+1} \neq \eta_n$, either +1 or -1. 

\begin{figure}[ht]
\centering
\includegraphics[width=85mm]{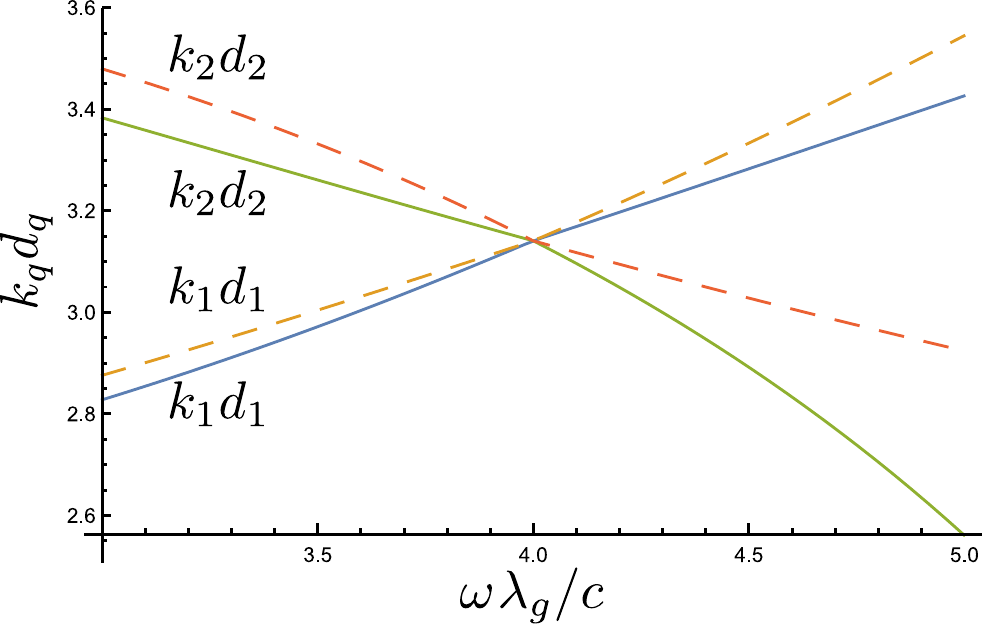}
\caption{Phase accumulations, $k_qd_q,\,q=1,2$, of the $n=2$ (solid curves) and $n=3$ (dashed curved) infinite-grating eigenmodes over the grating section $d_1$ or $d_2$ along the $x$ axis as the functions of the dimensionless frequency $\omega\lambda_g/c$. All of them become equal to $\pi = 3.14\ldots$ at the mode-crossing point $\omega = \omega_c = 4c/\lambda_g$; $\varepsilon_1 = 6.25, \varepsilon_2 = 3.9, \rho = 0.39183, k_x=0$.}
\label{Fig:PhaseAccumulations}
\end{figure}

The result (\ref{d(kd)/dv}) leads to the linear correction $\Delta k_{zn} = k_{zn}(\omega +\Delta \omega) - k_{zn}(\omega)$ to both infinite-grating eigenmode $z$-wavenumbers in the vicinity of the degenerate point: 
\begin{equation} \label{dkzn}
\Delta k_{zn} = \frac{\varepsilon_1 + \varepsilon_2 (d_2/d_1)^{(2-\eta_n)} }{1 + (d_2/d_1)^{(2-\eta_n)}} \frac{\omega \ \Delta \omega}{c^2 k_{zn}(\omega)}, \ \eta_n = \pm 1 .
\end{equation}
Note that the correction (\ref{dkzn}), albeit being linear, is different from Eq. (\ref{dk_zn}) and therefore cannot be given by Eq. (\ref{dk_zn}) since the latter does not apply due to the degeneracy $A_0 = A_1 = 0$.  

In the present paper we consider the intersection $D_{2,3}^0$,
\begin{equation} \label{kz2=kz3}
k_{z2}(\omega) = k_{z3}(\omega) \qquad \textrm{at} \quad \omega =\omega_c, 
\end{equation}
between the dispersion curves of the $n=2$ and $n=3$ infinite-grating eigenmodes (see Fig.~\ref{Fig:Dispersion}) which have the critical indices $\eta_n = (-1)^n \text{sign}(\omega - \omega_c)$ depending on the sign of the frequency detuning $\omega - \omega_c$ and possess the corresponding even or odd $x$-spatial parity of the main order $(m_{x1}=1,\,m_{x2}=1)$ as per Eq.~(\ref{kd=mpi}) (see Fig.~\ref{Fig:PhaseAccumulations}).

In fact, the leading asymptotics (\ref{b0}) provides not just the scaling, but an accurate analytical description of the parity symmetry breaking, i.e., restructuring of the spatial $x$-profile of the infinite-grating eigenmodes, in the vicinity of the mode-crossing point $D_{2,3}^0$ (Figs.~\ref{dispCurveMulti}, \ref{Fig:Dispersion}) if the frequency detuning $|\omega - \omega_c|$ significantly exceeds the longitudinal wavenumber $|ck_x|$.

To make the picture of main infinite-grating eigenmodes complete, we also refer to their dispersion curves at the crossing frequency $\omega_c$ as the function of the in-plane wave number $k_x$ in the presence of a BIC (see Fig.~\ref{Parabolas-extendedBrillouin}). In accord with Eq. (\ref{kz2=kz3}), the parameters of the grating are chosen in such a way that the gap between the $n=2$ and $n=3$ bands is closed at the center of the Brillouin zone $k_x=0$.

For the evaluation of radiation losses, we need to know the amplitude of the contribution from the zeroth spatial Fourier harmonic to the infinite-grating eigenmode $f_n$ in Eq.~(\ref{ticlde c_p^n}). At the center of the Brillouin zone, $k_x=0$, it can be found from Eq. (\ref{tildec_p^n}) in a very simple form
\begin{equation} \label{tildec_0^n}
\begin{split}
&\tilde{c}_0^n = \frac{8 \big[ \cos(k_1d_1) - \cos(k_2d_2) \big]}{\lambda_g k_1 k_2 \langle B_q^{\dagger}, B_q \rangle^{1/2}} \\
&\times \Big[ k_2^2 \sin\Big(\frac{k_1d_1}{2}\Big)e^{-ik_1d_1/2} +k_1^2 \sin\Big(\frac{k_2d_2}{2}\Big)e^{ik_2d_2/2} \Big] ,
\end{split}
\end{equation}
where the relations $b_1^+ -b_1^- =4k_2 [\cos(k_1d_1) - \cos(k_2d_2)]$ and $b_2^+ -b_2^- =4k_1 [\cos(k_1d_1) - \cos(k_2d_2)]$ were employed.
Since $\cos\alpha_1 - \cos\alpha_2 = 2\sin\frac{\alpha_1-\alpha_2}{2} \sin\frac{\alpha_1+\alpha_2}{2}$, we can approximate the contribution $\tilde{c}_0^n$ of the zeroth spatial Fourier harmonic to the $n=2$ and $n=3$ eigenmodes in the vicinity of the degenerate point $D_{2,3}^0$, where $|k_qd_q-\pi|\ll 1, q=1,2,$ as follows
\begin{equation} \label{tildec_0^n near pi}
\begin{split}
&\tilde{c}_0^n \approx \frac{16i(\varepsilon_2 - \varepsilon_1)\omega^2}{\lambda_g k_1 k_2 c^2 \langle B_q^{\dagger}, B_q \rangle^{1/2}} \\
&\times \sin\Big(\frac{k_2d_2-k_1d_1}{2}\Big) \sin\Big(\frac{k_2d_2+k_1d_1}{2}\Big) .
\end{split}
\end{equation}

By the way, in the vicinity of the higher-order $(m_{x1}=2,\,m_{x2}=2)$ degenerate point $D_{4,5}^0$ (see Fig.~\ref{dispCurveMulti} and Eq.~(\ref{kd=mpi})), where $|k_qd_q-2\pi|\ll 1, q=1,2,$ the contribution $\tilde{c}_0^n$ of the zeroth spatial Fourier harmonic to the $n=4$ and $n=5$ eigenmodes tends to zero even faster,
\begin{equation} \label{tildec_0^n near 2pi}
\begin{split}
&\tilde{c}_0^n \approx -\frac{8 \big[k_2^2 \sin(k_1d_1-2\pi) +k_1^2 \sin(k_2d_2-2\pi) \big]}{\lambda_g k_1 k_2 \langle B_q^{\dagger}, B_q \rangle^{1/2}} \\
&\qquad \times \sin\Big(\frac{k_2d_2-k_1d_1}{2}\Big) \sin\Big(\frac{k_2d_2+k_1d_1}{2}\Big),
\end{split}
\end{equation}
that suggests a different behavior of this BIC resonance in the relevant region of waveguide parameters.

A proper description of the vicinity of the degeneracy point $D^0_{2,3}$, where $k_{z2} = k_{z3}$ and $k_1 d_1 = k_2 d_2 = \pi$ at $\omega = \omega_c$, calls for the Taylor expansion of the characteristic equation (\ref{CharactEq}) for the crossing infinite-grating eigenmodes $n=2$ and $n=3$ up to the second order in the frequency detuning $\Delta \omega = \omega' - i\omega'' - \omega_c$ and in-plane wavenumber $k_x$, since the linear corrections vanish as per Eq.~(\ref{As}). Introducing relevant detunings of the phase accumulations over the grating sections along the $x$ axis, 
\begin{equation} \label{x-phase accum}
\chi_q = k_q d_q - \pi, \qquad q = 1, 2,
\end{equation}
we get this expansion in the following symmetric form
\begin{equation} \label{TaylorCE}
\chi_1^2 + 2\gamma \chi_1 \chi_2 + \chi_2^2 = (k_x\lambda_g)^2, \ \ \gamma = \frac{1}{2} \Big(\frac{d_2}{d_1} + \frac{d_1}{d_2} \Big) \geq 1.
\end{equation}
It yields an explicit expression for the second phase accumulation in terms of the first one,
\begin{equation} \label{chi2fromCE}
\chi_2 = -\gamma \chi_1 + (-1)^n \sqrt{(\gamma^2 -1)\chi_1^2 + (k_x\lambda_g)^2}.
\end{equation}
The sign in front of the square root is different for the considered $n=2$ and $n=3$ infinite-grating eigenmodes.

Those phase-accumulation detunings are related also via the identity $k_2^2 - \varepsilon_2 \frac{\omega^2}{c^2} = k_1^2 - \varepsilon_1 \frac{\omega^2}{c^2}$ which expresses the fact that an infinite-grating eigenmode has the same $z$-wavenumber $k_{zn}$ for both grating sections $d_{1,2}$, Eq. (\ref{CharactEq}). Within a linear approximation, it yields the second relation for the phase-accumulation detunings:
\begin{equation} \label{chi2vschi1}
\chi_2 \approx -\eta \frac{\Delta \omega \lambda_g}{c} + \frac{d_2^2}{d_1^2} \chi_1, 
\end{equation}
$$\eta = \frac{(\varepsilon_1 - \varepsilon_2)d_2^2 \omega_c}{\pi c \lambda_g} = \Big(\frac{1}{\rho} -1 \Big) \sqrt{(1-2\rho)(\varepsilon_1-\varepsilon_2)}.$$
Plugging in Eq. (\ref{chi2vschi1}) for $\chi_2$ into Eq. (\ref{chi2fromCE}) and solving the resultant equation for $\chi_1$, we get the phase-accumulation detunings $\chi_{1,2}$ and, hence, wavenumber $k_{zn}$ for both infinite-grating eigenmodes $n=2, 3$ in the vicinity of their crossing point $D^0_{2,3}$ as explicit functions of the frequency detuning $\Delta \omega$ and in-plane wavenumber $k_x$. We have
$$\chi_1 \approx \frac{\xi \eta \Delta \omega \lambda_g}{c\delta} - (-1)^n \sqrt{(\gamma^2-1)\Big(\frac{\eta \Delta \omega \lambda_g}{c\delta} \Big)^2 + \frac{(\lambda_gk_x)^2}{\delta}},$$
\begin{equation} \label{Deltakzn}
k_{zn}^2 = \varepsilon_q \frac{\omega^2}{c^2} - k_q^2 
= \varepsilon_q \frac{(\omega_c + \Delta \omega)^2}{c^2} - \frac{(\pi + \chi_q)^2}{d_q^2}, \ q=1,2.
\end{equation}
These functions involve the grating parameters: $\gamma, \eta$ and
\begin{equation}
\xi = \frac{d_2^2}{d_1^2} + \gamma, \quad \delta = \xi^2 - \gamma^2 +1 
\equiv \Big( \frac{d_2^3}{d_1^3} +1 \Big) \Big( \frac{d_2}{d_1} +1 \Big).
\end{equation}

The second-order Taylor expansion for the complex amplitudes (\ref{b+-}) of waves constituting the infinite-grating eigenmodes (\ref{f_n}), written in terms of the obtained phase-accumulation detunings $\chi_{1,2}$, Eqs. (\ref{chi2vschi1}), (\ref{Deltakzn}), follows
\begin{equation} \label{b+-Taylor}
\begin{split}
&b_1^{\pm} = -\frac{2i\pi}{d_2} e^{-\frac{ik_x\lambda_g}{2}} \Big[ \chi_1+\frac{d_2}{d_1}\chi_2 \pm k_x\lambda_g + \Big(\frac{d_2}{d_1}+1\Big)\frac{\chi_1\chi_2}{\pi}+ \\
&k_x\lambda_g \Big( i\chi_1 \pm \frac{\chi_2}{\pi} \Big) \pm\frac{i}{2} \Big( (k_x\lambda_g)^2 +\chi_1^2-\chi_2^2 \Big) \Big] , \\
&b_2^{\pm} = -\frac{2i\pi}{d_1} e^{\frac{ik_x\lambda_g}{2}} \Big[ \chi_2+\frac{d_1}{d_2}\chi_1 \pm k_x\lambda_g + \Big(\frac{d_1}{d_2}-1\Big)\frac{\chi_1\chi_2}{\pi}- \\
&k_x\lambda_g \Big( i\chi_2 \pm \frac{\chi_1}{\pi} \Big) \mp\frac{i}{2} \Big( (k_x\lambda_g)^2 +\chi_2^2-\chi_1^2 \Big) \Big] .
\end{split} 
\end{equation}
The corresponding Taylor expansion of the eigenmode norm squared, Eq. (\ref{norm}), can be found from the equation
\begin{equation} \label{normTaylor}
\begin{split}
&\langle B_q^{\dagger}, B_q \rangle =\sum_{q=1,2} \frac{d_q}{\lambda_g} \Big[ b_q^+b_q^{+*} + b_q^-b_q^{-*} + \\
&\frac{\chi_q}{\pi}(b_q^-b_q^{+*}+b_q^+b_q^{-*})
-i\frac{(-1)^q\chi_q^2}{\pi}(b_q^-b_q^{+*}-b_q^+b_q^{-*}) \Big],
\end{split}
\end{equation}
where frequency detuning $\Delta \omega$ is not subjected to complex conjugation as if it would be real-valued. 

Restructuring of the spatial parity of the infinite-grating eigenmodes $n=2,3$ in the vicinity of the crossing point $D_{2,3}^0$ and their radiation losses can be characterized by their coupling $\tilde{c}_0^n$, Eq. (\ref{tildec_p^n}), with the zeroth spatial Fourier harmonic
\begin{equation} \label{tildec_0^nTaylor}
\begin{split}
&{\tilde c}^n_0 = \frac{2ie^{-\frac{i}{2}k_x\lambda_g}}{\lambda_g\langle B_q^{\dagger}, B_q \rangle^{1/2}}
\sum_{q=1}^2 (-1)^{q+1}d_q\Big[ - \frac{k_xd_q(b_q^-+b_q^+)}{\pi^2} \\
&+\frac{b_q^--b_q^+}{\pi+\chi_q} \Big( -\frac{i}{2}(\chi_q+k_xd_q) +i(\chi_2-k_xd_1)\delta_{q,2} \Big) \Big].    
\end{split} 
\end{equation}

\begin{figure}[ht] 
    \centering
    \includegraphics[width=80mm]{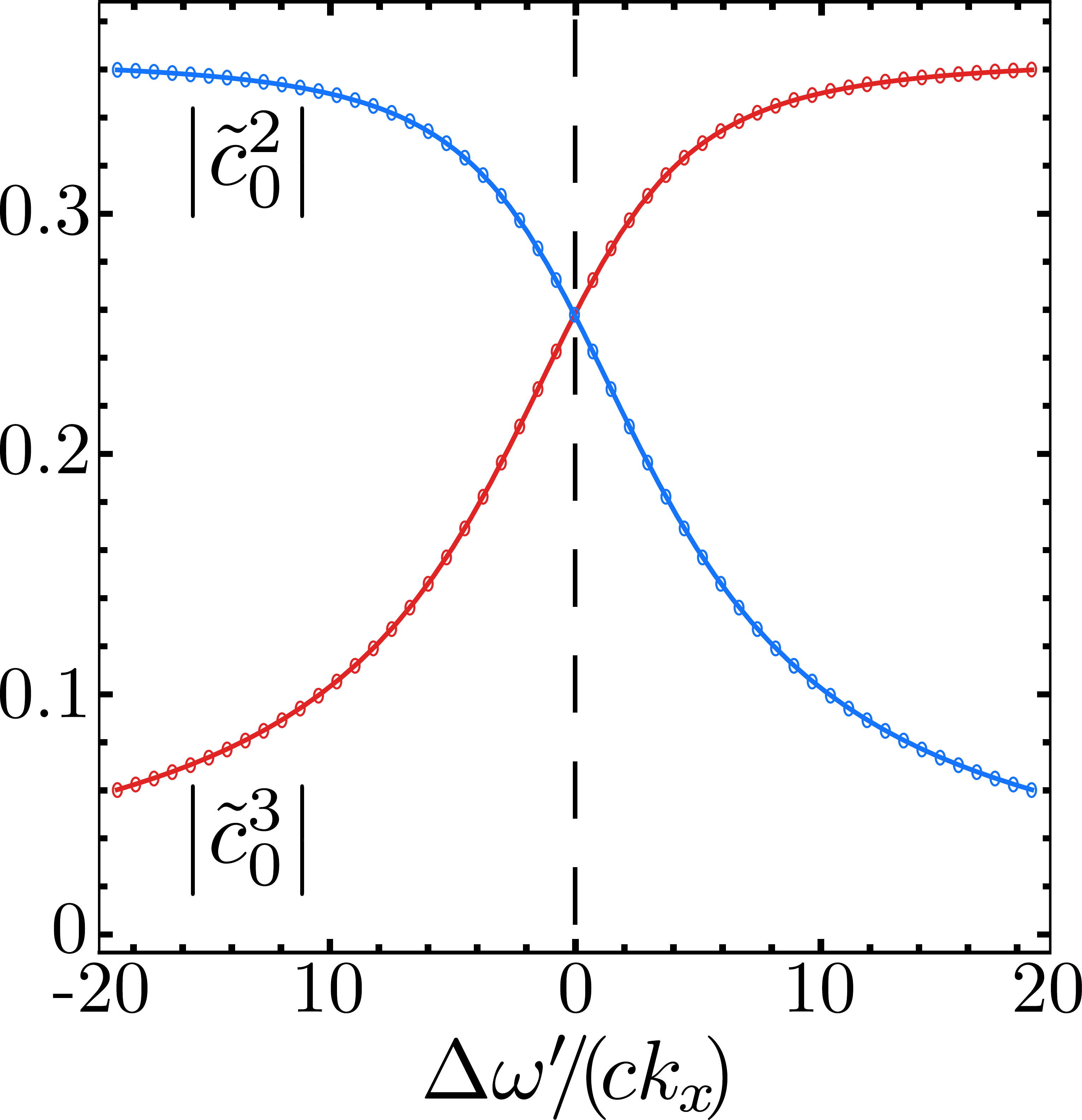}
    \caption{The amplitude $\tilde{c}_0^n$ of the $p=0$ spatial Fourier harmonic within the $n=2$ and $n=3$ infinite-grating eigenmodes in Eq.~(\ref{ticlde c_p^n}) as the universal function (\ref{tildec_0^n vs kx}) (solid curves) or the original Fourier integral (\ref{ticlde c_p^n}) (dots) plotted versus the scaled frequency detuning $(\omega - \omega_c)/(ck_x)$ at small in-plane wavenumbers, $|k_x\lambda_g| \ll 1$; $\omega'' = 0$. The grating parameters are $\varepsilon_1 = 6.25$, $\varepsilon_2 = 3.9, \rho = 0.39183$, hence, the spectral width parameter in Eqs.~(\ref{v}), (\ref{tildec_0^n vs kx}) is $w = 6.95$.}
    \label{Fig:c_0^nDip1}
\end{figure}

The result (\ref{tildec_0^nTaylor}) generalizes the asymptotics in Eq.~(\ref{tildec_0^n near pi}) to the case of nonzero in-plane wavenumbers, $k_x \neq 0$, and an arbitrarily small frequency detuning $\Delta \omega$. Its dependence on the scaled frequency detuning $\Delta \omega/(ck_x)$ at small in-plane wavenumbers, $|k_x\lambda_g| \ll 1$, is universal (see Fig.~\ref{Fig:c_0^nDip1}). Mode crossing at the degenerate frequency $\omega_c$ amounts to almost abrupt change of spatial parities of the $n=2,3$ infinite-grating eigenmodes from pure odd to pure even and vice versa over a very narrow frequency range proportional to the in-plane wavenumber $k_x$ and dimensionless spectral width parameter $w$,
\begin{equation} \label{v}
\Delta \omega \sim \Omega=wc k_x; \ w = 2 \Big[\frac{r^3+1}{(\varepsilon_1-\varepsilon_2)(r-1)^3}\Big]^{\frac{1}{2}}, r = \frac{d_2}{d_1}.
\end{equation}

Analytical results in Eqs. (\ref{b+-Taylor})-(\ref{tildec_0^nTaylor}) allow us to find the universal, self-similar structure of this narrow resonance in the vicinity of the mode-crossing point $D_{2,3}^0$ at small $x$-wavenumbers, $|k_x\lambda_g| \ll 1$, explicitly as follows
\begin{equation} \label{tildec_0^n vs kx}
\begin{split}
&\tilde{c}_0^n \approx \frac{2\sqrt{2}(r-1)(1+r)^{\frac{1}{2}} k_x\lambda_g}{\pi [(1+r^3)(k_x\lambda_g)^2 + (1+r)(\chi_1 + r\chi_2)^2]^{\frac{1}{2}}}\\
&=\frac{\big[4\sqrt{2}r/(\pi \eta)\big]}{\sqrt{w^2 + \Big[ \frac{\Delta \omega}{ck_x} +(-1)^n \sqrt{w^2+\left(\frac{\Delta \omega}{ck_x}\right)^2} \,  \,\Big]^2}} \ , \ n=2, 3.
\end{split}
\end{equation}
The result in Eq.~(\ref{tildec_0^n vs kx}) describes the universal resonant transition of each of the crossing infinite-grating eigenmodes $n=2, 3$ from even to odd or from odd to even parity at the mode-crossing point. It is reminiscent of a very steep step function, but has a finite spectral width determined by just one dimensionless parameter $w$, Eq.~(\ref{v}). It sets the spectral width of the degenerate BIC and depends on the combined geometrical ($d_1 \neq d_2$) and dielectrical ($\varepsilon_1 \neq \varepsilon_2$) asymmetry of the grating. In the next two sections V and VI we show that this universality of the parity symmetry breaking at the mode-crossing point leads to the universal shape of the degenerate BIC resonance since the zeroth-Fourier-harmonic contribution $\tilde{c}_0^n$ in Eq.~(\ref{tildec_0^n vs kx}) is directly responsible for the radiation losses of the infinite-grating eigenmodes.

Remarkably, the zeroth-Fourier-harmonic contribution at the degenerate frequency (that is, at zero detuning, $\Delta\omega=0$) is the same for both crossing infinite-grating eigenmodes $n=2, 3$ and has a definite finite value 
\begin{equation} \label{tildec0n(wc)}
\tilde{c}_0^n (\omega_c) = \frac{2(r-1)}{\pi} \sqrt{\frac{r+1}{r^3 +1}} \  
\end{equation}
which depends only on the grating asymmetry parameter $r=d_2/d_1$ and does not depend on a value of the in-plane wavenumber if it is relatively small, $k_x \ll \Delta \omega/(cw)$. It means that at the mode-crossing point both infinite-grating eigenmodes $n=2,3$ have a mixed, not purely odd or even, spatial parity.

One can verify that for the value of the asymmetry parameter $r = 1/\rho -1 \approx 1.55$, corresponding to the filling factor $\rho \approx 0.39183$ adopted in Fig.~\ref{Fig:c_0^nDip1}, the resonant, $\tilde{c}_0^n(\omega_c) \approx 0.258$, and asymptotic, $\tilde{c}_0^n \approx 0.365$, values of the zeroth-Fourier-harmonic contribution, calculated numerically in Fig.~\ref{Fig:c_0^nDip1} via its original, implicit integral representation in Eq. (\ref{ticlde c_p^n}), coincide with those given by the explicit analytical formulas (\ref{tildec_0^n vs kx}), (\ref{tildec0n(wc)}). Far from the degenerate frequency, $|\Delta \omega| \gg wc |k_x|$, the amplitude of the zeroth Fourier harmonic tends to zero on one side of the resonance and saturates at the finite value $\sqrt{2} \tilde{c}_0^n (\omega_c)$ on the other side. The latter is greater than the resonant value (\ref{tildec0n(wc)}) by a universal factor $\sqrt{2}$.

\section{Designing a waveguide that supports a degenerate bound state in the continuum at a given frequency}

Below we employ the analytical formulas of sections III and IV to exemplify the design, properties and mechanism for the formation of the BIC originating from the degenerate point $D_{2,3}^0$ which is the first in the hierarchy of degenerate points shown in Fig.~\ref{dispCurveMulti}. As is stated in sect. I, we focus on a particular case of the BIC eigenmode with almost zero in-plane wavenumber, $k_x \approx 0$, in the grating waveguide with just one central, zeroth spatial Fourier harmonic leaking into the cover and substrate.

Remarkably, such a BIC is supported by the grating waveguide as long as the crossing $D_{2,3}^0$ occurs at the degenerate frequency $\omega_c$ satisfying the condition in Eq.~(\ref{p=0emitted}), 
\begin{equation} \label{omega-crit}
\omega_c \lambda_g/c < 2\pi/\sqrt{\varepsilon^{\pm}} .
\end{equation}
The latter leaves only one radiation-loss channel open, occurring through emission of the $p=0$ spatial Fourier harmonic out of the grating layer into the cover and substrate, but prohibits radiation of all higher-order ($p \neq 0$) spatial Fourier harmonics, i.e., diffraction orders.

Recall that there are six dimensionless parameters of the waveguide (see Fig.~\ref{Fig:Geometry}). First, there are three dimensionless parameters of the lamellar grating: The high, $\varepsilon_1$, and low, $\varepsilon_2$, dielectric constants of the two grating sections and the fill factor of the high dielectric constant section, $\rho = d_1/\lambda_g$. All spatial dimensions are scaled by the period of the grating, $\lambda_g = d_1 + d_2$. Second, there are three other dimensionless parameters which set a global design of the planar grating waveguide: The dielectric constants of the cover, $\varepsilon^+$, and substrate, $\varepsilon^-$, materials enclosing the grating layer and the ratio of the thickness of the grating layer to the grating period, $L/\lambda_g$. 

To get the degenerate BIC at a given resonant frequency, it suffices to adjust any one or combination of the grating parameters (permittivities $\varepsilon_1, \varepsilon_2$, the fill factor $\rho$, and the period $\lambda_g$) so that the critical frequency of the mode-crossing point $D_{2,3}^0$ would coincide with that given frequency. This is because the degenerate BIC eigenfrequency $\omega_m$ is predetermined to be very close to the crossing frequency $\omega_c$. (Its dimensionless counterpart, $\omega_m \lambda_g/c$, is also scaled by the $\lambda_g$.) It can be achieved by tuning just one of the phase accumulations over the grating sections in Eq.~(\ref{kd=mpi}), say, $k_1d_1$, to $\pi$ since the second one, $k_2d_2$, will also become equal to $\pi$ automatically:
\begin{equation}  \label{kd}
k_1d_1 = \pi , \quad k_2d_2 = \pi \qquad \textrm{at} \quad \omega =\omega_c. 
\end{equation}

\begin{figure}[ht] 
    \centering
    \includegraphics[width=85mm]{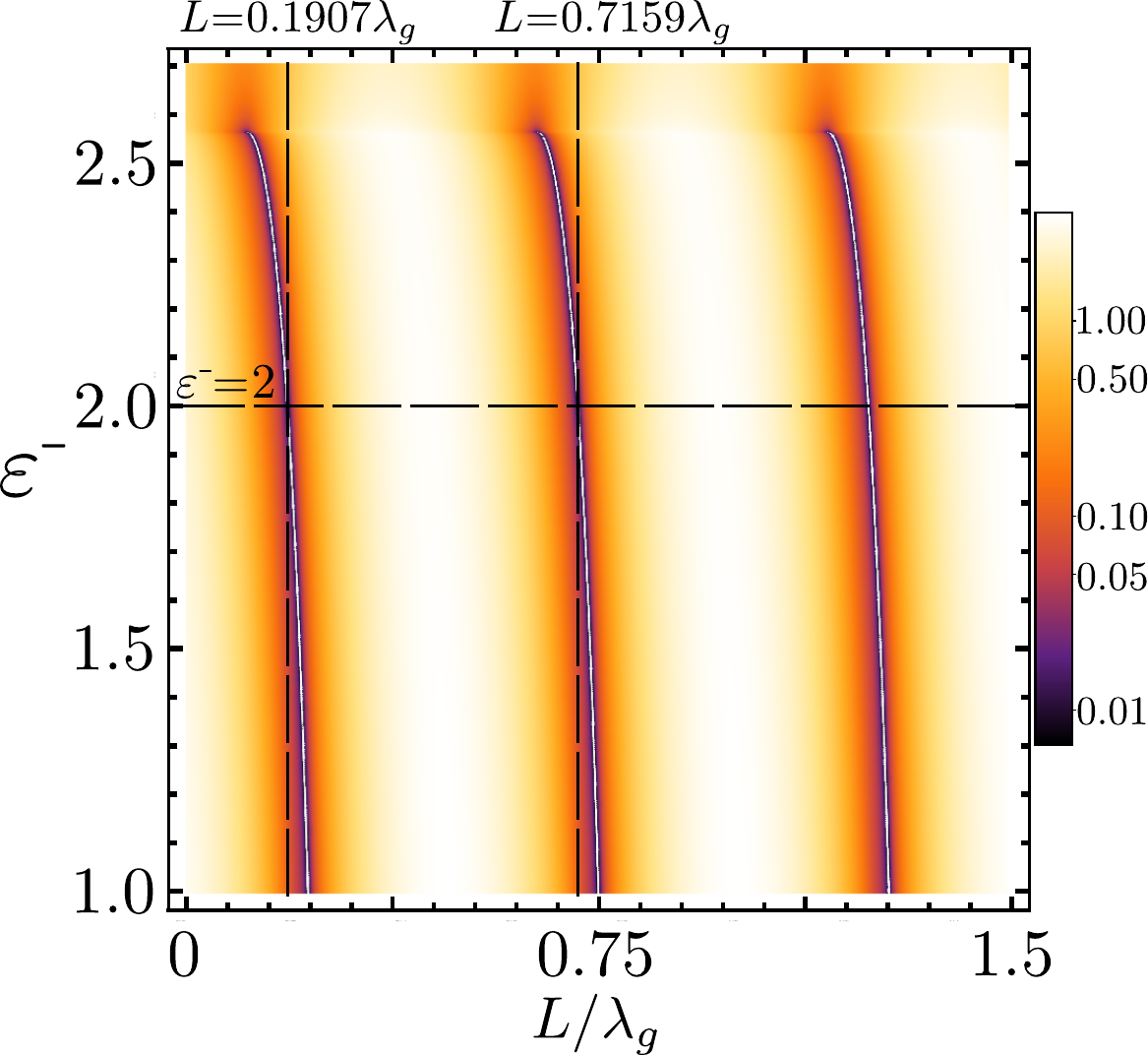}
    \caption{The absolute value of the characteristic function $\det (M-\mathbbm{1})$ of waveguide eigenmodes, Eq. (\ref{det}), as a function of the dimensionless grating layer thickness $L/\lambda_g$ and the permittivity of the substrate $\varepsilon^-$; $\varepsilon^+ = 1, \varepsilon_1 = 6.25, \varepsilon_2 = 3.9, \rho = 0.39183, \omega' = \omega_c = 4 c/\lambda_g, \omega'' = 0, k_x\lambda_g=10^{-5}$.}
    \label{Fig:detPlot-Eigenmode branches}
\end{figure}

For instance, let us take titanium oxide (TiO$_2$) as the material for the first section of the grating, that is, take the high dielectric constant to be $\varepsilon_1=6.25$, and choose the low dielectric constant of the second grating section to be, say, $\varepsilon_2 = 3.9$. Suppose we want to get the degenerate BIC with the dimensionless resonant frequency equalling exactly four, i.e., $\omega_c \lambda_g/c = 4$, as in Fig.~\ref{Fig:Dispersion}. We immediately achieve this goal by adjusting the fill factor to the value $\rho=0.39183$ that tunes the phase accumulation $k_1d_1$ at this frequency to $\pi$ as per Eqs.~(\ref{cpkz}), (\ref{kd}). 

Next, we need to select the other three waveguide parameters -- the thickness $L$ of the grating layer and the permittivities $\varepsilon^{\pm}$ of the cover and substrate. There is just one global constraint (\ref{omega-crit}) on the permittivities: They should be less than a certain threshold value above which the $p=\pm 1$ spatial Fourier harmonic starts to radiate out of the grating layer and into the cover or substrate. Such radiation would mean opening a radiation-loss channel, additional to the one provided by the zeroth Fourier harmonic, that would degrade the BIC to just a low-Q leaky resonance. For the chosen resonant frequency $\omega_c \lambda_g/c = 4$, the constraint (\ref{omega-crit}) requires $\varepsilon^{\pm} < (\pi/2)^2$. 

The only nontrivial constraint on the waveguide parameters is the fulfillment of the characteristic equation (\ref{det}) for the considered BIC, that is, the waveguide eigenmode comprising the $n=2$ and $n=3$ infinite-grating eigenmodes and demonstrating high Q factor near the selected frequency $\omega_c$. Suppose one sets the dielectric constant of the cover to be some reasonable number, say,  $\varepsilon^+ =1$. Then the choice of the other two parameters, $L$ and $\varepsilon^-$, can be guided by the corresponding 3D plot of the absolute value of the waveguide characteristic function $D(\omega = \omega_c - i\omega'') = \det [M(\omega = \omega_c - i\omega'')-\mathbbm{1}]$ in Eq. (\ref{det}) considered as a function of the variables $L$ and $\varepsilon^-$ at the mode-crossing frequency and a vanishing decay rate (see Fig. \ref{Fig:detPlot-Eigenmode branches}). 

The center line of the black branches on this plot corresponds to the curve on the $(L, \varepsilon^-)$-plane of parameters where the determinant in Eq. (\ref{det}) approaches zero. In other words, the black branches indicate the narrow ranges of the waveguide parameters for which the waveguide supports the high-Q waveguide eigenmodes with the eigenfrequency close to the chosen one $\omega_c -i \omega''$. Note that the entries of the waveguide master matrix $M$ in Eq. (\ref{MF=F}) are calculated by means of the aforementioned explicit, analytical formulas via the $z$-wavenumbers $k_{zn}(\omega'')$ of the infinite-grating eigenmodes which now include the first-order correction $\Delta k_{zn} \propto \omega''$, Eq. (\ref{dkzn}), along with the zeroth-order solution $k_{zn}(\omega''=0)$ to the transcendental characteristic equation (\ref{CharactEq}) at the given real-valued mode-crossing frequency $\omega_c$. 

\begin{figure}
\includegraphics[width=85mm]{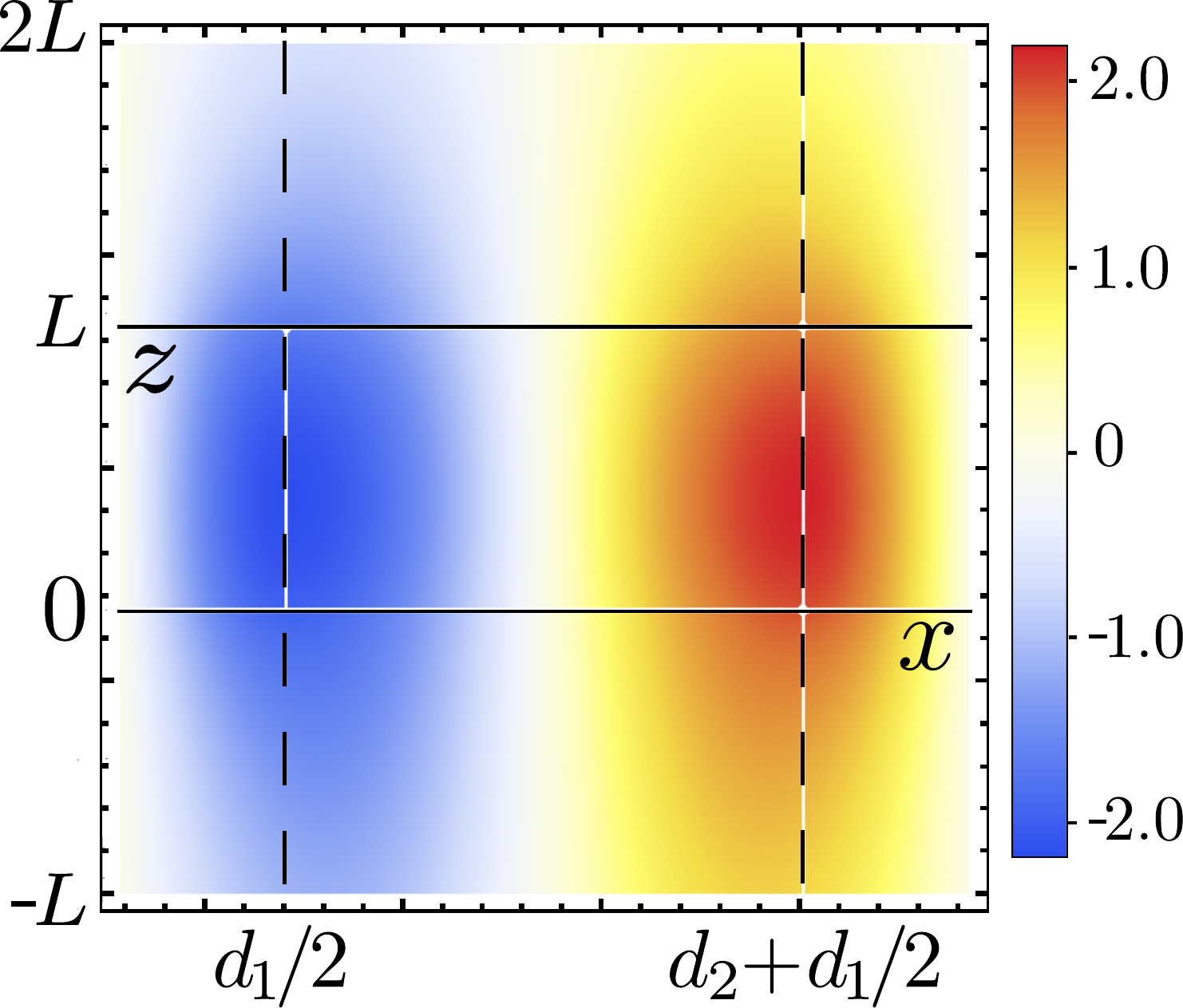}
\caption{Distribution of the electric field $\text{Re}(E_y^{(1,2)}(x,z))$, Eq. (\ref{E^(m)}), in the $xz$-plane of the cover, grating and substrate layers for the $m_z = 1, m_x = 2$ degenerate-BIC high-Q waveguide eigenmode. The grating layer thickness is $L = 0.1907 \lambda_g$. The permittivity of the cover and substrate is $\varepsilon^+ = 1$ and $\varepsilon^- = 2$, respectively; $\varepsilon_1 = 6.25, \varepsilon_2 = 3.9, \rho = 0.39183, k_x=0$.} 
\label{Fig:Field distribution m=1}
\end{figure}

In this paper we consider the BIC waveguide eigenmode associated with the arch-like branches shown in Fig.~\ref{Fig:detPlot-Eigenmode branches}. We enumerate them by the index $m_z = 1, 2, ...$ starting from the most left branch. In fact, the integer $m_z$ indicates approximately the number of half-wavelengths, stacked along the grating layer of thickness $L$ in the $z$ direction, in the transverse eigenmode's electric field distribution inside the grating layer. The corresponding distribution of the electric field $\text{Re}(E_y^{(m_z,m_x)}(x,z))$ in the $xz$-plane of the planar grating waveguide, computed by means of  Eq.~(\ref{E^(m)}), is shown in Fig.~\ref{2Dpatterns} and Fig.~\ref{Fig:Field distribution m=1} for the $m_z = 2$ and $m_z = 1$ waveguide eigenmodes with the grating layer thickness $L = 0.71587 \lambda_g$ and $L = 0.1907 \lambda_g$, respectively. The dashed vertical lines across Figs.~\ref{2Dpatterns}, \ref{Fig:Field distribution m=1} separate the two sections of the lamellar grating: the first section of length $d_1 = \rho \lambda_g$ and permittivity $\varepsilon_1$ and the second section of length $d_2 = \lambda_g - d_1$ and permittivity $\varepsilon_2$. In both cases the dielectric constants of the cover and substrate are chosen to be $\varepsilon^+ = 1$ and $\varepsilon^- = 2$, respectively. The latter choice ensures, according to Eq.~(\ref{omega-crit}), that all Fourier harmonics in the cover and substrate are evanescent except for the $p=0$ Fourier harmonic. Thus, only the zeroth Fourier harmonic freely propagates in the cover and substrate and, hence, provides the only channel of radiation losses for the BIC waveguide eigenmode. Figs.~\ref{2Dpatterns}, \ref{Fig:Field distribution m=1} confirm that the BIC slightly stands out from the grating layer due to the coupling of its infinite-grating eigenmodes with the evanescent spatial Fourier harmonics via the boundary conditions. Obviously, the fields penetrate inside the substrate deeper than inside the cover since the permittivity of the substrate is larger, $\varepsilon^- = 2 > \varepsilon^+ = 1$. 

Fig.~\ref{Fig:detPlot-Eigenmode branches} clearly shows that the BIC waveguide eigenmodes associated with the arch-like branches cease to exist as high-Q modes due to the opening of the second radiation-loss channel via the $p=\pm 1$ Fourier harmonics above the aforementioned cutoff value of the cover/substrate permittivity $\varepsilon^{\pm} = (\pi/2)^2 \approx 2.467$. (Note that very low-Q waveguide eigenmodes are not described within the linear or quadratic approximations in Eqs.~(\ref{dkzn}), (\ref{Deltakzn}) which are valid only if $\omega'' \ll \omega' \approx \omega_c$.) 

Hereinafter, all numerical calculations are easy and fast to perform since they are based on the explicit analytical formulas and a sufficient accuracy is achieved already in the approximation involving just a very few, namely, first five infinite-grating eigenmodes ($n=1,2,3,4,5;\,N=5$) and Fourier harmonics ($p\,=\,0,\,\pm 1,\,\pm 2; \ S=5$). In other words, only very small sized $5\times 5$ matrices are involved.

By knowing the exact location of a given waveguide eigenmode in the parameter space via the guidance provided by a 3D plot (such as in Fig. \ref{Fig:detPlot-Eigenmode branches}) of the determinant  $\det(M-\mathbbm{1})$ of the waveguide eigenmode problem, it is straightforward to calculate the real and imaginary parts of the waveguide eigenmode frequency $\omega_m = \omega'_m - i\omega''_m$ as functions of the grating layer thickness $L$ or other waveguide parameters. For instance, for any given set of waveguide parameters, one can use a 3D plot of the absolute value of the waveguide characteristic function $|\det [M(\omega' - i\omega'')-\mathbbm{1}]|$ over the complex $(\omega', \omega'')$-plane, magnify a small region around a predetermined position of the eigenfrequency, and then determine its value $\omega_m = \omega'_m - i\omega''_m$ corresponding to the zero of the characteristic function, $\det (M-\mathbbm{1}) = 0$. In this way, the complex eigenfrequency can be calculated quite accurately.

Equivalently, one can use a 3D plot of the norm $\| M{\bf F^{(m)}} - {\bf F^{(m)}} \|$ for a waveguide eigenmode ${\bf F^{(m)}}$, Eq.~(\ref{MF=F}), over the complex $(\omega', \omega'')$-plane. Its zeroes signify the complex eigenfrequencies $\omega_m$ of the waveguide eigenmodes. 

\begin{figure}[ht] 
    \centering
    \includegraphics[width=85mm]{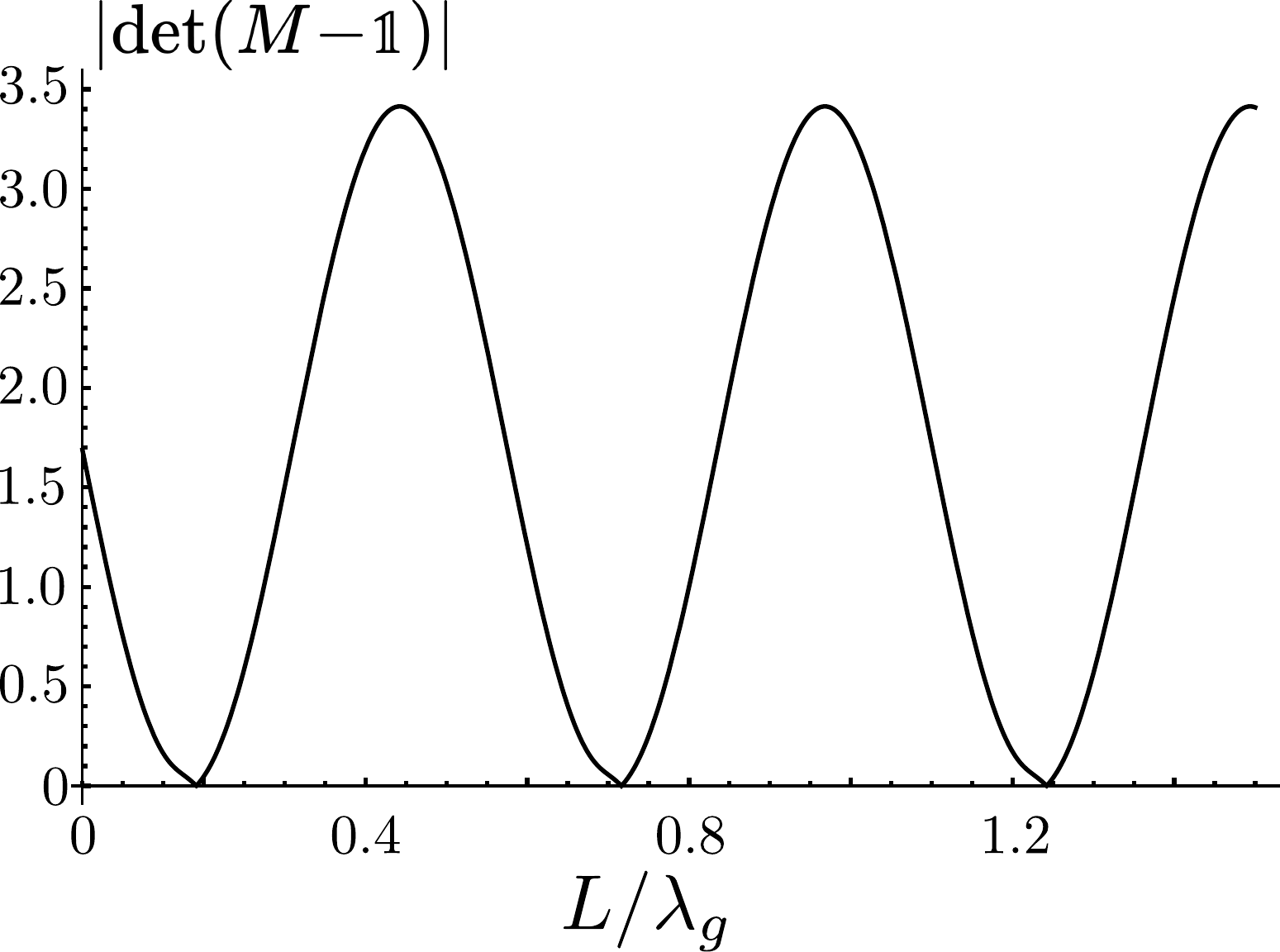}
    \caption{The absolute value of the waveguide characteristic function $\det (M-\mathbbm{1})$, Eq. (\ref{det}), versus the grating layer thickness, $L/\lambda_g$, at the degenerate frequency $\omega_c = 4 c/\lambda_g$, small decay rate $\omega''\approx 0$ and in-plane wavenumber \(k_x\lambda_g=10^{-5}\). Touching down the zero level at a discrete set of thicknesses definitively points to the series of transverse waveguide eigenmodes corresponding to the black branches along the cross section $\varepsilon^- = 2$ of the 3D plot in Fig.~\ref{Fig:detPlot-Eigenmode branches}.}
    \label{Fig:detPlot-L}
\end{figure}
Alternatively, one can plot a set of curves showing the absolute value of the waveguide characteristic function $|\det (M-\mathbbm{1})|$ as a function of one of the waveguide parameters, say, the grating layer thickness $L$, for a set of the real and imaginary parts of the frequency $\omega' - i \omega''$ in the vicinity of the BIC resonance.
A curve that approaches zero at some value of the waveguide parameter (say, the thickness $L$) signifies the presence of the BIC waveguide eigenmode of the corresponding eigenfrequency $\omega' - i \omega''$ at this value of the waveguide parameter.

Such a function plotted for the critical frequency $\omega_c$ and a given small decay rate $\omega''$ within a wider range of the waveguide parameter, like the one in Fig.~\ref{Fig:detPlot-L}, reveals an entire set of high-Q BIC and leaky waveguide eigenmodes. Each sharp tip that points to zero indicates an approximate value of the waveguide parameter for which the waveguide supports a high-Q eigenmode. In fact, Fig.~\ref{Fig:detPlot-L} is a cross section $\varepsilon^- = 2$ of the 3D contour plot in Fig.~\ref{Fig:detPlot-Eigenmode branches} and reveals the set of waveguide eigenmodes that is represented by black branches in Fig.~\ref{Fig:detPlot-Eigenmode branches}. 

\begin{figure}[ht] 
    \centering
    \includegraphics[width=85mm]{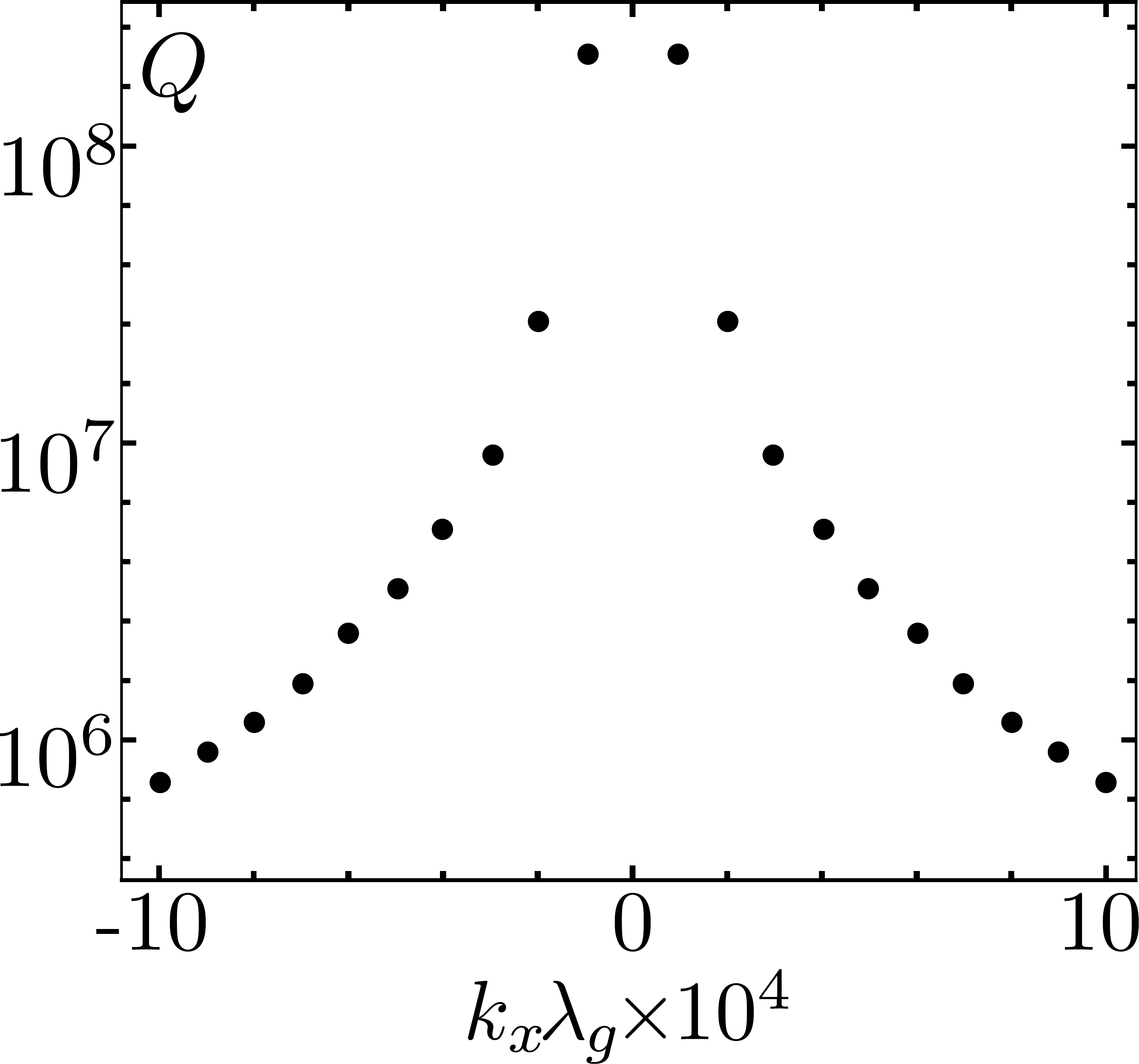}
    \caption{Narrow resonance in the Q factor of the degenerate BIC, formed by the high-Q (2,2)-eigenmode of the grating waveguide shown in Fig.~\ref{2Dpatterns}, as a function of the in-plane wavenumber $k_x$ around the $\Gamma$-point $k_x=0$.}
    \label{Fig:Q}
\end{figure}

Thus, one can easily calculate the quality factor $Q = \omega'_m/(2\omega''_m)$, and the eigenfrequency, $\omega_m$, of the BIC waveguide eigenmode as a function of any particular parameter of the waveguide, say, the grating layer thickness $L$. For example, the degenerate-BIC (2,2)-eigenmode of the grating waveguide, specified in Fig.~\ref{2Dpatterns}, has a very high Q factor, $Q \approx 0.8\times 10^{6},$ even at the relatively large in-plane wavenumber $k_x\lambda_g = 10^{-3}$ throughout the entire vicinity of the degeneracy point when the BIC's eigenfrequency scans across the degenerate frequency, $\omega'_m (L) \approx \omega_c -1.8c(L-L_c)/\lambda_g^2$, as one varies the grating layer thickness $L$ around the value $L_c = 0.71587\lambda_g$ corresponding to the mode-crossing frequency $\omega_c = 4c/\lambda_g$. 
The result for the Q factor of the degenerate BIC as a function of the in-plane wavenumber reveals a steep resonance at $k_x\lambda_g \ll 1$ as is shown in Fig.~\ref{Fig:Q}. Its scaling at $k_x \to 0$ approximately follows an inverse square law, $Q \sim 1/k_x^2$. 
Figs.~\ref{Fig:detPlot-L}, \ref{Fig:Q} clearly show how robust, definitive and accurate the determination of the waveguide eigenmode frequency by means of the characteristic equation (\ref{det}) is. 

Calculation of the BIC Q factor can be verified by a different method which is based on the energy conservation for the field of a $m$-th high-Q waveguide eigenmode,
\begin{equation} \label{EC}
-\frac{\partial \mathcal{E}^{(m)}}{\partial \textit{t}} = \mathcal{A} (S_z^{+} + S_z^{-}) .
\end{equation}
The energy of the electromagnetic field of the waveguide eigenmode within the substrate-grating-cover sandwich of a thickness $\mathcal{L} \sim 3L$ and cross-sectional area $\mathcal{A} \sim \lambda_g^2$,
\begin{equation} \label{Em}
\mathcal{E}^{(m)} = \frac{e^{-2\omega''t}}{16\pi}\int_V\left[\varepsilon(x,z)\left|E_y\right|^2 + \left|H_x\right|^2 + \left|H_z\right|^2 \right]\,\mathrm{dV},
\end{equation}
decays in time with the rate $2\omega''_m$ due to the radiation flux going away into the cover and substrate via the outgoing $z$-component of the Poynting vector, $\mathbf{S} = (c/(8\pi))\mathrm{Re} [\mathbf{E}\times \mathbf{H}^*]$, of the zeroth Fourier harmonic in the cover, $S_z^{+}$, and substrate, $S_z^{-}$, respectively. It is straightforward to calculate (a) the integral in Eq.~(\ref{Em}) via the known spatial distribution of the electric and magnetic fields $E_y^{(m)}(x,z), H_x^{(m)}(x,z) = (ic/\omega_m) (\partial E_y^{(m)}/\partial z), H_z^{(m)}(x,z) = (-ic/\omega_m) (\partial E_y^{(m)}/\partial x)$ (see Figs.~\ref{2Dpatterns}, \ref{Fig:Field distribution m=1}) and (b) the Poynting vector for a plane wave of the zeroth Fourier harmonic. Then the decay rate and Q factor are given by Eq.~(\ref{EC}) as follows 
\begin{equation} \label{DR}
\omega''_m = \frac{\mathcal{A} (S_z^{+} + S_z^{-})}{2\mathcal{E}^{(m)}}, \qquad Q = \frac{\omega'_m}{2\omega''_m} .
\end{equation}
We have verified that the energy conservation method in Eq.~(\ref{DR}) gives the same result for the Q factor of the degenerate BIC waveguide eigenmode as the method of the waveguide characteristic equation outlined above.

In reality the Q factor never reaches infinity but saturates at a finite maximum value which depends on the in-plane wavenumber $k_x \neq 0$ and is determined also by the finite size of the grating in the $x$ and $y$ directions, the impurities, defects and inhomogeneity of the waveguide material, roughness of the waveguide, cover and substrate surfaces, as well as other similar effects. Such a behavior is described by saying that, in reality, the BIC appears as a quasi-BIC \cite{Hsu-Nature2016,Rybin2017}.

\section{Origin of the degenerate BIC: Parity symmetry breaking and decoupling from the radiation-loss channel due to destructive interference at mode crossing}

The origin of the degenerate BIC described above can be explained as follows. 
The result of coupling and restructuring of the infinite-grating eigenmodes near the mode-crossing point on the radiation losses and Q factor of the $m$-th BIC waveguide eigenmode is determined by the amplitude of two zeroth Fourier harmonics, $g_0^{(m)\pm}$, emitted out of the grating layer into the cover and substrate. Their amplitudes are given by the $p=0$ component of the Fourier-harmonic column $S$-vectors, ${\bf g^{(m)\pm}} = (\{g_p^{(m)\pm} |p=0, \pm 1,...,\pm p^* \})^T$, each containing $S = 2p^* + 1$ amplitudes of the co- or counter-propagating Fourier harmonics emitted from the grating boundaries with the cover ($z=L$) or substrate ($z=0$), respectively. The Fourier harmonic vectors are given by the transmittance matrix in Eq. (\ref{T}) applied to the corresponding waveguide eigenmode vector, Eq. (\ref{MF=F}), of the amplitudes of the infinite-grating eigenmodes:
\begin{equation} \label{g}
{\bf g^{(m)+}} = T^+ \Lambda {\bf F^{(m)}}, \qquad {\bf g^{(m)-}} = T^- \Lambda R^+ \Lambda {\bf F^{(m)}} .
\end{equation} 

The radiation losses of the $m = (2,2)$ degenerate BIC, as measured by the amplitudes $g_0^{(m)\pm}$ of the outgoing (into the cover and substrate) zeroth Fourier harmonics, vary across the parity symmetry breaking region, that is, with a resonant restructuring of the contributions $\tilde{c}_0^n$ of the zeroth Fourier harmonic to the $n=2, 3$ infinite-grating eigenmodes shown in Fig.~\ref{Fig:c_0^nDip1}. 

Consider the same example of the grating waveguide as was discussed above: $L = 0.71587 \lambda_g, \ \varepsilon^- = 2, \ \varepsilon^+ = 1, \varepsilon_1 = 6.25, \varepsilon_2 = 3.9, \rho = 0.39183$. At the center of the degenerate BIC resonance, when the frequency $\omega_m$ of the BIC waveguide eigenmode approaches the mode-crossing frequency $\omega_c$, the decay rate $\omega''_m$ remains very small and the eigenvector of this $m = (2,2)$ BIC waveguide eigenmode at $k_x \lambda_g = 10^{-5}$ (see Fig.~\ref{2Dpatterns}) tends to
\begin{equation}\begin{split} \label{eigenvector}
{\bf F^{(m)}} = (0, 0.704, -0.704, -0.084-i0.039, 0)^T .
\end{split}\end{equation}

Its contents reveals an equal presence of the second and third infinite-grating eigenmodes with the amplitudes $F^{(m)}_{2,3} \approx \pm \frac{1}{\sqrt{2}}$ and a minor admixture of the fourth one, while the rest of the infinite-grating eigenmodes, the first, fifth, and higher, don't contribute significantly to the degenerate-BIC waveguide eigenmode at resonance.

Throughout the entire resonance, the degenerate-BIC waveguide eigenmode is decoupled from all even-parity infinite-grating eigenmodes. Also, one should take into account that the higher-order infinite-grating eigenmodes (such as the 4th, 5th and higher-order ones in this example) are usually highly evanescent (as per Figs.~\ref{dispCurveMulti}, \ref{Parabolas-extendedBrillouin}). So, even if they are present in the BIC waveguide eigenmode at the starting point (say, at $z=+0$) of their propagation path through the grating layer, they would contribute exponentially little to the radiation losses at the exit out of the grating layer (say, in the cover at $z=+L$).

The amplitudes of the Fourier harmonics constituting the $n$-th infinite-grading eigenmode for the chosen complex frequency and waveguide parameters are given by the $n$-th column of the $\tilde{C}$ matrix, 
\begin{equation} \label{C-matrix}\begin{split}
\{ \tilde{c}_p^n \} &=
  \left[ {\begin{array}{ccccc}
 0.027 & 0.068 & -0.025 & 0.545 & 0.441 \\
 -0.085 & -0.320 & 0.010 & 0.028 & -0.046 \\
 -0.931 & 0.258 & 0.258 & 0 & 0 \\
 -0.085 & 0.010 & -0.320 & -0.0286 & 0.047 \\
 0.027 & -0.025 & 0.068 & -0.545 & -0.444 \\
\end{array}} \right]
\\
&+i\left[
\begin{array}{ccccc}
 -0.022 & -0.055 & 0.020 & -0.441 & 0.548 \\
 -0.242 & -0.906 & 0.030& 0.081 & 0.016 \\
0 & 0 & 0 & 0 & 0.024 \\
 0.241 & -0.030 & 0.906 & 0.081 & 0.016 \\
 0.022 & -0.020 & 0.055 & -0.440 & 0.547 \\
\end{array}
\right] ,
\end{split}\end{equation}
calculated by means of the explicit formulas in Eqs.~(\ref{ticlde c_p^n}), (\ref{tildec_p^n}). The spatial Fourier harmonic of the $p$-th diffraction order is represented by the $p$-th row. The rows are enumerated from left to right by the integer $p = -2, -1, 0, 1, 2$. The degeneracy between the infinite-grating eigenmodes leads to the relation $f_2 = f_3^*$ and implies the symmetry $\tilde{c}^2_p = (\tilde{c}^3_{-p})^*$ for the 2-nd and 3-rd columns for all $p$ at the mode-crossing frequency and $k_x \approx 0$. According to Eq.~(\ref{C-matrix}), the $p=0$ Fourier harmonic is strongly present in both infinite-grating eigenmodes $n=2, 3$ intersecting at the mode-crossing point in agreement with the universal result in Eq.~(\ref{tildec_0^n vs kx}) (see Fig.~\ref{Fig:c_0^nDip1}), but is absent in the only other, odd-parity mode $n=4$ constituting, along with the $n=2, 3$ modes, the degenerate-BIC waveguide eigenmode (\ref{eigenvector}). The corresponding coefficients $\tilde{c}^n_0$ are crutially important since they directly determine the partial contribution of each infinite-grating eigenmode to the radiation field of the zeroth Fourier harmonic, emitted out of the grating layer into the cover, $g_0^{(m)+}$, and substrate, $g_0^{(m)-}$. The couplings $\tilde{c}^n_0$ with the radiation channel are given by Eqs.~(\ref{tildec_p^n}), (\ref{tildec_0^n}), (\ref{tildec_0^n near pi}), (\ref{tildec_0^n vs kx}).

\begin{figure}[ht] 
    \centering
    \includegraphics[width=85mm]{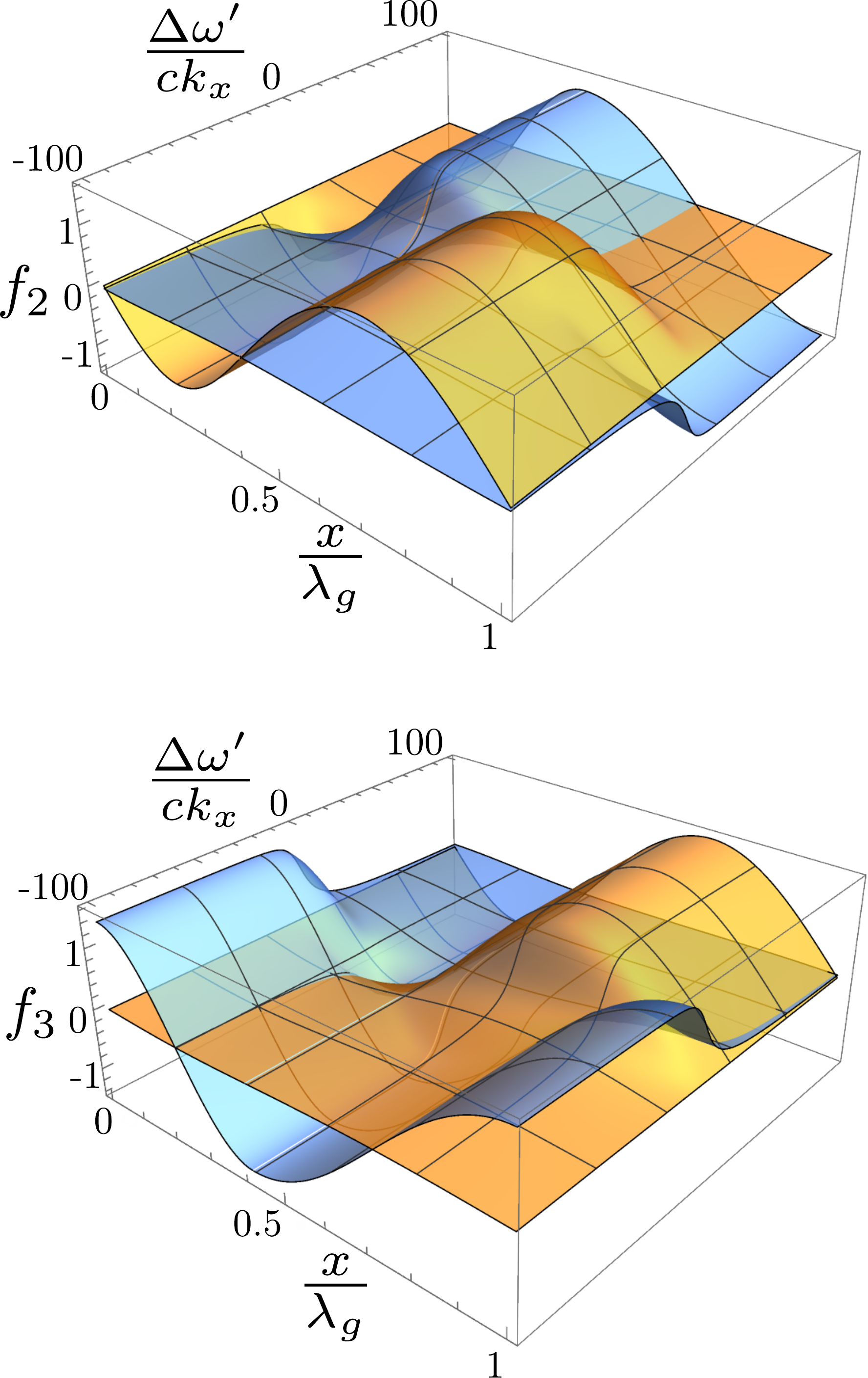}
    \caption{Parity symmetry breaking of the $n=2$ (top) and $n=3$ (bottom) infinite-grating eigenmodes: Nontrivial restructuring of their real $\rm{Re} [f_n(x)]$ (orange) and imaginary $\rm{Im} [f_n(x)]$ (blue) $x$-profiles in the complex plane as the frequency $\omega'$ varies across the mode-crossing frequency $\omega_c = 4 c/\lambda_g$; $\varepsilon_1 = 6.25, \varepsilon_2 = 3.9, \rho = 0.39183$, $k_x\lambda_g = 10^{-5}$.}
    \label{x-profiles_n=2}
\end{figure}

The essence of the degenerate BIC is that throughout the entire region of degeneracy it stays decoupled from the only available radiation channel, carried out by the $p=0$ Fourier harmonic, and possesses a very high Q factor. This phenomenon is possible only due to the destructive interference of the $n=2, 3$ infinite-grating eigenmodes since within the transition region of their restructuring, i.e., parity symmetry breaking, they possess mixed-parity $x$-profiles as per Fig.~\ref{x-profiles_n=2}. Tuning the waveguide parameters, say, the grating layer thickness $L$, in such a way that BIC waveguide (2,2)-eigenmode frequency $\omega'_m$ varies across the resonance with the degenerate, mode-crossing frequency $\omega_c$ greatly, by many orders of magnitude, changes the couplings $\tilde{c}_0^2, \tilde{c}_0^3$ as per Fig.~\ref{Fig:c_0^nDip1} and forces the $x$-profiles of the $n=2, 3$ infinite-grating eigenmodes to restructure from the pure even or odd on one side of the resonance to the opposite, pure odd or even, on the other side of the resonance. 
Radiation losses are determined by the $x$-profile of the waveguide-eigenmode field incident onto the substrate and cover from the grating layer which is given by the second and first lines in Eq.~(\ref{E^(m)}), respectively. The corresponding eigenvectors of the infinite-grating-eigenmode amplitudes, ${\bf F^{(m)-}} = {\bf \Lambda R^+ \Lambda F^{(m)}}$ and ${\bf F^{(m)+}} = {\bf \Lambda F^{(m)}}$, are given by the solutions of the waveguide master equation (\ref{MF=F}). For example, consider the degenerate-BIC high-Q (2,2)-eigenmode of the grating waveguide, specified in Fig.~\ref{2Dpatterns}, at $k_x\lambda_g = 10^{-5}$ and the grating layer thickness adjusted so that the frequency detuning from the mode-crossing resonance, $\Delta \omega = \omega'_m -\omega_c$, is set to the center ($\Delta \omega = 0$) or the wing ($\Delta \omega = -2ck_x$) of the resonance. Then, the corresponding eigenvectors of the waveguide-eigenmode field incident on the substrate are as follows
\begin{equation} \label{F w=0}
{\bf F^{(m)-}} (\Delta \omega = 0) \approx (0, 0.7, -0.7, 10^{-4}e^{-i0.44}, 0)^T, 
\end{equation}
\begin{equation} \label{F w=-2}
{\bf F^{(m)-}} (\Delta \omega = -2ck_x) \approx (0, 0.6, -0.8, 10^{-4}e^{-i0.44}, 0)^T .
\end{equation}

We see that the relative weights of the $n=2,3$ infinite-grating eigenmodes in the contents of the $m = (2,2)$ high-Q waveguide eigenmode at the resonance wing, $F^{(m)-}_2 \approx 0.6, F^{(m)-}_3 \approx -0.8$, significantly differ from their values at the center of the resonance, $F^{(m)-}_{2,3} \approx \pm 1/\sqrt{2} \approx \pm 0.7$. Far away from the BIC resonance the $m = (2,2)$ high-Q waveguide eigenmode is totally dominated by just one of the $n=2,3$ infinite-grating eigenmodes -- the one which has the odd parity of the $x$-profile: $n=3$ for the negative frequency detuning or $n=2$ for the positive detuning. 

This restructuring of the contents of the $m = (2,2)$ high-Q waveguide eigenmode follows variation of the couplings $\tilde{c}_0^2, \tilde{c}_0^3$ as per Fig.~\ref{Fig:c_0^nDip1} and occurs consistently with the parity symmetry breaking of the $n=2,3$ infinite-grating eigenmodes shown in Fig.~\ref{x-profiles_n=2}. In virtue of such a synchronous variation of the $F^{(m)\pm}_{2,3}$ and $\tilde{c}_0^{2,3}$, the $x$-profile of the field of the degenerate-BIC waveguide eigenmode incident onto the boundaries of the grating layer, $E_y^{(m)} \approx F^{(m)\pm}_2 f_2(x) + F^{(m)\pm}_3 f_3(x)$, remains of almost odd parity since the amplitude of its zeroth Fourier harmonic stays close to zero throughout the entire transition region, $F^{(m)\pm}_2 \tilde{c}_0^2 + F^{(m)\pm}_3 \tilde{c}_0^3 \approx 0$. As a result, the radiation losses stay always very weak and the degenerate-BIC waveguide (2,2)-eigenmode remains the high-Q BIC in the entire mode-crossing region. Outside the mode-crossing region, the degenerate BIC evolves into the conventional symmetry-protected BIC consisting mainly of a single odd-parity infinite-grating eigenmode $n=2\, \rm{or}\, 3$. 

Symmetry breaking of the $f_2(x)$ and $f_3(x)$ profiles is caused by the avoiding crossing of the dispersion curves $k_{z2}(\omega)$ and $k_{z3}(\omega)$ shown in Fig.~\ref{Fig:Dispersion}. Away from the mode-crossing point, the $n=2$ and $n=3$ infinite-grating eigenmodes have $x$-profiles of almost pure opposite parities in virtue of the $\pi$-phase accumulations over both grating sections as per Eqs.~(\ref{kd=mpi}), (\ref{kd}). Namely, at large negative detunings, the $x$-profiles of the $n=2$ and $n=3$ infinite-grating eigenmodes are even and odd, respectively, on each uniform segment of the grating relative to the center of this segment. However, due to the degeneracy of those two eigenmodes at the mode-crossing point, this parity symmetry gets inevitably broken. It happens very close to the point of degeneracy, within a narrow resonance region $\Delta \omega \sim \Omega = wck_x$, which is determined by the dimensionless spectral width parameter $w$ in Eq.~(\ref{v}) and scales proportionally to the in-plane wavenumber $k_x$, and can be described by the self-similar variable $\Delta \omega /(ck_x)$ as per Eq.~(\ref{tildec_0^n vs kx}) (see Fig.~\ref{Fig:c_0^nDip1}). 

It is worth noting that the parity symmetry breaking occurs gradually by expanding the $x$-profiles $f_n(x), n=2, 3,$ into the complex-valued $x$-profile functions, that is, via a gradually increasing admixture of the real-valued odd function $\rm{Re} [f_n(x)]$ to the imaginary-valued even function $\rm{Im} [f_n(x)]$, as is clearly illustrated in Fig.~\ref{x-profiles_n=2}. 

The reason for such dramatic parity symmetry breaking of the $n=2, 3$ infinite-grating eigenmodes is related to the existence of two different universality classes $\eta_n = \pm 1$ of the $x$-profile asymptotics (scaling) when approaching the degenerate point $D_{2,3}^0$ as per Eq.~(\ref{b0}). 

\begin{figure}[ht]   
    \centering
    \includegraphics[width=5cm]{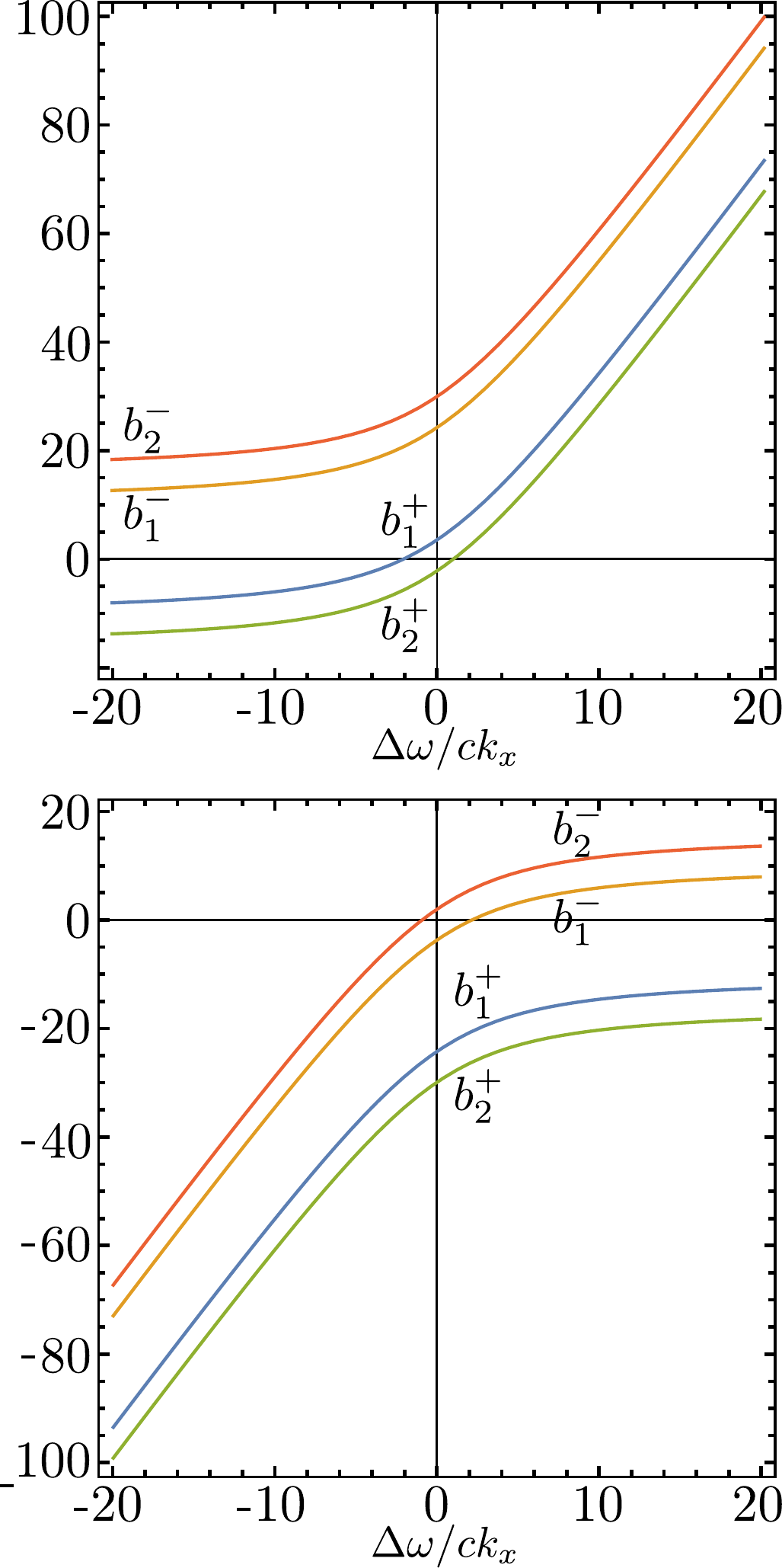}
    \caption{Parity symmetry breaking expressed via the universal behavior (\ref{b+-Taylor}) of the amplitudes $b_q^{\pm}, q=1, 2,$ of the waves in the $x$-profile, Eq.~(\ref{f_n}), of the $n=2$ (top plot) and $n=3$ (bottom plot) infinite-grating eigenmodes: $-ib_q^{\pm}/(k_x\lambda_g)$ as a function of the scaled frequency detuning $(\omega - \omega_c)/(ck_x)$ in the vicinity of the mode-crossing frequency $\omega_c = 4c/\lambda_g$; $\varepsilon_1 = 6.25, \varepsilon_2 = 3.9, \rho = 0.39183, \omega''=0, k_x \lambda_g = 10^{-5}$.}
    \label{bq+-_n=2}
\end{figure}

\begin{figure}[ht]  
    \centering
    \includegraphics[width=5cm]{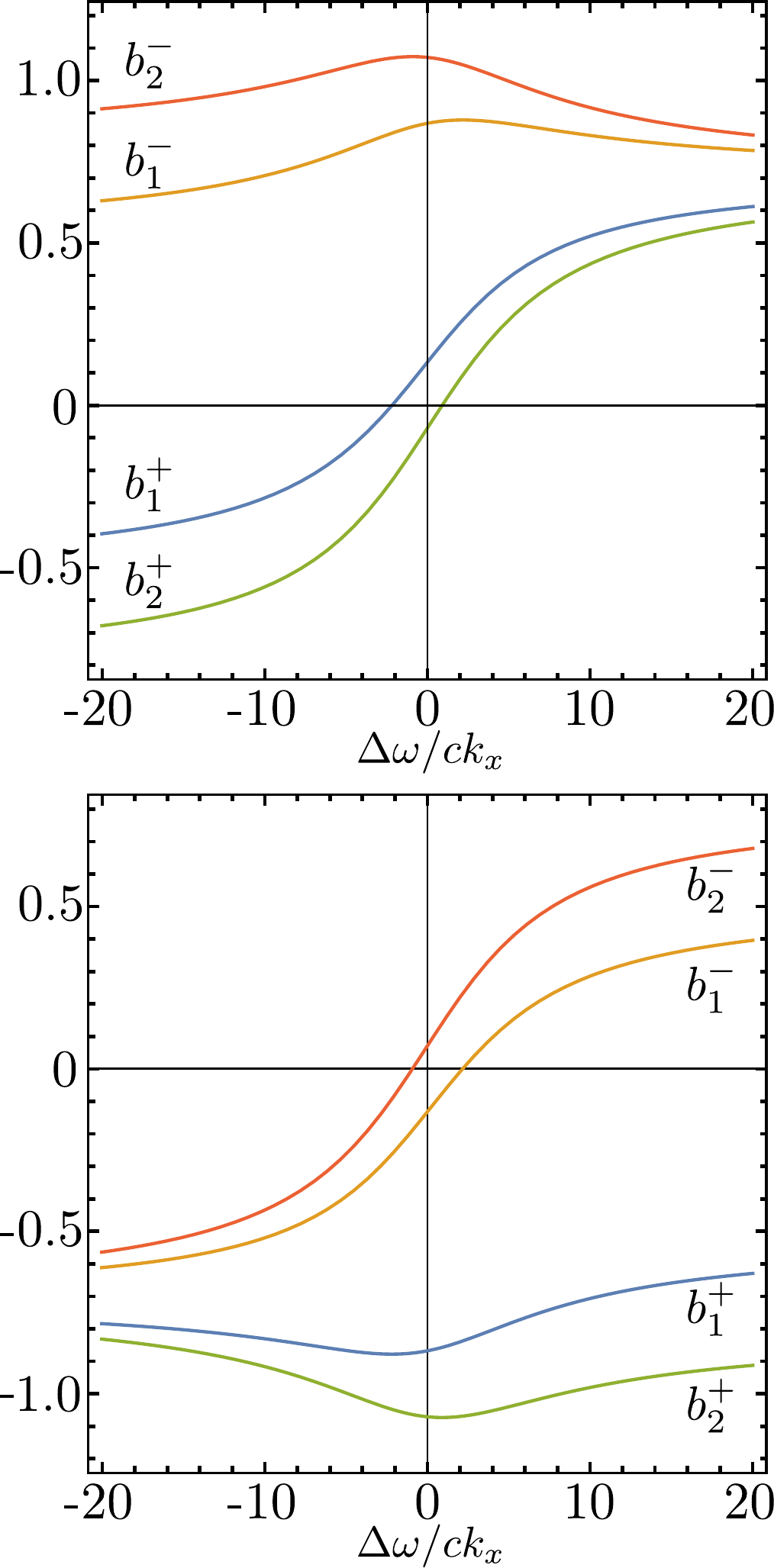}    
    \caption{Universal scaling of parity symmetry breaking in a vicinity of the degenerate frequency: The same as in Fig.~\ref{bq+-_n=2} amplitudes $b_q^{\pm}$ in the $x$-profile (\ref{f_n}) of the $n=2$ (top plot) and $n=3$ (bottom plot) infinite-grating eigenmodes but scaled now with their norms in Eqs.~(\ref{norm}), (\ref{normTaylor}), $-ib_q^{\pm}/\sqrt{\langle B_q^{\dagger}, B_q \rangle}$.}
    \label{pigeon_bq+-_n=2}
\end{figure}

The explicit self-similar resonant behavior revealed in Eq.~(\ref{tildec_0^n vs kx}) also manifests itself in the exact solution (\ref{b+-}) for the amplitudes $b_q^{\pm},\,q=1,2$, of the waves in the $x$-profile of the infinite-grating eigenmodes, Eq.~(\ref{f_n}). As a result, we get a full detailed description of the universal scaling of the parity symmetry breaking in the vicinity of the degenerate point $D_{2,3}^0$, that is, near the degenerate BIC high-Q resonance. The nontrivial universal behavior of the amplitudes $b_q^{\pm}$ for both sections of the grating ($q=1,2$) in the $n=2, 3$ infinite-grating eigenmodes is determined by the scale of the in-plane wavenumber $k_x$ and the decay rate $\omega''$ and is given by the explicit approximate solution in Eq.~(\ref{b+-Taylor}) illustrated in Figs.~\ref{bq+-_n=2}, \ref{pigeon_bq+-_n=2}. In the vicinity of the degeneracy point the approximate solution (\ref{b+-Taylor}) perfectly matches the exact result in Eq.~(\ref{b+-}).     

There are no other high-Q BICs in the range of parameters associated with the aforementioned narrow resonance. In particular, this is a consequence of the fact that if the $n=2\, \rm{or}\, 3$ infinite-grating eigenmode has an even parity, then it is generally present in any waveguide eigenmode due to reflection at the boundaries with the cover and substrate. Yet any waveguide eigenmode associated with the even-parity infinite-grating eigenmode has a low Q factor everywhere outside this resonance due to the strong presence of the zeroth Fourier harmonic freely leaking outside the grating layer into the cover and substrate.

The analytical theory of the degenerate BIC outlined above explains in detail the paradox stated in sect. II, Fig.~\ref{fig:paradox}, and shows a clear path to verifying an existence of the degenerate BIC numerically, say, via RCWA.

\begin{figure}
    \includegraphics[width=85mm]{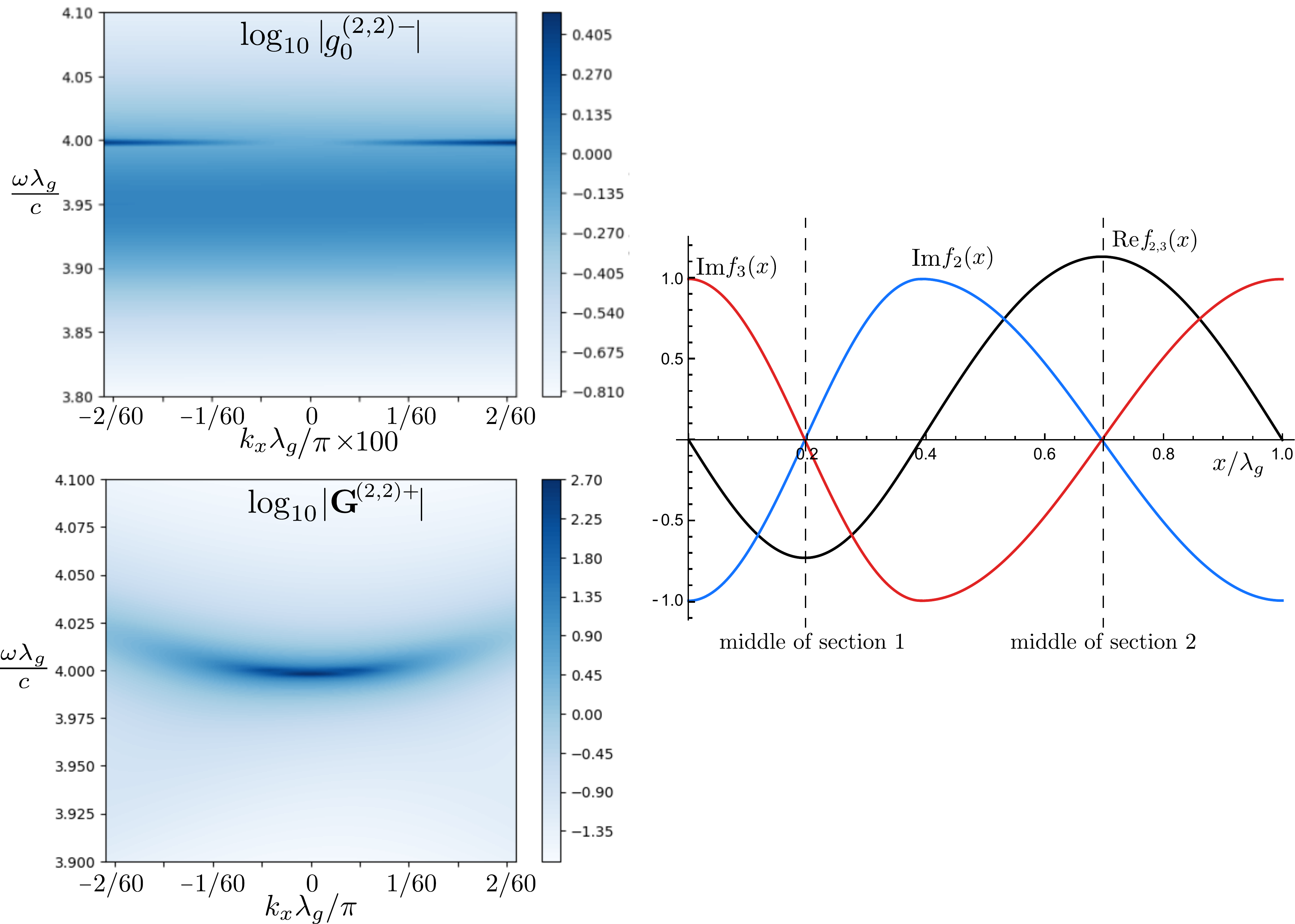}
\caption{Numerical rigorous coupled wave analysis (RCWA): Resonance response of the grating waveguide to a monochromatic source near the degenerate frequency $\omega_c = 4\lambda_g/c$. The top plot shows the intensity of the radiation emitted into the substrate if the unity-amplitude Fourier harmonics $p = 1$ and $p = -1$ are incident onto the grating waveguide from the cover at $z=L$. The upper branch (dark blue line) represents the high-Q waveguide (2,2)-eigenmode shown in Fig.~\ref{2Dpatterns} and has a break at $k_x \approx 0$ that reveals the existence of the degenerate BIC that was hidden in Figs.~\ref{fullRCWA}, \ref{fig:paradox}. The lower, wider and weaker, branch (moderately blue stripe) represents the low-Q waveguide (2,2)-eigenmode which overlaps the high-Q one but cannot fully hide the degenerate BIC. 
The bottom plot shows the intensity of the degenerate BIC at $z=L$ if a source excites the anti-phased superposition $f_2(x)-f_3(x)$ of the second and third infinite-grating eigenmodes inside the grating layer with the unity amplitude at $z=0$. 
The insert on the right shows the real and imaginary parts of x-profiles for the infinite-grating eigenmodes ($f_2(x)$, $f_3(x)$) constituting the $x$-profiles $f_2(x)-f_3(x)$ and $f_2(x)+f_3(x)$ of the high-Q (degenerate BIC) and low-Q (leaky) waveguide eigenmodes, respectively. 
The grating waveguide is the same as in Fig.~\ref{fullRCWA}.}
\label{fig:DegenBIC}
\end{figure}

\begin{figure}
\includegraphics[width=85mm]{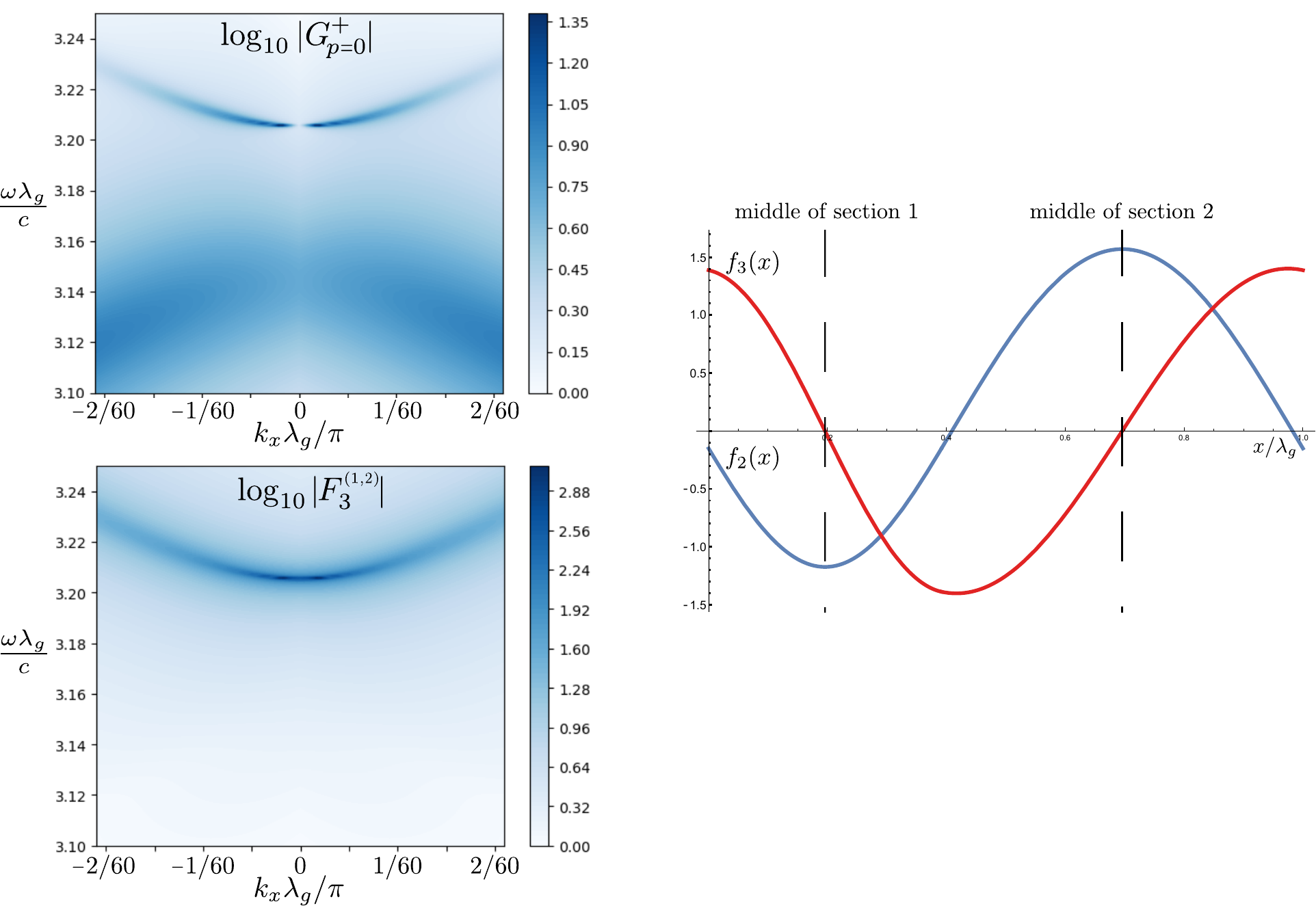}
\caption{Numerical rigorous coupled wave analysis (RCWA): Density plots (with the logarithmic $\log_{10}$ color scaling) of the intensity of the \(p=0\) Fourier harmonic emitted into the substrate at $z=0$ (top plot) and the intensity of the third infinite-grating eigenmode inside the grating at $z=L$ (bottom plot) computed as a resonance response to a monochromatic source consisting of the unity-amplitude Fourier harmonics $p=1$ and $p=-1$ incident onto the grating waveguide from the cover at $z=L$. The plots resolve in detail the avoiding-crossing resonance branches shown in Fig.~\ref{fullRCWA} near the frequency $\omega \approx 3.2\lambda_g/c$ and reveal the high-Q (the usual, solitary symmetry-protected BIC at \(k_x=0\)) and low-Q waveguide (1,2)-eigenmodes shown in Fig.~\ref{2Dpatterns}. 
The insert on the right shows the longitudinal x-profiles for the infinite-grating eigenmodes ($f_3(x)$, $f_2(x)$) constituting the high-Q (BIC) and low-Q (leaky) waveguide eigenmodes at $k_x \approx 0$, respectively. 
The grating waveguide is the same as in Fig.~\ref{fullRCWA}.}
\label{fig:SymProtBIC}
\end{figure}

Recognizing that the degenerate BIC is the extremely high-Q waveguide eigenmode associated with the anti-phased superposition $f_2(x)-f_3(x)$ of the second and third infinite-grating eigenmodes at $k_x = 0$, we entail a monochromatic source which drives this mode, at least, as strongly as the low-Q neighbour mode. Then, due to a higher Q factor, the degenerate BIC will acquire much larger amplitude than the low-Q mode and dominate in the resonance response despite spectral overlapping of the BIC (high-Q) and leaky (low-Q) waveguide eigenmodes. A commonly used choice of the source in the form of a single plane wave ($p=0$ Fourier harmonic) incident onto the grating layer from outside (say, substrate) is improper since such a source is decoupled from the BIC at $k_x \approx 0$ and predominantly excites only the leaky, low-Q waveguide eigenmode. Another drawback of the RCWA in Fig.~\ref{fig:paradox} is a popular choice of the intensity of that same, $p=0$ Fourier harmonic reflected from the grating layer as the response function. At the resonance with the BIC at $k_x \approx 0$, when the BIC gets decoupled from the radiation-loss channel, the reflection of the $p=0$ Fourier harmonic reaches maximum. Hence, the telltale feature of the BIC -- the disappearance of radiation of the $p=0$ Fourier harmonic -- becomes completely obscured.

Surely, with the proper choice of the source (say, two Fourier harmonics $p=1$ and $p=-1$ incident onto the grating layer from the cover at $z=L$) and the response function (say, the intensity of the $p=0$ Fourier harmonic outgoing from the grating layer into the substrate) the paradox of the RCWA in Fig.~\ref{fig:paradox} gets resolved as is shown in the top panel of Fig.~\ref{fig:DegenBIC}. We clearly see that the narrow spectral (dark blue) line of the resonant radiation losses of the high-Q waveguide eigenmode at the degenerate frequency $\omega_c = 4c/\lambda_g$ disappears in a very narrow vicinity of the zero $x$-wavenumber, $|k_x| \sim 10^{-4} k_g$. This fact definitely proves the appearance of the degenerate BIC. We see also that the wide but weaker spectral (moderately blue) stripe of the resonant radiation losses of the low-Q waveguide eigenmode extends over the BIC spectral line. Note that, for clarity of demonstrating the nature of the degenerate BIC, we intentionally choose the waveguide parameters in such a way that there is a slight frequency splitting of these crossing modes.

Thus, the degenerate BIC (high-Q) and leaky (low-Q) waveguide (2,2)-eigenmodes can be easily addressed separately since they constitute completely independent entities. They differ by the binary index ($s=0$ and $s=1$), have different (odd and even) parities of the field $x$-profile at $k_x = 0$, respectively, and strongly mismatched 2D spatial patterns (see sect. II and Fig.~\ref{2Dpatterns}). In particular, if the source excites just the superposition $f_2(x)-f_3(x)$ of the second and third infinite-grating eigenmodes at $z=0$ and one measures the sum of intensities of all infinite-grating eigenmodes at $z=L$, then an extremely strong (stronger than the source by many orders of magnitude) resonance response of the degenerate-BIC waveguide eigenmode is the only signal present in the RCWA output as is exemplified in the bottom panel of Fig.~\ref{fig:DegenBIC}. In this case the response from the leaky, low-Q waveguide eigenmode is absent since its $x$-profile $f_2(x)+f_3(x)$ is orthogonal to the source field distribution $f_2(x)-f_3(x)$ as per the right panel in Fig.~\ref{fig:DegenBIC}. Moreover, the degenerate BIC manifests itself by a strong response signal even at zero $x$-wavenumber $k_x=0$. As is explained in sections IV-VI, such behavior and restructuring of the BIC and leaky waveguide eigenmodes near the degenerate frequency occur due to parity symmetry breaking of the infinite-grating eigenmodes at mode crossing.

For comparison, we present in Fig.~\ref{fig:SymProtBIC} the similar numerical RCWA plots for the resonance response of the high-Q and low-Q waveguide (1,2)-eigenmodes (see the lower row in Fig.~\ref{2Dpatterns}) associated with the infinite-grating eigenmodes whose dispersion curves are far from the degenerate frequency, are separated by a relatively wide gap, and do not experience mode crossing. This example shows a conventional symmetry-protected BIC at $k_x=0$ on the upper branch of the resonance. In this case each of the high-Q (BIC) and low-Q (leaky) waveguide eigenmodes is primarily associated with just one infinite-grating eigenmode, $f_3(x)$ and $f_2(x)$, respectively. These infinite-grating eigenmodes have a definite, odd or even, $x$-profile parity and do not experience parity symmetry breaking.

Thus, the origin and all properties of the degenerate BIC are explained by taking into account the parity symmetry breaking and destructive interference of the infinite-grating eigenmodes whose dispersion curves intersect at the mode-crossing frequency.

\section{Conclusions}

We present the analytical eigenmode approach to the description of BICs in the grating waveguide based on the analytical theory of the infinite-grating eigenmodes \cite{PRA2019}. 
It includes the overview of the origin and interpretation of the BIC as a waveguide eigenmode (Figs.~\ref{2Dpatterns}, \ref{fullRCWA}) formed by a symbiosis of (a) the guided eigenmodes of a homogeneous slab waveguide, which consist of various plane waves in the substrate, slab and cover, (Fig.~\ref{slab-modes}) and (b) the partial infinite-grating eigenmodes (Fig.~\ref{Parabolas-extendedBrillouin}).
We focus on the presentation of the analytical eigenmode concept of BICs per se (not aiming at the most general theory, complex photonic-crystal structures or BICs of new types) as well as on the analysis of the particularly nontrivial case of the degenerate BIC (Fig.~\ref{fig:DegenBIC}).

We reveal a general mechanism for the formation of the degenerate BIC in a planar waveguide with a lamellar grating. The degenerate BIC is a complicated version of the conventional symmetry-protected BIC and emerges near the mode-crossing frequency $\omega_c$ where two infinite-grating eigenmodes have (i) intersecting dispersion curves and (ii) mixed, neither purely odd or even, in-plane spatial parity so that both infinite-grating eigenmodes are coupled to the zeroth spatial Fourier harmonic leaking out of the grating waveguide layer and into the cover and/or substrate. Nevertheless, in the entire region around such a mode-crossing point, the waveguide eigenmode dominated by those two coupled infinite-grating eigenmodes gets disconnected from the radiation-loss channel and constitutes a BIC of a very high Q factor. 

Even in the ideal case when the waveguide parameters are tuned precisely to the values making the eigenfrequency of the BIC waveguide eigenmode equal to the mode-crossing frequency, $\omega_m = \omega_c$, the BIC's leakage rate becomes zero $\omega_m'' \to 0$ in the limit of small in-plane wavenumbers $k_x \to 0$. Hence, the ideal BIC becomes completely decoupled from radiation waves, i.e., invisible from outside the waveguide, but at the same time it cannot be excited by continuum of waves incident from outside. (Of course, a laser or other source emitting waves which is located inside the grating layer will excite the field of the BIC eigenmode.) 
If the thickness of the grating layer and/or other waveguide parameters are detuned from those ideal values 
such that the eigenfrequency $\omega_m$ of the BIC waveguide eigenmode becomes detuned far away from the mode-crossing frequency  
$\omega_c$, then the degenerate BIC evolves into the conventional symmetry-protected BIC consisting, mainly, of just one infinite-grating eigenmode of purely odd spatial parity. 
The existence of the degenerate BIC, despite the parity symmetry breaking of two crossing infinite-grating eigenmodes, is due to the additional mechanism of BIC's formation entering the scene in a vicinity of the mode-crossing point -- the destructive interference of the coupled infinite-grating eigenmodes. 

If the in-plane wavenumber $k_x$ is increased, then the degenerate BIC acquires nonzero coupling with the radiation waves and an increased leakage rate $\omega_m'' \neq 0$. As a result, the degenerate BIC is characterized by a very narrow, sharp resonance with the Q factor scaling approximately as $1/k_x^2$ (Fig.~\ref{Fig:Q}). Such a perturbed, leaky BIC becomes visible and excitable by continuum of waves incident from outside the grating waveguide. This leaky BIC, as a usual damped harmonic oscillator, has a standard response to the excitation by an incident monochromatic wave in the form of the Lorentzian resonance curve. It should not be mixed with the BIC's hallmark - the steep resonant profile of the BIC Q factor shown in Fig.~\ref{Fig:Q}. The latter describes a certain internal property of the leaky BIC expressed by the function $Q(k_x) = \omega_m'/(2\omega_m'')$, and has nothing to do with the BIC's excitation by an external source of a variable frequency shown in Figs.~\ref{fullRCWA},\ref{fig:DegenBIC}.   

The existence of such a resonance phenomenon is explicitly evinced by numerical plots in Figs.~\ref{Fig:detPlot-Eigenmode branches}, \ref{Fig:detPlot-L} calculated via the analytical formulas for the degenerate BIC which is formed by the high-Q (2,2)-eigenmode shown in Fig.~\ref{2Dpatterns} and associated with the degenerate point $D_{2,3}^0$ (see Fig.~\ref{dispCurveMulti}). The parameters of the planar grating waveguide are chosen on the basis of a well-known optical material, titanium oxide (TiO$_2$). The sharp resonance is illustrated in Fig.~\ref{Fig:Q} by the Q-factor profile calculated within the $5\times 5$ approximation of the analytical waveguide-eigenmode theory. Remarkably, such a simple low-dimensional approximation (explained in the beginning of sect. III and after Eq.~(\ref{T})), which includes just five ($N=5$) first infinite-grating eigenmodes and five ($S=5$) spatial Fourier harmonics of the diffraction order $p =0, \pm1, \pm 2$, provides an accurate description of the degenerate BIC and main details of the mechanism of its formation. 
The point is that all of the higher-order infinite-grating eigenmodes and Fourier harmonics are evanescent and contribute very little to the selected BIC. Numerical calculations within the $7\times 7$ or $9\times 9$ approximation, which includes more infinite-grating eigenmodes and Fourier harmonics, do not lead to any appreciable corrections. 
Moreover, in the close vicinity of the mode-crossing point the main qualitative properties of the degenerate BIC are reproduced already within the approximation involving just those two infinite-grating eigenmodes whose dispersion curves intersect at the degenerate point. Also, we verified that the analytical results exemplified above for the specific 1D-grating waveguide, such as the BIC waveguide eigenmodes' dispersion curves and field patterns shown in Figs.~\ref{2Dpatterns}, \ref{fullRCWA}, \ref{fig:DegenBIC}, \ref{fig:SymProtBIC} are in excellent agreement with the {\it ad hoc} numerical calculations based on the widely used code of the rigorous coupled wave analysis (RCWA) \cite{RCWA}.

Various BICs and mechanisms of their formation in different photonic structures had been widely discussed in the literature (see, for example,
\cite{Yang2014,Quaranta2018,Wang1990,PRB2019,Hsu-Nature2016,Kildishev2021,Friedrich1985,Rybin2017,Sadrieva2017,Doeleman2018,Gomis-Nature2017,BulgakovPRA2017,Boyd2021} and references therein). The formation mechanism revealed in the present paper for the degenerate BIC in the grating waveguide shares features of most those mechanisms. The reasons for that are, first, an insight provided by the eigenmode approach (which addresses the ultimate physical entities living in the photonic system and, hence, is the most logical physically) and, second, a basic nature of the planar lamellar-grating waveguide. This waveguide is the simplest photonic structure (among nontrivial ones) which shares many common principal features, including the existence of BICs, with general photonic crystals, but still admits an analytical solution via eigenmodes and is just one further step of complication from the trivial case of the planar dielectric waveguide which has an elementary solution in terms of plane waves \cite{Kogelnik1975,Marcuse1991,Young2021} and does not support BICs.   

Below we briefly comment on these standard (marked below in {\it italic}) BIC formation mechanisms and related physical interpretations. Of course, the BIC is connected with {\it an interference} between partial modes of different structure or nature \cite{Feshbach1958,Friedrich1985,Hsu-Nature2016,PRB2019,Rybin2017,Sadrieva2017,Doeleman2018,Yang2014}. In the present case, those partial modes include two infinite-grating eigenmodes which possess the opposite, even and odd, in-plane spatial parity far from the degeneracy point. However, these modes experience the phenomenon of parity symmetry breaking and become of a mixed parity for the waveguide parameters corresponding to the vicinity of an intersection of their dispersion curves. Mutual reflection and interference of these two infinite-grating eigenmodes is responsible for the main features of the degenerate BIC. Besides, the interference due to Bragg diffraction on the grating, which is infinite in the $xy$-plane, is responsible for "discretization" of the radiation continuum into a discrete set of the potential radiation-loss channels (each of them being the leaking Fourier harmonic of the diffraction order $p = 0,\pm 1,\ldots$) as well as the very formation of the infinite-grating eigenmodes. In the end, the destructive interference of radiation for different partial waves constitutes a physical explanation for the BIC's existence and, in particular, is the ultimate reason for emission cancellation of the zeroth Fourier harmonic at the degenerate BIC resonance in the grating waveguide considered in the present paper. 

Similar to the conventional BICs in a photonic crystal slab at the center of the first Brillouin zone, for instance, the BIC shown in Fig.~\ref{fig:SymProtBIC}, the degenerate BIC in the grating waveguide at the origin of the momentum space, $k_x=0$, is {\it symmetry-protected} since it exists due to a certain spatial symmetry of the field profile within the BIC's waveguide eigenmode. Yet, the degenerate BIC can also be called {\it accidental} or {\it interference-based BIC through parameter tuning} since it is based on vanishing of the waveguide eigenmode coupling with the radiation continuum as per Fig.~\ref{fig:DegenBIC} due to destructive interference taking place at the values of the waveguide parameters specially tuned in such a way that the BIC eigenfrequency meets the mode-crossing frequency of the infinite grating. In other words, as is discussed in sect. V and illustrated in Fig.~\ref{Fig:detPlot-Eigenmode branches}, formation of the degenerate BIC is achieved through {\it parameter tuning} within the system of coupled partial infinite-grating eigenmodes and such a BIC is known as {\it an accidental BIC}. At the same time, from the waveguide-eigenmode point of view, the degenerate BIC appears as a so-called {\it single-resonance parametric BIC}. 

Moreover, since the degenerate BIC as the waveguide eigenmode is formed out of more than one partial infinite-grating eigenmodes via their coupling, its mechanism is reminiscent of {\it the Friedrich–Wintgen mechanism of the BIC formation} \cite{Friedrich1985} that also acts near frequency crossings of the uncoupled resonances. A Fabry–Pérot BIC, originated from spatially separate cavity modes coupled through a semitransparent mirror, is so special that it is not directly relevant to the grating-waveguide BIC.

Note also that some analogy can be traced with the BICs in the conventional open cavities and waveguides \cite{Sadreev2020}, outside the realm of photonic crystals. For instance, {\it the accidental or single-resonance parametric BIC} can exist in the open Sinai billiard in which a cavity eigenmode gets decoupled from the continuum of the waveguide by an appropriate smooth adjusting of the position of the intracavity disc. In the open cavities, the BIC frequency is also located in a vicinity to the degenerate-point frequency of eigenmodes of the closed cavity \cite{Sadreev2020,Sadreev2006}.

There is a principle feature of the degenerate BIC clearly seen in Figs.~\ref{2Dpatterns}, \ref{Fig:Field distribution m=1}. The BIC's field slightly extends out from the grating layer into the cover and substrate because of the coupling with the evanescent Fourier harmonics due to boundary conditions. Such coupling with the evanescent waves is crucially important for the very existence of the BIC. 

The presence of the degenerate, mode-crossing points $D_{n,n+1}^l$ (see Fig.~\ref{dispCurveMulti}) and the corresponding spatial parity, Eqs.~(\ref{kd=mpi}), (\ref{kz2=kz3}), is a topological, robust property associated with the map of the solutions to the characteristic equation (\ref{CharactEq}) over the space of the waveguide parameters and frequency. The existence and qualitative properties of the infinite-grating eigenmodes as well as the topology of the intersection of their dispersion curves predetermine how perturbations in the grating, cover, substrate and waveguide parameters affect the BIC. Such perturbations continuously, smoothly modify, but do not destroy neither the eigenmodes and mode-crossing points nor the BIC waveguide eigenmode (see Fig.~\ref{fullRCWA}). In that sense, {\it the degenerate BIC is topologically protected} (cf. \cite{Hsu-Nature2016}). 

The BIC shows a very sharp (compared to that in a conventional optical cavity) increase in the Q factor that ideally tends to infinity in a vicinity of the mode-crossing frequency, Fig.~\ref{Fig:Q}. However, such an ideal behavior implies a 1D or 2D infinite size of the photonic structure that should be reminiscent of an infinite diffraction lattice discretizing the radiation continuum $\omega = ck/\sqrt{\varepsilon^{\pm}}$ into the diffraction orders (spatial Fourier harmonics). There is a rigorous theorem \cite{Colton1998} that forbids appearance of a BIC in a finite-size dielectric structure. 
In reality, the Q factor is saturated at some maximum value that is determined by finite-size effects, nonzero in-plane wavenumber, structural disorder, fabrication imperfections, defects, impurities and losses in the waveguide materials, roughness of the layers' surfaces, and similar scattering and dissipation effects. They say that the ideal BIC actually manifests itself as {\it a quasi-BIC} due to the above factors \cite{Hsu-Nature2016,Kildishev2021,Rybin2017,Sadrieva2017,Yang2014}.

A general analysis of the emergence of degenerate BICs via the mechanism of the destructive interference and parity symmetry breaking due to mode crossing as well as an analysis of the entire spectrum of all BICs and leaky resonances in the grating waveguide and similar photonic structures can be done along the lines sketched in sect. II (see Figs.~\ref{Parabolas-extendedBrillouin}, \ref{2Dpatterns}, \ref{fullRCWA}) and will be given elsewhere. It is based on the hierarchy of the eigenmode-crossing points $D_{n,n+1}^l$ in the infinite grating illustrated in Fig.~\ref{dispCurveMulti}. 
In particular, the row of crossing points $D_{2,3}^0, D_{4,5}^0, D_{6,7}^0$, $D_{8,9}^0,\ldots$ corresponds to the intersections $D_{2m_x,2m_x+1}^0$ between the dispersion curves of the $n=2m_x$ and $n+1=2m_x+1$ infinite-grating eigenmodes of the $x$-spatial parity of the order $(m_{x1},m_{x2}) = (m_x,m_x),\, m_x= 1, 2,\ldots$, introduced in Eq.~(\ref{kd=mpi}). All such mode-crossing points have the same $z$-wavenumber (see Eq.~(\ref{cpkz})) which is exactly the one emerged in Figs.~\ref{dispCurveMulti} and \ref{Fig:Dispersion}, $(k_{zn}c/\omega_c)^2 \approx 2.232$. A sequence of the crossing points $D_{n,n+1}^l$ for a given pair of the intersecting infinite-grating eigenmodes $n=2m_x$ and $n+1=2m_x+1$ is finite and runs over the $x$-spatial parities of order $(m_{x1},m_{x2}) = (m_x+l, m_x-l);\, l= 0, 1,\ldots, m_x-1$. For example, the series $D_{6,7}^0, D_{6,7}^1, D_{6,7}^2$ in Fig.~\ref{dispCurveMulti} contains three crossing points of spatial parities $(m_{x1},m_{x2}) = (3,3), (4,2), (5,1)$ corresponding to phase accumulations (\ref{kd=mpi}) over the grating sections $d_1, d_2$ equal to $(3\pi,3\pi), (4\pi,2\pi), (5\pi,\pi)$, respectively.  

To demonstrate the degenerate BIC within the model of the planar grating waveguide studied in the present paper, one has to specify the values of eight physical quantities. They include four parameters of the lamellar grating (the high, $\varepsilon_1$, and low, $\varepsilon_2$, dielectric constants of the two grating sections of a length $d_1$ and $d_2$, the grating period $\lambda_g = d_1 + d_2$, and the fill factor of the high dielectric constant section, $\rho = d_1/\lambda_g = 1-d_2/\lambda_g$), three parameters of the waveguide (the thickness of the grating layer $L$ and the dielectric constants of the cover, $\varepsilon^+$, and substrate, $\varepsilon^-$), and the resonant frequency $\omega_m$. The relevant design procedure is explained in sect. V. 

The general conclusion is that the planar grating waveguide supports degenerate BICs for a very wide range of parameters if three simple, physically logical conditions are met. They are related to (i) the emergence of the parity symmetry breaking due to mode crossing at the corresponding mode-crossing frequency $\omega_c$, (ii) turning off the radiation-loss channels via undue spatial Fourier harmonics by making them evanescent, and (iii) choosing the waveguide parameters ensuring an existence of the solution to the characteristic equation (\ref{det}) for the high-Q waveguide-eigenmode frequency close to the critical frequency $\omega_c$ of a mode-crossing point $D_{n,n+1}^l$. 

In the case of the BIC at the mode-crossing point $D_{2,3}^0$, those conditions are explained in sect. V. The first of them requires the existence of crossing of the dispersion curves of two neighboring, $n=2$ and $n=3$, infinite-grating eigenmodes with the opposite, even and odd, respectively, spatial parities. In fact, the existence of crossings is a generic property of the solutions to the characteristic equation (\ref{CharactEq}) for the infinite-grating eigenmodes as is illustrated in Fig.~\ref{dispCurveMulti}. 
The second condition, stated in Eq.~(\ref{omega-crit}), limits the number of radiation-loss channels to just one provided by the zeroth spatial Fourier harmonic propagating into the cover and substrate.
The only nontrivial, third condition is the existence of the solution to the characteristic equation (\ref{det}), $\det\,[M(\omega'_m - i\omega''_m) - \mathbbm{1}] = 0$, for the waveguide-eigenmode frequency $\omega_m = \omega_m' - i\omega_m''$ with a very small decay rate $\omega_m'' \ll \omega'_m \approx \omega_c$. The solution can be conveniently traced by plotting the contour plots such as in Fig.~\ref{Fig:detPlot-Eigenmode branches}. The position of the degenerate BIC eigenfrequency is known in advance since it coincides with the critical frequency $\omega_c$ of the mode-crossing point $D_{2,3}^0$ given by Eq.~(\ref{cpkz}). 

Thus, most of the eight grating and waveguide parameters remain free. For implementing degenerate BICs in more complex photonic waveguides, crystals or structures the number of free parameters could be even larger. 

Anyway, understanding of the degenerate BIC formation mechanism described above and the presence of such free parameters could help targeting various applications of the BIC phenomenon in laser cavities \cite{Rybin2017,Kodigala2017}, sensors \cite{Liu2021,Mesli2021,Maksimov2020,Zhou2019,ZhouOptExpress2019}, filters \cite{Wang1993,Sang2007,Magnusson2015,Foley2015,Zhao2022,Zong2022}, polarizers \cite{Magnusson2004}, reflectors \cite{Mateus2004,Chang2012,Magnusson2014,Moitra2015,Yu2015}, photodetectors \cite{Zhu2014}, spectroscopy, precision measurement and control in photonics and metamaterial fabrication technologies, optical communications, e.g., low-loss fibres, etc. For a discussion of such applications, we refer to a recent review \cite{Kildishev2021} and references therein. 
\vspace{3mm}

\section*{Acknowledgements}
We acknowledge support from both the William Robba Graduate Study Scholarship and the Texas A$\&$M Triads for Transformation program, Round Four T3 project 1750, for C. B. Reynolds.

\end{document}